\documentclass[10pt,preprint,prd,aps,superscriptaddress,nofootinbib,showpacs,a4paper]{revtex4}
\usepackage{color}
\usepackage{graphicx}
\usepackage{epsfig}
\usepackage{amsmath}
\usepackage{amssymb}
\newcommand{\noi}{\noindent}
\newcommand{\be}{\begin{equation}}
\newcommand{\ee}{\end{equation}}
\newcommand{\bqa}{\begin{eqnarray}}
\newcommand{\eqa}{\end{eqnarray}}

\begin{document}

\title{Dalitz plot studies of  \mbox{\boldmath $D^0 \to K^0_S \pi^+ \pi^-$} decays in a factorization approach}

\author{\bf {J.-P.~Dedonder}}
\affiliation{Sorbonne Universit\'es, Universit\'e Pierre et Marie Curie, Sorbonne Paris Cit\'e, Universit\'e
 Paris Diderot, et IN2P3-CNRS, UMR 7585, Laboratoire de Physique Nucl\'eaire et de Hautes \'Energies,  
4 place Jussieu, 75252 Paris, France}

\author{\bf {R.~Kami\'nski}}
\affiliation{Division of Theoretical Physics, The Henryk Niewodnicza\'nski Institute of Nuclear Physics,
                  Polish Academy of Sciences, 31-342 Krak\'ow, Poland\\}
 \author{\bf {L.~Le\'sniak$^2$}}

 \author{\bf {B.~Loiseau$^1$}}


\date{\today }

\begin{abstract}
\noindent

The presently available high-statistics data of the $D^0 \to K^0_S \pi^+ \pi^-$ processes measured by 
the Belle and BABAR Collaborations are analyzed within a quasi two-body factorization framework.
Starting from the weak effective Hamiltonian, tree and annihilation amplitudes build up the $D^0 \to K^0_S \pi^+ \pi^-$ decay amplitude. 
Two of the three final-state mesons are assumed to form a single scalar, vector or tensor state originating
 from a quark-antiquark pair so that the factorization hypothesis can be applied.
The meson-meson final state interactions are described by $K\pi$ and $\pi \pi$ scalar and vector form 
factors for the $S$ and $P$ waves and by relativistic Breit-Wigner formulae for the $D$ waves.
A combined $\chi^2$ fit to a Belle Dalitz plot density distribution, to the total experimental branching fraction and to the $\tau^- \to K^0_S \pi^- \nu_\tau$ decay data is carried out to fix the 33 free parameters.
These are mainly related to the strengths of the scalar form factors and to unknown meson to meson 
transition form factors at a large momentum transfer squared equal to the $D^0$ mass squared.
A good overall agreement to the Belle Dalitz plot density distribution is achieved.
Another set of parameters fits equally well the BABAR Collaboration Dalitz plot model. The parameters of both fits 
are close, following from similar Dalitz density distribution data for both collaborations.
The corresponding  one-dimensional effective mass distributions display the contributions of the ten quasi 
two-body channels entering our $D^0 \to K^0_S \pi^+ \pi^-$ decay amplitude.
The branching fractions of the dominant channels compare well with those of the isobar Belle or BABAR models.
The lower-limit values of the branching fractions of the annihilation amplitudes 
are significant.
Built upon experimental data from other processes, the unitary $K \pi$ and $\pi\pi$ scalar form factors, entering our decay amplitude and satisfying analyticity and chiral symmetry constraints, 
are furthermore constrained by the present Dalitz plot analysis.
Our $D^0 \to K^0_S \pi^+ \pi^-$ decay amplitude could be a useful input for determinations of
$D^0$-$\overline{D}^0$ mixing parameters and of the CKM angle $\gamma$ (or $\phi_3$).

\pacs{13.25.Hw, 13.75.Lb}
\end{abstract}

\maketitle

\section{Introduction}
\label{Introduction}

Measurements of the $D^0$-$\overline{D}^0$ mixing parameters, through Dalitz-plot time dependent amplitude 
analyses of the the weak process $D^0 \to K^0_S \pi^+ \pi^-$, have been performed by 
the Belle~\cite{ZhangPRL99_131803} and BABAR~\cite{SanchezPRL105_081803} Collaborations.
These studies could help in the understanding of the origin of mixing and may indicate the possible presence of new physics contribution.
No $CP$  violation  in these $D^0$ decays~\cite{D.M.AsnerPRD70_0911018CLEO, T.Aaltonen_CDF2012} has yet been found, in agreement with the very small value predicted by the standard model in the charm sector.
The Cabibbo-Kobayashi-Maskawa, CKM, angle $\gamma$ (or $\phi_3$) has been evaluated from the analyses of the  $B^\pm \to D^0 K^\pm, D^0 \to K^0_S \pi^+ \pi^-$ decays~\cite{J.Libby_PRD82_CLEO, R.Aaij_LHCb2012, H.Aihara_Belle2012, SanchezPRL105121801, J.P.Lees_PRD87_052015_BABAR, A.Poluektov_PRD81_112002_Belle}. 
A good knowledge of the final state meson interactions is important to reduce the uncertainties in the determination of  the $D^0$-$\overline{D}^0$ mixing parameters and of the angle $\gamma$.
The very rich structures seen in the Dalitz plot spectra point out to  the complexity of these final state strong interactions.

The experimental analyses~\cite{ZhangPRL99_131803,SanchezPRL105_081803} rely mainly on the use of the isobar model.  In this approach one can take into account the many existing resonances coupled to the interacting pairs of mesons. 
However, the corresponding decay amplitudes are not unitary and unitarity is not preserved in the three-body decay channels; it is also violated in the two-body sub-channels.
Furthermore, it is  difficult  to differentiate the $S$-wave amplitudes from the non-resonant background terms. 
Their interferences are 
noteworthy and then some two-body branching fractions, extracted from the data, could be unreliable.
The isobar model is tractable but it has many free parameters: at least two fitted parameters
for each amplitude and for example, 
the Belle Collaboration in Ref.~\cite{ZhangPRL99_131803} has used 
40 fitted parameters  and BABAR Collaboration 43  in Ref.~\cite{SanchezPRL105_081803}.

Imposing unitarity for three-body strong interactions in a wide range of
meson-meson effective masses is difficult. 
Some three-body unitarity corrections have been evaluated in Ref.~\cite{KamanoPRD84} for 
$D^0 \to \pi^0\pi^+ \pi^-$ decays  and in Ref.~\cite{MagalhaesPRD84} for $D^+\rightarrow K^- \pi^+ \pi^+$.
In a unitary coupled-channel model Ref.~\cite{KamanoPRD84} has shown that two-body rescattering terms could
 be important.
They find that the decay amplitudes of the unitary model can be rather different from those of the isobar
 model.
In Ref.~\cite{MagalhaesPRD84} the three-body unitarity is formulated with an integral equation inspired by 
the Faddeev formalism. 
There,  they sum up a perturbation series and find that three-body effects important close to threshold 
fade away at higher energies.
In the present work, as a first step, we  require two-body unitarity in the $D$-decay amplitudes with 
$K^0_S \pi^{\pm}$ final state in $S$ wave and with the $\pi^+\pi^-$ final state in $S$ and $P$ waves.
According to the experimental works~\cite{ZhangPRL99_131803,SanchezPRL105_081803}, the sum of the branching
 fractions corresponding to these amplitudes yields an important part of the total branching fraction of 
the $D^0\rightarrow K^0_S \pi^+ \pi^-$ decay.

The two-body QCD factorization has been applied with success to charmless nonleptonic $B$ decays~
(see \textit{e.g.} Ref.~\cite{Beneke2003}).
For the $D$ meson the charm quark mass $m_c$ is lighter than the bottom quark mass by about a factor of three.
The $c$ quark mass is too high to apply chiral perturbation theory and too light to use heavy quark expansion approaches.
One expects nonperturbative $D$-decay contributions of order $\Lambda_{QCD}/m_c$ to be more important than in $B$ decays.
Consequently the factorization hypothesis could be less reliable.
Nevertheless, following the initial articles of Bauer, Stech and Wirbel~\cite{Wirbel1985, Bauer1987} the assumption of factorization has been applied successfully to $D$ decays in several recent papers~\cite{BoitoPRD79_034020, El-Bennich_PRD79, BoitoPRD80_054007, Fu-Sheng_Yu_PRD84_074019}.
The Wilson coefficients are treated as phenomenological parameters to 
 account for possible important non-factorizable corrections~\cite{BurasNPB434_606}. An alternative diagrammatic approach  for the description of  hadronic charmed meson decays into two body has been the support of the works presented in Refs.~\cite{ChengPRD81_074021} and~\cite{ChengPRD81_074031}. 

In the framework of the quasi two-body factorization approximation~\cite{Beneke2003} and 
of the extension of a program devoted to the understanding of rare three-body $B$ decays~\cite{fkll,El-Bennich2006,Bppk, Leitner_PRD81,DedonderPol} we analyze the presently available $D^0 \to K^0_S \pi^+ \pi^-$ data.
So far no factorization scheme has been worked out for three-body decays.
Then, as in our previous studies, we assume that two of the three final-state mesons forms a single state which originates from a quark-antiquark, $q\bar{q}$, pair.
Such an hypothesis leads to quasi two-body final states to which the factorization procedure is applied.
The three-meson final states $K^0_S \pi^+ \pi^-$ are here supposed to  
be formed by the following quasi two-body pairs, $[K^0_S \pi^+]_{L} \  
\pi^-$, $[K^0_S \pi^-]_{L} \ \pi^+$ and $K^0_S \ [\pi^+{\pi^-}]_{L}$ where
two of the three mesons form a state in $L= S, P$ or $D$ wave.
The $D^0 \to K^0_S \pi^+ \pi^-$ decay amplitudes, derived from the weak effective Hamiltonian, have contributions from tree diagrams  but none from penguin or $W$-loop diagrams.
There are also annihilation amplitudes arising from $W$-meson exchange between the $D^0$ quark constituents.
The amplitudes corresponding to the $c \to su\bar d$ transition are Cabibbo favored (CF) while those with $c \to du\bar s$ are doubly Cabibbo suppressed (DCS).

In the factorization approach, the CF and DCS amplitudes are expressed as superpositions of appropriate effective
 coefficients and two products of two transition matrix elements.
For the CF tree amplitudes, the first and second product correspond to the transition matrix element between  the $D^0$ and $[\overline {K}^0 \pi^-]_{L}$ or $[\pi^+ \pi^-]_{L}$ state multiplied by 
the transition matrix element between the
 vacuum and the $\pi^+$ (proportional to the pion decay constant) or the $\overline {K}^0$ (proportional to the kaon decay constant), respectively.
For the DCS tree amplitude these products correspond to the transition between  
the $D^0$ and $\pi^-$ or $[\pi^- \pi^+]_{L}$ state multiplied by the transition between the
 vacuum and the $[K^0 \pi^+]_L$ (proportional to the kaon-pion form factor) or the ${K}^0$ (proportional to the kaon decay constant), respectively. In the latter case, in the $K^0 \pi$ center of mass frame, the bilinear quark current involved forces the $[K^0 \pi^+]$ pair to be in a $L=S$ or $P$ wave.
For the CF (DCS) annihilation amplitudes the products correspond to the transition between  the $\pi$ or $\overline {K}^0$(${K}^0$) and $[\overline {K}^0 \pi^-]_{L}$($[{K}^0 \pi^+]_{L}$) or $[\pi^+ \pi^-]_{L}$ state, multiplied by the transition between the
 vacuum and the $D^0$ (proportional to the $D^0$ decay constant), respectively.

 We presume that the transition of the $D^0$  to the meson pairs $[\overline {K}^0 \pi^-]_{L}$ or $[\pi^+ \pi^-]_{L}$ goes first through the dominant intermediate resonance $R_L$ of these  pairs.
For the $[\overline {K}^0 \pi^-]_{L}$ pair, we take, $R_S[\overline {K}^0 \pi^-]=K^*_0(1430)^-$, $R_P[\overline {K}^0 \pi^-]=K^*(892)^-$, $R_D[\overline {K}^0 \pi^-]=K^*_2(1430)^-$ and for the $[\pi^+ \pi^-]_{L}$ pair, $R_S[\pi^+ \pi^-]=f_0(980)$, $R_P[\pi^+ \pi^-]=\rho(770)^0$ and $R_D[\pi^+ \pi^-]=f_2(1270)$. 
We further calculate the $D^{0} \to \overline {K}^0 \pi^-$ or $\pi^+ \pi^-$ matrix elements as products of the  $D^0 \to R_L[\overline {K}^0 \pi^-]$ or $R_L[\pi^+ \pi^-]$  transition form factors by the relevant vertex function describing the decay of the $[\overline {K}^0 \pi]_{L}$ or $[\pi \pi]_{L}$ states into the final meson pair.
The vertex functions are in turn expected to be proportional to the kaon-pion or pion scalar form factor  for the $S$ wave, to the vector form factor for the $P$ wave and to a relativistic Breit-Wigner formula for the $D$ wave.
For the CF (DCS) annihilation amplitudes we follow the same steps as for the tree amplitudes but for the replacement of $D^0$ by $\pi$ or $\overline {K}^0$(${K}^0$). 

The meson-meson final state interactions for the $S$ and $P$ waves are then described in terms of 
experimentally and theoretically constrained $K\pi$ and $\pi \pi$ scalar and vector form factors.
Using unitarity, analyticity and chiral symmetry constraints, the scalar form factors have been been derived
 in Ref.~\cite{Bppk} for the  
$K\pi$ case and in Ref.~\cite{DedonderPol} for the pion one.
They are single unitary functions describing the two scalar resonances $K^*_0(800)$ (or $\kappa$),
 $K^*_0(1430)$  and the three scalar resonances, $f_0(500)$, $f_0(980)$ and $f_0(1400)$ present in the $K^0_S \pi^\pm$ and  $\pi^+ \pi^-$ interactions, respectively.
The vector form factors are 
based on the Belle analyses of the  $\tau^- \to K^0_S \pi^- \nu_\tau$ ~\cite{EpifanovPLB654} and of the  $\tau^- \to  \pi^- \pi^0 \nu_\tau$~\cite{Belletau2008} decay processes. 
We also include the amplitude describing the $D^0 \to \omega(782) K^0_S$ channel followed by  the $\omega(782) \to \pi^+\pi^-$ decay. 
Relativistic Breit-Wigner formulae are introduced to describe the final state $D$ wave meson-meson interactions. 
The undetermined parameters of our $D^0 \to K^0_S \pi^+ \pi^-$ decay amplitudes, mainly related to the
 strength of the $[{K} \pi]_{S}$ and $[\pi \pi]_{S}$ scalar form factors and to the unknown meson to
 meson transition form factors, are obtained through a $\chi^2$ fit to the Dalitz plot data sample of 
the 2010 Belle Collaboration analysis~\cite{A.Poluektov_PRD81_112002_Belle,A.Poluektov_private2013}.
We also fit the Dalitz plot density of the BABAR Collaboration  
model~\cite{F.Martinez_private2013}. \\

The paper is structured as follows. Section II describes formally  the amplitudes calculated in the 
framework 
 of the quasi two-body factorization approach. Section III provides a practical formulation of these amplitudes by
 introducing combinations of some of them more amenable to numerical calculations. A discussion of the
 branching fractions is also given there. Section IV lists the necessary input for the evaluation
 of the amplitudes.  Results are presented and discussed in Section V while Section VI summarizes the
 outcome of this analysis and proposes some conclusions and perspectives.

\section{The $D^0 \to K^0_S \pi^+ \pi^-$ decay amplitudes in factorization framework}
\label{amplitudes}

The decay amplitudes for  the $D^0 \to K^0_S \pi^+ \pi^-$ process 
 can be evaluated as matrix elements of the effective weak Hamiltonian \cite{Buchalla1996}

\be \label{Heff}
H_{eff}=\frac{G_F}{\sqrt{2}} V_{CKM} \Big[ C_1(\mu) O_1(\mu)+ C_2(\mu) O_2(\mu) \Big] + h.c., 
\ee
 where the coefficients $V_{CKM}$ are given in terms of Cabibbo-Kobayashi-Maskawa quark-mixing matrix 
elements  and $G_F$ denotes the Fermi coupling constant. The $C_i(\mu)$ are the Wilson coefficients of 
the  four-quark operators  $O_i(\mu)$ at a renormalization scale $\mu$, chosen to be equal to the 
$c$-quark mass $m_c$. The left-handed current-current operators $O_{1,2}(\mu)$ arise from $W$-boson exchange.

The transition matrix elements that occur in the present work require two specific values of  the $V_{CKM}$ coupling matrix elements:
 \be
\label{lambda}
\Lambda_1\equiv V^*_{cs} V_{ud}\hspace{1cm}{\rm and}\hspace{1cm} \Lambda_2\equiv V^*_{cd} V_{us}.
\ee

\noi The amplitudes are functions of the Mandelstam invariants 
 \be \label{3sa}
 s_{\pm}= m_{\pm}^2 =(p_0+p_\pm)^2,\hspace{2cm}  s_{0}=m_0^2=(p_+ + p_-)^2,
 \ee
\noi where $p_0$, $p_+$ and $p_-$ are the four-momenta of the $K_S^0$, $\pi^+$ and $\pi^-$ mesons, 
respectively. Energy-momentum conservation implies  
\be \label{pD0}
 p_{D^0}=p_0+p_++p_-\hspace{1cm}
{\rm and} \hspace{1cm}
s_0+s_++s_-=m_{D^0}^2+m_{K^0}^2 + 2 m_\pi^2,\ee
where $p_{D^0}$ is the $D^0$ four-momentum and  $m_{D^0}$, $m_{K^0}$ and $m_\pi$ denote the  $D^0$, $K^0$  and charged pion masses.

The full amplitude is the superposition of two tree Cabibbo favored and doubly Cabibbo suppressed amplitudes, $T^{CF}(s_0,s_-,s_+)$ and $T^{DCS}(s_0,s_-,s_+)$ and of two annihilation (i.e., exchange of W meson between the $c$ and $\overline  u$ quarks of the $D^0$) CF and DCS amplitudes, $A^{CF}(s_0,s_-,s_+)$ and $ A^{DCS}(s_0,s_-,s_+)$. Thus,  one writes the full amplitude as
\bqa
\label{fulamp}
\mathcal{M}(s_0,s_-,s_+)&=&
\left \langle K^0_S(p_0)\ \pi^+(p_+){\pi^-}(p_-) \vert H_{eff}\vert D^0(p_{D^0}) \right \rangle \nonumber \\
&=& T^{CF}(s_0,s_-,s_+) + T^{DCS}(s_0,s_-,s_+)+ A^{CF}(s_0,s_-,s_+) + A^{DCS}(s_0,s_-,s_+),
\eqa
where the CF amplitudes are proportional to $\Lambda_1$ and the DCS ones  to $\Lambda_2$.
Although the three variables $s_0,s_-,s_+$ appear as arguments of the amplitudes,
because of the relation (\ref{pD0}) all amplitudes depend only on two of them.

\begin{figure}[h] 
\begin{center}
\includegraphics[scale = 0.6]{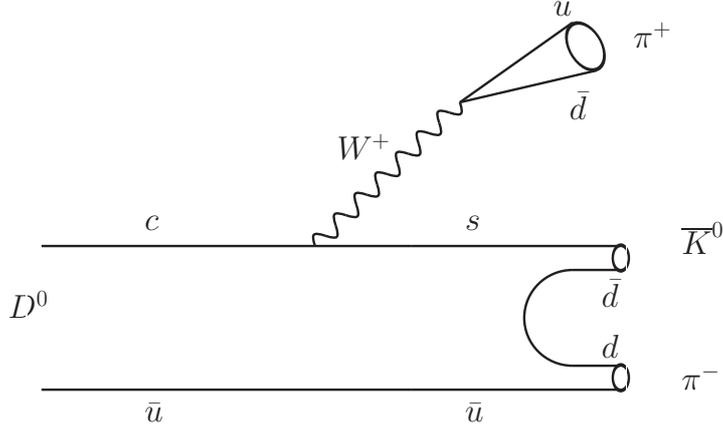}
\caption{ Tree diagram for Cabibbo favored amplitudes  with  
$\left [\overline{K}^0\pi^-\right ]\pi^+$ final states.}\label{K2}
\end{center}
\end{figure}

\begin{figure}[h] 
\begin{center}
\includegraphics[scale = 0.6]{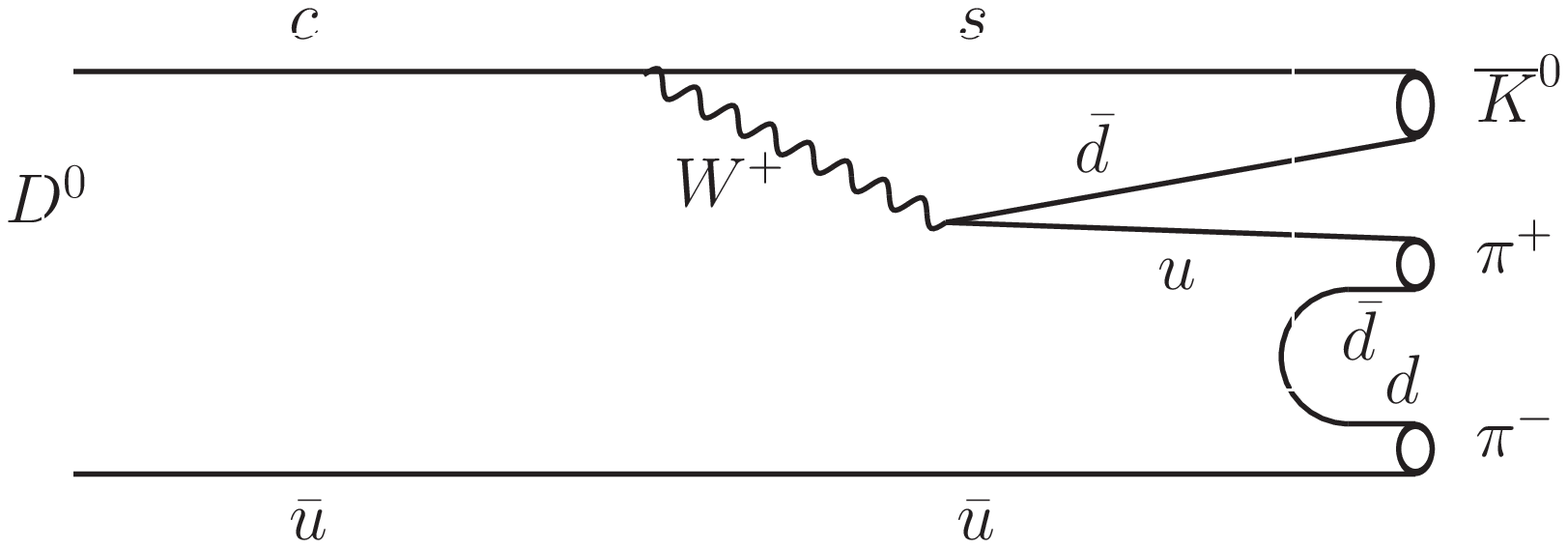}
\caption{ As in Fig. \ref{K2} but for  $\overline{K}^0\left [\pi^+\pi^-\right ]$ final states.}
\label{f2}
\end{center}
\end{figure}

Assuming that the factorization approach \cite{Beneke2003,BurasNPB434_606,Buchalla1996,AliPRD58} 
with quasi two-body $[K \pi]_{L} \pi$ or $K [\pi \pi]_{L}, L=S, P, D$, states  holds, the tree CF amplitudes  
read, with $\vert 0 \rangle$ indicating the vacuum state, 
\begin{eqnarray}
\label{TCF}
T^{CF}(s_0,s_-,s_+)
&\simeq & \frac{G_F}{2}\ \Lambda_1 \sum_{L={S,P,D}} \Big [ a_1(m_c) 
 \langle  [\overline K^0 (p_0){\pi^-}(p_-) ]_L \vert (\overline  s \ c)_{V-A}\vert D^0(p_{D^0})\rangle
   \nonumber \\ 
& & \cdot \ \langle \pi^+(p_+)\vert (\overline  u \ d)_{V-A}\vert 0  \rangle 
 +a_2(m_c) \langle \overline K^0(p_0)\vert (\overline  s \ d)_{V-A} \vert 0 \rangle  \nonumber \\
& & \cdot \ \langle  [\pi^+(p_+){\pi^-}(p_-) ]_L \vert (\overline  u \ c)_{V-A} \vert D^0(p_{D^0})\rangle \Big ] \nonumber \\ 
&=&\sum_{L={S,P,D}} T^{CF}_{[\overline{K}^0\pi^-]_L\pi^+}(s_0,s_-,s_+)
 +  \sum_{L={S,P,D}} T^{CF}_{\overline{K}^0 [\pi^+\pi^-]_L }(s_0,s_-,s_+)\nonumber \\
&=&T^{CF}_{[\overline{K}^0\pi^-]\pi^+}(s_0,s_-,s_+) + T^{CF}_{\overline{K}^0[\pi^+\pi^-]}(s_0,s_-,s_+).
\end{eqnarray}
 In deriving Eq. (\ref{TCF}) small $CP$ violation effects in $K^0_S$ decays are neglected and we use 
 \be
\vert K^0_S \rangle \approx \frac{1}{\sqrt{2}}\  \left (\vert K^0 \rangle+ \vert \overline{K}^0 \rangle \right ).
\ee
\noi At leading order in the strong coupling constant $\alpha_S$, the effective QCD factorization 
coefficients $a_1(m_c)$ and $a_2(m_c)$ are expressed as 
 \be
 \label{a12}
 a_1(m_c)=C_1(m_c)+\frac{C_{2}(m_c)}{N_C},\hspace{1cm} a_2(m_c)=C_2(m_c)+\frac{C_{1}(m_c)}{N_C},
 \ee
 where  $N_C=3$ is the number of colors. 
Higher order vertex and hard scattering corrections are not discussed in the present work and
we introduce effective values for these coefficients (see Sec. \ref{input}). From now on, the 
simplified notation 
 $a_1 \equiv  a_1(m_c)$ and $a_2 \equiv  a_2(m_c)$ will be used.
 In Eq. (\ref{TCF}), we have introduced the short-hand notation 
\be
(\overline {q}\ q)_{V-A}=\overline {q} \gamma\ (1 -\gamma_5)\ q 
\ee
which will be used throughout the text.
\noi The amplitudes  $T^{CF}_{[\overline{K}^0\pi^-]\ \pi^+}(s_0,s_-,s_+)$ and  $T^{CF}_{\overline{K}^0\ [\pi^+\pi^-]}(s_0,s_-,s_+)$ are illustrated diagrammatically  in
 Figs. \ref{K2} and  \ref{f2}. 

\vspace{.8cm}

\vspace{.8cm}

\begin{figure}[h] 
\begin{center}
\includegraphics[scale = 0.6]{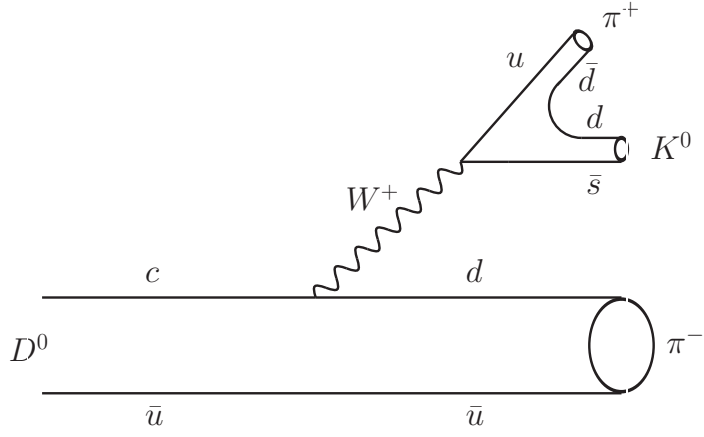}
\caption{Tree diagram for the doubly Cabibbo suppressed amplitude with   
$\left [{K}^0\pi^+\right ]\pi^-$ final states.}\label{diag1}
\end{center}
\end{figure}

\begin{figure}[h] 
\begin{center}
\includegraphics[scale = 0.6]{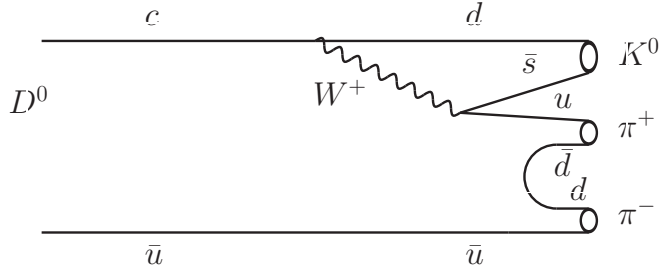}
\caption{ As in Fig. \ref{diag1} but for ${K}^0\left [\pi^+\pi^-\right ]$ final states.}\label{diag2}
\end{center}
\end{figure}

Similarly, the DCS tree amplitudes,  illustrated by the diagrams shown in Figs. \ref{diag1} and \ref{diag2}, read 

\bqa
\label{TDCS}
T^{DCS}(s_0,s_-,s_+) 
&\simeq & \frac{G_F}{2}\ \Lambda_2 \sum_{L={S,P,D}}  \Big [ a_1  \langle [K^0 (p_0) 
\pi^+(p_+) ]_L\vert (\overline  u \ s)_{V-A}\vert 0   \rangle  \nonumber \\
& & \cdot \ \langle \pi^-(p_-) )\vert (\overline  d \ c)_{V-A}\vert D^0(p_{D^0}) \rangle 
 +  a_2 \langle K^0(p_0)\vert (\overline  d \ s)_{V-A} \vert 0  \rangle \nonumber \\
& &\cdot \ \langle
   [\pi^+(p_+)\pi^-(p_-) ]_L \vert (\overline  u \ c)_{V-A} \vert D^0(p_{D^0}) \rangle \Big ] \nonumber \\ 
  &=&\sum_{L={S,P,D}} T^{DCS}_{[{K}^0\pi^+]_L\pi^-}(s_0,s_-,s_+)
 + \sum_{L={S,P,D}} T^{DCS}_{{K}^0[\pi^-\pi^+]_L}(s_0,s_-,s_+)\nonumber \\
&=&  T^{DCS}_{[{K}^0\pi^+] \pi^-}(s_0,s_-,s_+) +  T^{DCS}_{{K}^0 [\pi^-\pi^+]}(s_0,s_-,s_+). 
\eqa

\begin{figure}[h] 
\begin{center}
\includegraphics[scale = 0.6]{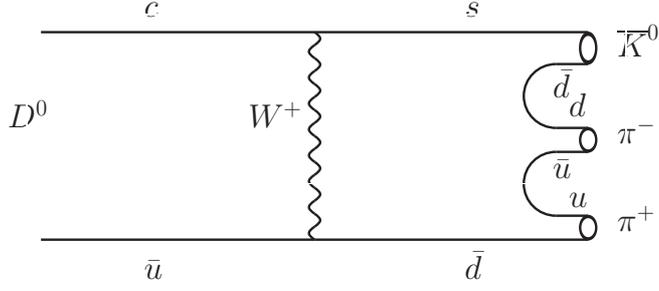}
\caption{Diagram for the Cabibbo favored annihilation ($W$-exchange) amplitudes.}\label{ann1}
\end{center}
\end{figure}

A similar derivation for  the CF  annihilation amplitudes, illustrated by the diagram  in Fig.~\ref{ann1},  yields
\bqa
\label{ANNCF}
A^{CF}(s_0,s_-,s_+)&\approx&
 \frac{G_F}{2}\ \Lambda_1\ a_2
  \sum_{L={S,P,D}}  \left [  \langle [\overline K^0 (p_0) \pi^-(p_-) ]_L\ \pi^+(p^+)
  \vert (\overline  s \ d)_{V-A}\vert 0  \rangle \right. \nonumber \\
& +& \left.  \langle \overline K^0(p_0) \ [ \pi^-(p_-) \pi^+(p^+)] \vert  (\overline  s \ d)_{V-A} \vert 0  \rangle \right  ]
 \cdot \langle 0\vert (\overline  c \ u)_{V-A} \vert D^0(p_{D^0}) \rangle   \nonumber \\ 
&=&\sum_{L={S,P,D}} A^{CF}_{[\overline{K}^0\pi^-]_L\pi^+}(s_0,s_-,s_+)
 +  \sum_{L={S,P,D}} A^{CF}_{\overline{K}^0 [\pi^+\pi^-]_L}(s_0,s_-,s_+)  \nonumber \\
 &=&   A^{CF}_{[\overline{K}^0\pi^-] \pi^+}(s_0,s_-,s_+) + A^{CF}_{\overline{K}^0 [\pi^+\pi^-]}(s_0,s_-,s_+). 
\eqa

\vspace{.8cm}

\vspace{.8cm}
\begin{figure}[h] 
\begin{center}
\includegraphics[scale = 0.6]{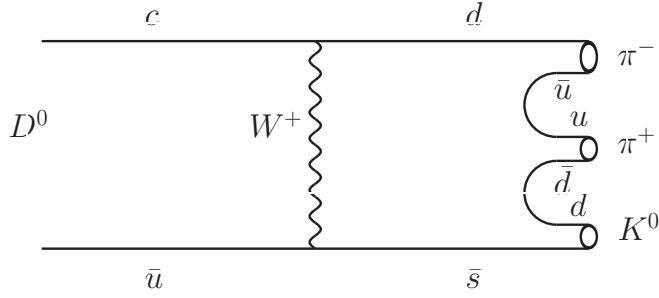}
\caption{As in Fig.~\ref{ann1} but for the doubly Cabibbo suppressed annihilation ($W$-exchange) 
amplitudes.}\label{ann1DCS}
\end{center}
\end{figure}

\noi The corresponding  DCS annihilation amplitudes (see Fig.~\ref{ann1DCS}), obtained 
from Eq.~(\ref{ANNCF}) with the substitutions 
$\Lambda_1 \Longrightarrow \Lambda_2$,  $\pi^+  \Longleftrightarrow \pi^-$, 
$\overline{K}^0  \Longrightarrow K^0$ and 
$d \Longleftrightarrow s$, read  
 
\bqa
\label{ANNDCS}
A^{DCS}(s_0,s_-,s_+)&\approx&
 \frac{G_F}{2}\ \Lambda_2\ a_2
  \sum_{L={S,P,D}}  \left [  \langle [K^0 (p_0) \pi^+(p_+) ]_L\ \pi^-(p^-)
  \vert (\overline  d \ s)_{V-A}\vert 0  \rangle \right. \nonumber \\
& +&  \left.  \langle  K^0(p_0) \ [ \pi^+(p_+) \pi^-(p^-)] \vert  
(\overline  d \ s)_{V-A} \vert 0  \rangle\right  ] \cdot \langle 0\vert 
(\overline  c \ u)_{V-A} \vert D^0(p_{D^0}) \rangle  \nonumber \\ 
&=& \sum_{L={S,P,D}} A^{DCS}_{[{K}^0\pi^+]_L \pi^-}(s_0,s_-,s_+)
 +  \sum_{L={S,P,D}} A^{DCS}_{{K}^0 [\pi^-\pi^+]_L}(s_0,s_-,s_+)  \nonumber \\
 &=&  A^{DCS}_{[{K}^0\pi^+]\pi^-}(s_0,s_-,s_+) + A^{DCS}_{{K}^0 [\pi^-\pi^+]}(s_0,s_-,s_+). 
\eqa

\noi Let us now review in detail the  
28 amplitudes that build up the total $D^0~\to~K^0_S \pi^+ \pi^-$ amplitude defined in 
Eq.~(\ref{fulamp}). Indeed, for each amplitude in Eq.~(\ref{fulamp}) there are 
three ($L=S$, $P$, $D$) contributions for the $[K\pi]\pi$ states and three for the $K[\pi\pi]$ ones as can 
be seen from
Eqs.~(\ref{TCF}), (\ref{TDCS})-(\ref{ANNDCS}).
To these 24 amplitudes one has to add the four contributions
 in which the $[\pi\pi]_P$ pair  in the $K[\pi\pi]$ final state originates from the $\omega(782) \to \pi^+\pi^-$ decay. 

\subsection{Cabibbo favored amplitudes}{\label{CFA}}

 \noi {\it The  
$[K_S^0\pi^-]_S\ \pi^+$ and $K_S^0\ [\pi^+\pi^-]_S$
amplitudes}
\\
 
Starting from  Eq.~(\ref{TCF}) we build now the expression of the different CF amplitudes following a 
derivation similar to that described in details in Ref.~\cite{DedonderPol} (see, in particular, 
Appendix A of Ref.~\cite{DedonderPol} and Sec.~\ref{2c} of this paper where an analogous explicit 
derivation for the annihilation amplitudes  is presented). 
The  
$[K_S^0\pi^-]_S\ \pi^+$
 amplitude is  
\bqa \label{A1S}
T^{CF}_{[\overline{K}^0 \pi^-]_S \ \pi^+}(s_0, s_-,s_+) 
&=& -\frac{G_F}{2}\ a_1 \Lambda_1\ \chi_1 \ \left (m_{D^0}^2-s_-\right ) \ f_\pi \ F_0^{D^0 R_S [\overline{K}^0\pi^- ]}(m_\pi^2)\  F_0^{\overline{K}^0\pi^-}(s_-)\nonumber \\
& \equiv &T_1.
\eqa
The transition form factor $ F_0^{D^0 R_S[\overline{K}^0\pi^-]}(m_\pi^2)$ is dominated by the $K_0^*(1430)^{-}$ resonance. It is real in the kinematical range considered here. The form factor $F_0^{\overline{K}^0\pi^-}(s_-)$ includes the contribution of the $ K_0^*(800)^-$ (or $\kappa^-)$ and $K_0^*(1430)^{-}$ resonances. 

The  
$K_S^0\ [\pi^+\pi^-]_S$ 
amplitude reads 
\bqa \label{A2S}
 T^{CF}_{\overline{K}^0 \ [\pi^+ \pi^-]_S}(s_0, s_-,s_+) 
&=& - \frac{G_F}{2}\ a_2\ \Lambda_1\ \chi_2\ (m_{D^0}^2-s_0)\  f_{K^0} \ F_0^{D^0 R_S[\pi^+\pi^-]} (m_{K^0}^2) \ F_0^{\pi^+\pi^-}(s_0)\nonumber \\ &\equiv& T_2,
\eqa
\noi where the  transition form factor $F_0^{D^0 R_S[\pi^+\pi^-]} (m_{K^0}^2)$  is assumed to be dominated by the $f_0(980)$ resonance. It is also purely real. 

In the equations above, $f_\pi$ and $f_{K^0}$ represent the pion and $K^0$ decay constants. The 
[$\pi\pi$] $S$-wave form factor $F_0^{\pi^+\pi^-}(s_0)$ includes 
the contribution of the  $ f_0(500)$ (or $\sigma), f_0(980)$ and   
$f_0(1400)$ resonances. The $K\pi$  and $\pi \pi$ scalar  form factors
$ F_0^{\overline{K}^0\pi^-}(s_-)$ and
$F_0^{\pi^+\pi^-}(s_0)  = \sqrt{\frac{2}{3}} \ \Gamma^{n*}_1(s_0)$ 
will be built following the methods discussed in Refs.~\cite{Bppk} and \cite{DedonderPol}.

In Eqs.~(\ref{A1S}) and (\ref{A2S}) the factors $\chi_1$ and $\chi_2$ are related to the strength of the $[K\pi]_S$ and $[\pi\pi]_S$ scalar form factors, respectively. As just mentioned these form factors receive contributions from different resonances.  If a resonance $R_S[K\pi]$ or $R_S[\pi\pi]$ was dominant  $\chi_1$ and $\chi_2$ could be evaluated in terms of the decay constant of these resonances. 
As shown in Eq.~(A.8) of Ref.~\cite{DedonderPol} and as discussed in Sec.~\ref{results} of the present
 paper, their values could be estimated from the dominating resonance decay properties. Here,
there is no dominant resonance then $\chi_1$ and $\chi_2$ are taken
as complex constants to be fitted. 
\\
\newpage
\noi {\it  The 
$[K_S^0\pi^-]_P\ \pi^+$ and $K_S^0\ [\pi^+\pi^-]_P$
 amplitudes}
\\

The 
$[K_S^0\pi^-]_P \ \pi^+$ 
 amplitude reads, with $K^{*-}\equiv K^*(892)^-$, 
\bqa \label{A1P}
T^{CF}_{[\overline{K}^0\pi^-]_P \ \pi^+}(s_0, s_-,s_+)
&=& \frac{G_F}{2}\ a_1 \ \Lambda_1\  \frac{f_\pi}{f_{{K^{*-}}}} 
\left (s_0 - s_+ + (m_{K^0}^2 - m_\pi^2) \ \frac{m_{D^0}^2- m_\pi^2}{s_-}  \right ) \nonumber\\  
 & & \times \ A_0^{D^0 R_P[\overline{K}^0\pi^-]} (m_\pi^2 ) \ F_1^{\overline{K}^0\pi^-}(s_-)
 \equiv T_3,
 \eqa
 where $A_0^{D^0 R_P[\overline{K}^0\pi^-]} (m_\pi^2 )$  denotes the form factor describing the $D^0$ to
 $[\overline{K}^0\pi^-]_P$  transition, largely dominated by the  $ K^*(892)^-$ resonance. 
  The form factor $F_1^{\overline{K}^0\pi^-}(s_-)$  includes a priori the contribution of the 
$K^*(892)^-$, $K_1(1410)^-$ and $K^*(1680)^-$ resonances~\cite{EpifanovPLB654} (see Sec.~\ref{input}).
 It has been discussed  notably  in Refs.~\cite{Bppk}, \cite{mouss2008} and
~\cite{BoitoEPJC}.\\

The  
$K_S^0\ [\pi^+\pi^-]_P$
amplitude is given by 
\be
\label{A2Prho}
T^{CF}_{\overline{K}^0 \ [\pi^+ \pi^-]_P} (s_0, s_-,s_+)
=  \frac{G_F}{2}\ a_2\ \Lambda_1\ \frac{f_{K^0}}{f_{{\rho}}} \ (s_- - s_+) \ A_0^{D^0 R_P[\pi^+\pi^-]}
(m_{K^0}^2) \ F_1^{\pi^+\pi^-}(s_0) \equiv T_4,
\ee
where the transition form factor $A_0^{D^0\\ R_P[\pi^+\pi^-]}(m_{K^0}^2) $ is dominated by the
  $\rho(770)^0$ resonance. 
The form factor $F_1^{\pi^+\pi^-}(s_0)$ which includes a priori the contributions of the $ \rho(770)^0$,
 $\rho(1450)^0$ and $\rho(1700)^0$
is the same as that introduced in Ref.~\cite{DedonderPol}, following the analysis in 
Ref.~\cite{Belletau2008} based on a Gounaris-Sakurai form with parameters extracted from
third column of their Table VII. 
Alternatively we also use one of the unitary parametrizations derived by  
Hanhart in Ref.~\cite{Hanhart}. Since the $K^{*-}$ and $\rho(770)^0$ are 
dominating resonances, we use in
Eqs.~(\ref{A1P}) and (\ref{A2Prho}), $f_{{K^{*-}}}$ and $f_{{\rho}}$ to represent the $R_P{[\overline{K}^0\pi^-]}$ and $R_P{[\pi^+\pi^-]}$ decay constants (here, $f_{\rho}$ denotes the charged $\rho$ decay constant). 

The $D^0 \to \overline{K}^0[\pi^+ \pi^-]_P$ decay can also proceed through the two-step process 
 $D^0 \to  \overline{K}^0\ \omega$ followed by  the decay $ \omega \to  \pi^+ \pi^-$; it yields an 
amplitude similar to that of the  $D^0 \to  \overline{K}^0\ [\pi^+\pi^-]_P$ process with the 
replacement of the $[\pi^+\pi^-]_P$ pair by the $\omega$ and the subsequent decay 
$\omega \to  \pi^+ \pi^-$, which violates isospin conservation.
 Thus, this term has to be added to the $P$-wave amplitude. Defining
\be 
\langle \overline{K}^0(p_0)\  [\pi^+(p_+)\pi^-(p_-)]_{\omega} \vert H_{eff} \vert D^0(p_D) \rangle = 
 T^{CF}_{\overline{K}^0 [\pi^+\pi^-]_\omega} (s_0, s_-,s_+),
\ee
one has, in the quasi two-body factorization, 
\bqa
 T^{CF}_{\overline{K}^0 [\pi^+ \pi^-]_\omega}(s_0, s_-,s_+)&=& 
\frac{G_F}{\sqrt{2}}\ \Lambda_1 \ a_2\  \langle \overline{K}^0(p_0) \vert (\overline {s}d)_{V-A} \vert 0 \rangle\nonumber \\
&& \cdot \ \langle [\pi^+(p_+)\pi^-(p_-)]_{\omega} \vert (\overline {u} c)_{V-A} \vert  D^0(p_{D^0}) \rangle  \eqa
with
\be  \label{f_K}
\langle \overline{K}^0(p_0)   \vert (\overline {s}d)_{V-A} \vert 0 \rangle = i \ f_{K^0} \ p_0,
\ee
and
\bqa
\label{ompipi} \langle [\pi^+(p_+)\pi^-(p_-)]_{\omega} \vert  (\overline {u} c)_{V-A} \vert  D^0(p_{D^0}) \rangle &=& \frac{1}{\sqrt{2}}\ G_{\omega\pi^+\pi^-}(s_0) \ \epsilon \cdot (p^+-p^-) \nonumber  \\
&\times & \ \langle \omega (p_++p_-) \vert \ (\overline {u} c)_{V-A} \vert  D^0(p_{D^0}) \rangle. \eqa
where $\epsilon$ represents the four-vector polarization of the $\omega$ meson. 
The matrix element in the above equation reads (see, {\it e.g.}, Eq.~(24) of
 Ref.~\cite{AliPRD58})
\be 
 \langle \omega (s_0) \vert \ (\overline {u} c)_{V-A} \vert  D^0(p_{D^0}) \rangle = - i \ \frac{2\ m_{\omega} \ (\epsilon^* \cdot p_D)}{p_0^2}\ p_0 \  A_0^{D^0\omega}(p_0^2) + {\rm ``other \ terms"}, \ee
\noi where  the ``other terms" do not contribute when they are multiplied by Eq.~(\ref{f_K}). 
The $\omega \pi^+\pi^-$ vertex function is given by
\be
 G_{\omega\pi^+\pi^-}(p_+ + p_-) = \frac{g_{\omega\pi\pi}}{m_{\omega}^2 - s_0 -i \ m_{\omega} \
 \Gamma
_{\omega}}, \ee
where the expression of the coupling coefficient  $g_{\omega\pi\pi}$  is given in Sec.~\ref{input} and 
$\Gamma
_{\omega}$ is the $\omega$ total width. One eventually arrives at 
\be \label{aomega}
T^{CF}_{\overline{K}^0 \ [\pi^+  \pi^-]_\omega}(s_0, s_-,s_+)
 =  \frac{G_F}{2}\ a_2 \ \Lambda_1\  \frac{f_{K^0}}{\sqrt{2}} \ m_{\omega} \  (s_- - s_+) \ \frac{g_{\omega\pi\pi}
 \ A_0^{D^0 \omega}(m_{K^0}^2)}{m_{\omega}^2 - s_0 - i \ m_{\omega} \ \Gamma_{\omega}
} \equiv T_5.
\ee \\

\noi {\it  The $[K_S^0\pi^-]_D\ \pi^+$ and $K_S^0\ [\pi^+\pi^-]_D$
 amplitudes}
\\

One has finally to evaluate the $[K_S^0\pi^-]_D\ \pi^+$  amplitude associated to the 
$K_2^{*-}\equiv K_2^{*-}(1430)$ resonance  for the $[K_S^0 \pi^-]_D$ states  and the 
$K_S^0 \ [\pi^+\pi-]_D$
 one related to the  $f_2\equiv f_2(1270)$ for the $[\pi^+ \pi^-]_D$ states. With the  notation
 $m_{K_2^*} \equiv  m_{K_2^{*-}(1430)} $, the  amplitude related to the $K_2^{*-}$ resonance reads
\bqa \label{aK2}
 T^{CF}_{[{\overline{K}^0}\pi^-]_D \ \pi^+}(s_0, s_-,s_+)
 & =& -\frac{G_F}{2}\ a_1\  \Lambda_1 \ f_{\pi} \ F^{{D^0} R_D[\overline{K}^0\pi^-]}(s_-,m_{\pi}^2)
 \ \frac{g_{K_2^{*-}  K^0_S\pi^-} \ D(\bf{p_{1}},\bf{p_{+}})}
{m_{K_2^*}^2 - s_- - i \ m_{K_2^*} \ \Gamma_{K_2^*}
} \nonumber \\
  &\equiv& T_6,
 \eqa 
 where $g_{K_2^{*-}  K_S^0\pi^-} $ is the  $K_2^{*-}$ coupling constant to the ${{K}^0_S}\pi^-$ pair 
since the width $\Gamma_{K_2^*}
$ will be considered as constant [see Eqs.~(\ref{gk2*})-(\ref{gamak2kpi})]. The function  
$D({\bf {p_{1}}},{\bf p_{+}})$
 is expressed in terms of the momenta in the $[{{K}^0_S}\pi^-]$ center-of-mass system defined in 
Appendix~\ref{kine}
\be \label{dss} 
D({\bf p_{1}},{\bf p_{+}})= \frac{1}{3} \ ( \vert {\bf p_{1}} \vert \ \vert 
{\bf p_{+}}
 \vert )^2 - ({\bf p_{1}}
 \cdot {\bf p_{+}} )^2.
\ee
The transition form factor $F^{D^0 R_D[\overline{K}^0\pi^-]}(s_-,m_{\pi}^2) $ follows from
 Ref.~\cite{KimPRD67} (see their Eq.~(10a)), and depends on three distinct functions of the four 
momentum transfer squared at 
 $m_{\pi}^2$, 
$ k^{D^0K_2^{*-}}(m_{\pi}^2)$, $b^{D^0K_2^{*-}}_+(m_{\pi}^2)$ and $b^{D^0 K_2^{*-}}_-(m_{\pi}^2)$,
 such that
\be \label{FD0K2}
F^{{D^0} R_D[\overline{K}^0\pi^-]}(s_-,m_{\pi}^2) = k^{D^0K_2^{*-}}(m_{\pi}^2) + b^{D^0K_2^{*-}}_+
(m_{\pi}^2) \ (m_{D^0}^2 - s_-) +b^{D^0 K_2^{*-}}_-(m_{\pi}^2) \ m_{\pi}^2.
\ee

For the amplitude related to  the $f_2$ meson with mass $m_{f_2}\equiv m_{f_2(1270)}$ one has 
\bqa \label{af2}
T^{CF}_{\overline{K}^0 \ [\pi^+\pi^-]_D}(s_0, s_-,s_+) 
 &=&  - \frac{G_F}{2}\ a_2 \  \Lambda_1 \ \frac{f_{K^0}}{\sqrt{2}} \ F^{D^0 R_D[\pi^+\pi^-]}
(s_0,m_{K^0}^2) \  \frac{g_{f_2  \pi^+\pi^-} 
\ D({\bf p_{2}},{\bf p_{0}})}
{m_{f_2}^2 - s_0 - i \ m_{f_2} \ \Gamma_{f_2}
(s_0)} \nonumber \\ &\equiv&  T_7,
\eqa
where $ g_{f_2  \pi^+\pi^-} $ characterizes  the strength of the $f_2\to \pi^+\pi^-$ transition
[see Eqs.~(\ref{gf2pipi}) and (\ref{gamaf2pipi})]. 
Here, because of the rather large width of the $f_2$ meson, the total width 
$\Gamma_{f_2}
(s_0)$ depends on the invariant mass squared $s_0$. 
The function $ D({\bf p_{2}},{\bf p_{0}})$ is given by the same expression as 
in Eq.~(\ref{dss}) replacing ${\bf p_{1}}$ by ${\bf p_{2}}$ and  ${\bf p_+}$ by ${\bf p_{0}}$,
the corresponding momenta and scalar product defined in Eqs.~(\ref{cmpipipp})-(\ref{cmpipipopp}).
In Eq.~(\ref{af2}), the $D^0$ to $f_2$ transition form factor, $F^{D^0 R_D[\pi^+\pi^-]}(s_0,m_{K^0}^2) $ 
depends on three distinct functions of the four momentum transfer squared at  $m_{K^0}^2$ 
\be \label{FD0f2}
F^{D^0 R_D[\pi^+\pi^-]}(s_0,m_{K^0}^2) = k^{D^0f_2}(m_{K^0}^2) + b_+^{D^0f_2}(m_{K^0}^2) \ (m_{D^0}^2 - s_0) +b_-^{D^0f_2}(m_{K^0}^2) \ m_{K^0}^2.
\ee

\subsection{The  doubly Cabibbo suppressed amplitudes}{\label{DCSA}}

To the Cabbibo favored amplitudes of the preceding subsection must now be added the doubly Cabibbo 
suppressed tree amplitudes  which are derived from Eq.~(\ref{TDCS}) in a similar way  to that used for
 the CF amplitudes. For the $[K^0_S\pi^+]_S \pi^-$ amplitude, we have
\be
\label{A1SDCS}
T^{DCS}_{[{K^0}\pi^+]_S \ \pi^-}(s_0, s_-,s_+)=  \frac{G_F}{2}\ a_1 \ \Lambda_2 \ (m_{D^0}^2-m_{\pi}^2) \ \frac{m_{K^0}^2 - m_{\pi}^2}{s_+} \ F_0^{D^0 {\pi^-}}(s_+) \ F_0^{K^0 \pi^+}(s_+)\equiv T_8,
\ee
while  the $K^0_S[\pi^-\pi^+]_S$ amplitude reads
\be
\label{A2SDCS}
 T^{DCS}_{K^0 \ [\pi^-\pi^+]_S} \ (s_0, s_-,s_+)  = \frac{\Lambda_2}{\Lambda_1}
 \ T^{CF}_{ \overline K^0 \ [\pi^+\pi^-]_S}  (s_0, s_-,s_+) = \frac{\Lambda_2}{\Lambda_1} \ T_2.
\ee

\noi For the $[K^0_S  \pi^+]_P \ \pi^-$ amplitude we obtain 
\bqa
\label{A1PDCS}
T^{DCS}_{[{K^0}\pi^+]_P \ \pi^-}(s_0, s_-,s_+) &=&  -  \frac{G_F}{2}\ a_1\ \Lambda_2 \  \left [ s_0 - s_- + (m_{D^0}^2 - m_{\pi}^2) \ \frac{m_{K^0}^2 - m_{\pi}^2}{s_+}\right ] \nonumber \\
&&\times \  F_1^{D^0 \pi^-}(s_+) \ F_1^{K^0\pi^+}(s_+) \equiv T_9. 
\eqa

\noi For the $K^0_S \ [\pi^-\pi^+]_P$ amplitude, one has two contributions, associated 
mainly to the $\rho(770)^0$ and to the $\omega(782)$. They read 
\be
\label{A2PDCS}
T^{DCS}_{{K}^0 \ [\pi^-\pi^+]_P} (s_0, s_-,s_+)  = \frac{\Lambda_2}{\Lambda_1} \ 
 T^{CF}_{{\overline K}^0 \ [\pi^+\pi^-]_P}(s_0, s_-,s_+)= \frac{\Lambda_2}{\Lambda_1} \ T_4
 \ee
 and
 \be \label{dcsomega}
 T^{DCS}_{{K^0} \ [\pi^-\pi^+]_{\omega}}(s_0, s_-,s_+) = \frac{\Lambda_2}{\Lambda_1} \  
 T^{CF}_{\overline K^0 \ [\pi^+\pi^-]_{\omega}}(s_0, s_-,s_+) = \frac{\Lambda_2}{\Lambda_1} \ T_5,
 \ee
 respectively. Associated to the $[K\pi]$ and $[\pi\pi]$ $D$- states, there is only one non-zero amplitude, that related to the $f_2$ meson,
\be \label{Af2cs}
T^{DCS}_{{K^0} \ [\pi^-\pi^+]_D}(s_0, s_-,s_+)     =  \frac{\Lambda_2}{\Lambda_1}
 \ T^{CF}_{\overline{K}^0 \ [\pi^+\pi^-]_D}(s_0, s_-,s_+) = \frac{\Lambda_2}{\Lambda_1} \  T_7.
\ee
\noi No contribution comes from  the $[K\pi]$ $D$-wave since one has $<0 \vert (\overline {u} \  s)_{V-A} \vert K_2^{*+} > = 0, $
so that
\be  \label{aK2cs}
T^{DCS}_{[{K}^0\pi^+]_D \ \pi^-}(s_0, s_-,s_+) \propto T_{10}=0. \ee

The expressions of the CF and DCS \textquotedblleft emission" amplitudes of the $D^0$ to 
pseudoscalar-vector meson decays, given in the Appendix of Ref.~\cite{Fu-Sheng_Yu_PRD84_074019}, 
agree with our CF [see Eqs.~(\ref{A1P}), (\ref{A2Prho}), (\ref{aomega})] and DCS 
[see Eqs.~(\ref{A1PDCS})-(\ref{dcsomega})] 
tree amplitudes for the dominant resonance  $K^*(892)$, $\rho(770)^0$ and $\omega$ part, respectively. 

\subsection{The annihilation ($W$-exchange) Cabibbo favored amplitudes} \label{2c}

Let us sketch a    
systematic derivation for these amplitudes defined in Eq.~(\ref{ANNCF}) and illustrated diagrammatically by Fig.~\ref{ann1}  (see, {\it e.g.},  Sec.~V.C in Ref.~\cite{AliPRD58}). Denoting by $M_1(p_1)$ and $M_2(p_2)$ the quasi two-meson final state, we may write, in the quasi two-body factorization, for the CF amplitudes 
\bqa
 \label{M1M2ann}
\langle M_1(p_1) M_2(p_2) \vert H_{eff} \vert D^0(p_{D^0}) \rangle  &=&   \frac{G_F}{\sqrt{2}} \ a_2 
\  \Lambda_1 \ \langle  M_1(p_1) M_2(p_2) \vert (\overline {s} d)_{V-A} \vert 0\rangle \nonumber \\
& &\cdot \  \langle 0 \vert (\overline {u}c)_{V-A}\vert D^0(p_{D^0}) \rangle \eqa
\noi The second term in the right hand side of Eq. (\ref{M1M2ann}) corresponds to the annihilation of the $D^0$ that goes through the W exchange between the $c\overline{u}$ quark pair that builds the $D^0$ (see Ref.~\cite{AliPRD58}).  In Eq.~(\ref{M1M2ann})  the possible quasi-two-meson pairs are (see Eq.~(\ref{ANNCF})):  
\be \label{M1KPI}
M_1(p_0+p_-) \equiv [\overline{K}^0(p_0) \pi^-(p_-)]_{L}, \hspace{2cm} M_2(p_+) \equiv \pi^+(p_+), \ee
\be \label{M1PIPI}
M_1(p_++p_-) \equiv [\pi^+(p_+) \pi^-(p_-)]_{L}, \hspace{2cm} M_2(p_0) \equiv \overline{K}^0(p_0). \ee
\noi The meson pairs are assumed to originate from a pair of quarks: a $s\overline{u}$ pair in the first case and a $d\overline{d}$ one in the second. For the $D^0$ decay constant, $f_{D^0}$ one takes (the phase is chosen 
in accordance with the choice made in Eq.~(A.3) of Ref.~\cite{DedonderPol})
\be\label{ubarcfD}
\langle 0 \vert (\overline {u} \  c)_{V-A} \vert D^0(p_{D^0}) \rangle  = - i \ f_{D^0} \ p_{D^0}. \ee

\noi Thus, all annihilation amplitudes will be proportional to the $D^0$ decay constant $f_{D^0}$.
  The form factor $\langle M_1(p_1) M_2(p_2) \vert (\overline {s} d)_{V-A} \vert 0 \rangle$ is evaluated in 
terms of the transition form factors between the pseudoscalar $M_2(-p_2)$ 
and the meson pair  $[m_1(p_3) m_2(p_4)]_L$   in scalar, vector or tensor state, with respective four-momenta $p_3$ and $p_4$.
 We introduce the hypothesis that the transitions of the pseudoscalar meson $M_2(-p_2)$ to the $[m_1(p_3) m_2(p_4)]_{L}$ states go through intermediate resonances $M_1(p_1)$ where the four-momentum $p_1$ fulfills the energy-momentum 
conservation relation  $p_1~=~p_3~+~p_4$;  these intermediate resonances then decay into the $[m_1(p_3), m_2(p_4)]$ pairs. In the case of Eq.~(\ref{M1KPI}) one identifies $m_1(p_3)$ with the $\overline{K}^0$ meson with four momentum $p_0$ and $m_2(p_4)$ with the $\pi^-$ meson with four-momentum $p_-$ whereas, in the case of Eq.~(\ref{M1PIPI}) one identifies $m_1(p_3)$ with the $\pi^+$ meson with four momentum $p_+$ and $m_2(p_4)$ with the $\pi^-$ meson with four-momentum $p_-$. The resonance decays are described by vertex functions $ G_{R_L[m_1m_2]}(p_1^2)$ modeled assuming them to be proportional  to the  scalar  $R_S[m_1 m_2]$
 or vector $R_P[m_1 m_2]$ form factor for the $S$ and $P$ amplitudes or to a relativistic Breit-Wigner function for the $R_D[m_1 m_2]$  states. The model thus yields the following contributions. 

For $[m_1m_2]_S$ waves
\bqa 
<M_1(p_1) M_2(p_2) \vert &\! \! (\overline {s} d)_{V-A} \!  \! &\vert 0  > =  G_{R_S[m_1m_2]}(s_{34}) \ 
 \langle R_S[m_1(p_3) m_2(p_4)] \vert (\overline {s} d)_{V-A} \vert M_2(-p_2)\rangle \nonumber \\
&= & \! \! -i \ G_{R_S[m_1m_2]}(s_{34}) \ 
\Big \{ \big [ -p_2 + p_3 +p_4 + \frac{p_2^2 -(p_3+p_4)^2}{m_{D^0}^2} \ p_{D^0} \big ] \ \nonumber \\
& \times & \! \! F_1^{M_2 R_S[m_1m_2]}(m_{D^0}^2) - \frac{p_2^2 -(p_3+p_4)^2}{m_{D^0}^2} \ p_{D^0} \ F_0^{M_2 R_S[m_1m_2]}(m_{D^0}^2) \Big \}\!,
\eqa
where $ F_0^{M_2 R_S[m_1m_2]}(m_{D^0}^2)$ and $ F_1^{M_2 R_S[m_1m_2]}(m_{D^0}^2)$ denote the $M_2 R_S[m_1m_2]$ scalar and vector form factors. The vertex function $G_{R_S[m_1m_2]}(s_{34})$ is modeled according to 
\be
G_{R_S[m_1m_2]}(s_{34}) = \chi_{R_S[m_1m_2]} \ F_0^{m_1m_2} (s_{34}), \hspace{1cm}{\rm with}\hspace{1cm} s_{34} = p_1^2=(p_3+p_4)^2, \ee
$F_0^{m_1m_2} (s_{34})$ being the $[m_1m_2]$ scalar form factor and $ \chi_{R_S[m_1m_2]}$ 
characterizing the strength of the $S$-state form factor contribution as discussed in Sec.~\ref{CFA}. 
With $\chi_{R_S[\overline{K}^0\pi^-]}\equiv \chi_1$ [see Eq.~(\ref{A1S})]
the CF $[\overline{K}^0(p^0) \pi^-(p_-)]_S\  \pi^+(p_+)$ annihilation amplitude is
\bqa \label{ann1S}
 A^{CF}_{[\overline{K}^0\pi^-]_S \ \pi^+} (s_0, s_-,s_+) 
&=&  - \frac{G_F}{2}\ a_2\ \Lambda_1\ \chi_1  \ (m_{\pi}^2 - s_-) \ f_{D^0} \ F_0^{\pi^+R_S[\overline{K}^0\pi^-]}(m_{D^0}^2) \  F_0^{\overline{K}^0\pi^-}(s_-)\nonumber \\
&\equiv& A_1 .\eqa
\noi For the  $[\pi^+(p_+) \pi^-(p_-)]_S$  pair, we have, with  $\chi_{R_S[\pi^+\pi^-]}\equiv \chi_2$, [see Eq.~ (\ref{A2S})],
\bqa \label{ann2S}
 A^{CF}_{ \overline{K}^0 \ [\pi^+\pi^-]_S}(s_0, s_-,s_+) &=& 
-  \frac{G_F}{2}\ a_2\ \Lambda_1\ \chi_2  \ (m_{K^0}^2 - s_0)\ f_{D^0} \ F_0^{\overline{K}^0R_S[\pi^+\pi^-]} (m_{D^0}^2) \ F_0^{\pi^+\pi^-}(s_0)\nonumber \\ &\equiv& A_{2}.
\eqa
Since the $D^0$ mass is larger than the masses of the two-meson thresholds $m_\pi + m_{K_0^*(800)}$ 
and  $m_{\overline{K}^0} + m_{f_0(500)}$,  
 the transition form factors $F_0^{\pi^+R_S[\overline{K}^0\pi^-]}(m_{D^0}^2)$ and 
$F_0^{\overline{K}^0R_S[\pi^+\pi^-]} (m_{D^0}^2)$ appearing in these equations are 
unknown complex parameters to be fitted. \\

For the $[m_1m_2]_P$ wave contributions, denoting for simplicity the vector meson resonances as 
$$V_R \equiv R_P[m_1m_2],$$
 we may write 
\be \langle [m_1(p_3)m_2(p_4)]_P\ M_2(p_2) \vert (\overline {d}s)_{V-A} \vert 0 \rangle = 
G_{V_R}(p_1^2) \ \epsilon \cdot (p_3-p_4) \ \langle V_R(p_1^2) \vert (\overline {s} d)_{V-A} \vert M_2(-p_2) \rangle, \ee
$\epsilon$ being the polarization of the vector resonance  and $G_{V_R}$  the $V_R$ decay vertex function. One 
has~\cite{AliPRD58}
\be\label{VRS34}
\langle  V_R(p_1^2)\vert (\overline {s} d)_{V-A} \vert M_2(-p_2)\rangle = - i \ \frac{2 m_{V_R} \ 
(\epsilon^*\cdot p_2)}{p^2_{D^0}} \ p_{D^0} \ A_0^{M_2 V_R}(m_{D^0}^2) + {\rm ``other \ terms"}. \ee
Here 
$p_{D^0}= p_1+p_2$. The ``other terms" do not contribute when multiplying the matrix element~(\ref{VRS34}) by that of Eq.~(\ref{ubarcfD}). The $P$ states
 being characterized by  dominant resonances, one writes 
$$G_{V_R}(p_1^2) = \frac{1}{m_{V_R} \ f_{V_R}} \ F_1^{m_1m_2}(p_1^2),$$ where $f_{V_R}$ is the $V_R$ decay constant. One thus arrives at the following expressions
\bqa \label{ann1P}
 A^{CF}_{[\overline{K}^0\pi^-]_P \ \pi^+}(s_0, s_-,s_+) &=&
 - \frac{G_F}{2}\ a_2\ \Lambda_1\  \frac{f_{D^0}}{f_{K^{*-}}} \left [ s_0 - s_+ + \frac{(m_{D^0}^2- m_{\pi}^2)(m_{K^0}^2- m_{\pi}^2)}{s_-} \right ]  \nonumber \\
 && \times  \ A_0^{\pi^+ R_P[\overline{K}^0\pi^-]}(m_{D^0}^2) \  F_1^{\overline{K}^0\pi^-}(s_-)
 \equiv A_{3},\eqa
and 
\bqa \label{annrho}
 A^{CF}_{ \overline{K}^0 \ [\pi^+\pi^-]_P} (s_0, s_-,s_+) &=&  \frac{G_F}{2}\ a_2\ \Lambda_1\  \frac{f_{D^0}}{f_{\rho}} \ (s_- - s_+) \ A_0^{\overline{K}^0 R_P[\pi^+\pi^-]} (m_{D^0}^2) \ F_1^{\pi^+\pi^-}(s_0)
\nonumber \\ &\equiv& A_4,\eqa
 and, if  the  $[\pi^-\pi^+]_P$  originates from the $\omega$ resonance, 

\be \label{annomega} 
A^{CF}_{\overline{K}^0 \ [\pi^+\pi^-]_{\omega}} (s_0, s_-,s_+)=   - \frac{G_F}{2}\ a_2\ \Lambda_1\  \frac{f_{D^0}}{\sqrt{2}} \ m_{\omega} \  (s_- - s_+) \ 
\frac{g_{\omega\pi\pi} \ A_0^{\overline{K}^0 [\pi^+\pi^-]_{\omega}}(m_{D^0}^2)}{m_{\omega}^2 - s_0 - i \ m_{\omega} \ 
\Gamma_{\omega}
}\equiv A_5,
\ee
Since we are in the $\overline{K}^0 V_R$ scattering region, the values of the form factors 
$A_0^{\overline{K}^0 R_P [\pi^+\pi^-]} (m_{D^0}^2)$ and $A_0^{\overline{K}^0  [\pi+\pi^-]_{\omega}}(m_{D^0}^2)$
are complex numbers. 

Finally, for the $[m_1m_2]_D$ wave contributions, denoting for simplicity the tensor meson resonances as 
$$T_R \equiv T_R[m_1m_2]$$ 
and  the polarization tensor of the D-wave resonance as $\mathbf{\epsilon}_{\alpha\beta}(\lambda) $, $\lambda$ being the spin
 projection, one can write 
\bqa 
\hspace{-1cm}\langle  [m_1(p_3) m_2(p_4)]_{D}\ M_2(p_2) \vert \  (\overline {s} d)_{V-A}  \vert 0 \rangle &=& 
G_{T_R}(p_1^2) \  \nonumber \\ 
& & \times\  \sum_{\lambda = -2}^{\lambda = +2} \epsilon_{\alpha\beta}(\lambda) \ p_3^{\alpha} \ p_4^{\beta}  \ \langle T_R^{\lambda}(p_1^2) \ M_2(-p_2) \vert (\overline {s} d)_{V-A} \vert 0 \rangle.
 \eqa
Reformulating the matrix element  for the $M_2 \ T_R$ to vacuum transition 
\be
-i \ f_{D^0}\ p_{D^0}\cdot \langle T_R^{\lambda}(p_1^2)\ M_2(p_2) \vert (\overline {s} d)_{V-A} \vert 0 \rangle = f_{D^0} \ F^{M_2 T_R}(p_1^2,m_{D^0}^2) \ \epsilon^*_{\mu\nu}(\lambda)\  p_2^{\nu} \  p_2^{\mu}. \ee
where (see Ref.~\cite{KimPRD67})
\be  \label{TRFF}
-i \ F^{M_2 T_R}(p_1^2,m_{D^0}^2) = k^{M_2 T_R}(m_{D^0}^2) + b_+^{M_2 T_R}(m_{D^0}^2) \ (m_{M_2}^2 - p_1^2) + b_-^{M_2 T_R}(m_{D^0}^2) \ m_{D^0}^2.\ee
Here, $ k^{M_2 T_R}$, $b_+^{M_2 T_R}$ and $ b_-^{M_2 T_R}$ are complex transition form factors since $m_{D^0}^2 > (m_{M_2} + m_{T_R})^2$. 
\noi Assuming then, for these cases, Breit-Wigner representations of the resonance vertex functions 
$G_{T_R}(p_1^2)$ and summing over the spin projections $\lambda$,
 one arrives at the following expressions
\bqa \label{annK2}
A^{CF}_{[\overline{K}^0\pi^-]_D \ \pi^+}(s_0, s_-,s_+) &=& \frac{G_F}{2}\ a_2\ \Lambda_1\ f_{D^0}
  \ F^{R_D[\overline{K}^0\pi^-]\pi^+ }(s_-,m_{D^0}^2) \nonumber \\
&&\times \ g_{K_2^{*-}  K^0_S\pi^-} \  
\frac{D({\bf p_{1}},{\bf p_{+}})}
{m_{K_2^{*}}^2 - s_- - i \ m_{K_2^{*}}
 \ \Gamma_{K_2^{*}}
}\equiv A_{6},
\eqa
\bqa \label{annf2} 
A^{CF}_{\overline{K}^0 \ [\pi^+\pi^-]_D} (s_0, s_-,s_+) 
&=&  \frac{G_F}{2}\ a_2\ \Lambda_1\  \frac{f_{D^0}}{\sqrt{2}}  \ F^{\overline{K}^0  R_D[\pi^+\pi^-]}(s_0,m_{D^0}^2) \nonumber \\
& & \times \  g_{f_2  \pi^+\pi^-} \  \frac{D({\bf p_{2}},{\bf p_{0}})}
{m_{f_2}^2 - s_0 - i \ m_{f_2} \ \Gamma_{f_2}
(s_0)} \equiv A_{7},\eqa
where the expressions of  $g_{K_2^{*-}  K^0_S\pi^-}$,  $g_{f_2  \pi^+\pi^-}$ and of the resonance 
widths are discussed in Sect.~\ref{input}. 
 
 \subsection{The annihilation ($W$-exchange) doubly Cabibbo suppressed amplitudes} \label{2d}

One has to evaluate the corresponding Cabbibo suppressed amplitudes. One obtains for the 
$[K^0_S\pi^+]_S \ \pi^-$ amplitudes
\bqa \label{csann1S}
 A^{DCS}_{[{K^0}\pi^+]_S \ \pi^-}(s_0, s_-,s_+) &=& - \frac{G_F}{2}\ a_2\ \Lambda_2\ \chi_1  \ (m_{\pi}^2 - s_+) \ f_{D^0} \ F_0^{\pi^- R_S[K^0\pi^+]}(m_{D^0}^2) \  F_0^{{K^0}\pi^+}(s_+)\nonumber \\
& \equiv& A_{8},
\eqa
\noi and 
\be \label{csann2S}
A^{DCS}_{ {K^0} \ [\pi^-\pi^+]_S}(s_0, s_-,s_+)=  A^{CF}_{\overline{K}^0\ [\pi^+\pi^-]_S}(s_0, s_-,s_+)=  \frac{\Lambda_2}{\Lambda_1}\  A_{2},\ee\\
for the $K^0_S \ [\pi^-\pi^+]_S$ amplitude, having assumed the charge symmetry relation for the form factors 
\be
F_0^{K^0R_S[\pi^-\pi^+]}(m_{D^0}^2) = F_0^{\overline{K}^0 R_S[\pi^+\pi^-]}(m_{D^0}^2).
\ee

\noi For the $[K^0_S\pi^+]_P \ \pi^-$ amplitudes, one has with 
$K^{*+}\equiv K^{*+}(892)$ [compare with Eq.~(\ref{ann1P})]
\bqa \label{csann1P}
 A^{DCS}_{[{K^0}\pi^+]_P \ \pi^-}(s_0, s_-,s_+)&=& -  \frac{G_F}{2}\  a_2\  \Lambda_2\ \frac{f_{D^0}}{f_{K^{*+}}}  \left [ s_0 - s_- + \frac{(m_{D^0}^2- m_{\pi}^2)(m_{K^0}^2- m_{\pi}^2)}{s_+} \right ]  \nonumber  \\
& & \times \ A_0^{R_P[K^0\pi^+] \pi^- }(m_{D^0}^2) \  F_1^{{K^0}\pi^+}(s_+) \equiv A_{9},
\eqa
while for the $K^0_S \ [\pi^-\pi^+]_P$ amplitudes, assuming the charge symmetry relations 
\bqa \label{A0K0RPpi+pi-}
A_0^{K^0 R_P[\pi^-\pi^+]}(m_{D^0}^2) &=& A_0^{\overline{K}^0 R_P[\pi^+\pi^-]}(m_{D^0}^2)\\ 
\label{A0K0omega} A_0^{K^0  [\pi^-\pi^+]_{\omega}}(m_{D^0}^2) &=&  A_0^{\overline{K}^0 [\pi^+\pi^-]_{\omega}}(m_{D^0}^2),
 \eqa
one obtains respectively
\bqa \label{csannrho}
A^{DCS}_{K^0 \ [\pi^-\pi^+]_P}(s_0, s_-,s_+)&=&\frac{\Lambda_2}{\Lambda_1}\ 
A^{CF}_{\overline{K}^0 \ [\pi^+\pi^-]_P}(s_0, s_-,s_+)= \frac{\Lambda_2}{\Lambda_1}\   A_{4},\\
 \label{csannomega} 
 A^{DCS}_{{K^0} \ [\pi^-\pi^+]_{\omega}}(s_0, s_-,s_+)&=&\frac{\Lambda_2}{\Lambda_1}\ 
 A^{CF}_{\overline{K}^0 \ [\pi^+\pi^-]_{\omega}}(s_0, s_-,s_+)= \frac{\Lambda_2}{\Lambda_1}\ A_{5}.\eqa

Finally, the $[K^0_S \pi^+]_D \ \pi^-$ amplitude reads
 \bqa \label{csann1D}
A^{DCS}_{[{K^0}\pi^+]_D \ \pi^-}(s_0, s_-,s_+) &=&\frac{G_F}{2}\ a_2\ \Lambda_2\  f_{D^0} \ F^{{R_D[K^0\pi^+] \pi^-}}(s_+,m_{D^0}^2) \nonumber \\
 &&\times \ g_{K_2^{*+}  K^0_S\pi^+} \  \frac{D({\bf p_{3}},{\bf p_{-}})}
 {m_{K_2^{*}}^2 - s_+ - i \ m_{K_2^{*}} \ 
\Gamma_{K_2^{*}}
}\equiv  A_{10},\eqa
where $\bf{p_{3}}$ and $\bf{p_{-}}$ are defined in Appendix A, 
and with the charge symmetry relation 
\be \label{FK0RDpi+pi-}
F^{K^0R_D[\pi^-\pi^+]}(s_0,m_{D^0}^2)=  F^{\overline{K}^0 R_D[\pi^+\pi^-]}(s_0,m_{D^0}^2),
\ee
the $K^0_S \ [\pi^-\pi^+]_D$ amplitude is
 \be \label{csann2D}
A^{DCS}_{K^0 \ [\pi^-\pi^+]_D}(s_0, s_-,s_+) =  \frac{\Lambda_2}{\Lambda_1}\ A^{CF}_{\overline{K}^0[\pi^+\pi^-]_D}(s_0, s_-,s_+)= \frac{\Lambda_2}{\Lambda_1}\ A_{7}.\ee

To summarize, of the 28 amplitudes describing the $D^0 \to K_S^0 \pi^+\pi^-$ decays, only 20 are 
independent among which one, $T^{DCS}_{[{K}^0\pi^+]_D\pi^-}(s_0, s_-,s_+) $ or $T_{10}$, is zero 
(Eq. (\ref{aK2cs})).

\section{Quasi two-body channel amplitudes and branching fractions}
 
This section is devoted to the construction of amplitudes suited for numerical computations. 
This aim leads us to  build specific combinations out of the amplitudes formally derived in the 
preceding section. The full decay  amplitude given in Eq.~(\ref{fulamp}) can be written as a 
superposition of ten partial amplitudes  $\mathcal{M}_i$ which are each made of a tree 
$\mathcal{T}_i$ and of an annihilation (W-exchange) $\mathcal{A}_i$ contributions

\be \label{T+A}
\mathcal{M}  \equiv \sum_{i=1}^{10} \mathcal{M}_i \equiv \sum_{i=1}^{10} (\mathcal{T}_i+\mathcal{A}_i).
\ee 

\subsection{Amplitudes recombined}\label{amprecom}
From Eqs.~(\ref{TCF}), (\ref{ANNCF}), (\ref{A1S}) and (\ref{ann1S}), the summed  
$[K^0_S\pi^-]_S\pi^+$ CF amplitudes read
\bqa
\label{S111}
\mathcal{M}_1&\equiv& \mathcal{T}_1+\mathcal{A}_1= T_1 + A_{1} = -\frac{G_F}{2} \ \Lambda_1 \ \chi_1 \ F_0^{\overline{K}^0\pi^-}(s_-)\nonumber \\
&& \left [a_1 \  (m_{D^0}^2-s_-) \ f_\pi \ F_0^{D^0R_S[\overline{K}^0\pi^-]}(m_\pi^2)\ + 
 a_2 \ (m_{\pi}^2 - s_-) \ f_{D^0} \ F_0^{R_S[\overline{K}^0\pi^-] \pi^+}(m_{D^0}^2) \right ].
\eqa

Recombining the tree amplitudes defined  in Eqs.~(\ref{TCF}),~(\ref{TDCS}) and given by 
Eqs.~(\ref{A2S}),~(\ref{A2SDCS}), and the annihilation amplitudes defined in 
Eqs.~(\ref{ANNCF}),~(\ref{ANNDCS}), and given by Eqs.~(\ref{ann2S}),~(\ref{csann2S}) yields the complete 
$K^0_S[\pi^+\pi^-]_S$ amplitude, 

 \bqa
\label{S212}
\mathcal{M}_{2} &\equiv& \mathcal{T}_2+\mathcal{A}_2 
=\ \left (1 + \frac{\Lambda_2}{\Lambda_1} \right ) \ (T_2 + A_{2}) \nonumber \\
&=& 
  -  \frac{G_F}{2} \  a_2 \  (\Lambda_1 + \Lambda_2 ) \ \chi_2 \   F_0^{\pi^+\pi^-}(s_0) \nonumber \\
& \times&
\left [ (m_{D^0}^2-s_0)\  f_{K^0} \ F_0^{D^0R_S[\pi^+\pi^-]}(m_{K^0}^2) \ +  (m_{K^0}^2 - s_0)\ f_{D^0} \ F_0^{\overline{K}^0R_S[\pi^+\pi^-]} (m_{D^0}^2)
\right ].\eqa

For the $P$ states, the summed $[K^0_S\pi^-]_P \ \pi^+$ CF amplitudes  from Eqs.~(\ref{TCF}), (\ref{ANNCF}), (\ref{A1P}) and (\ref{ann1P}), yield 
\bqa
\label{S313}
\mathcal{M}_{3} &\equiv&  \mathcal{T}_3+\mathcal{A}_3= T_3 + A_{3} = \frac{G_F}{2} \ \Lambda_1\left [ s_0 - s_+ + \frac{(m_{D^0}^2- m_{\pi}^2)(m_{K^0}^2- m_{\pi}^2)}{s_-} \right ] \  F_1^{\overline{K}^0\pi^-}(s_-)   \nonumber\\
& &\hspace{1.5cm} \times \left [ a_1 \ \frac{f_\pi}{f_{{K^{*-}}}} \ A_0^{D^0R_P[\overline{K}^0\pi^-]}(m_\pi^2 ) \  - \ a_2  \ \frac{f_{D^0}}{f_{K^{*-}}} \ A_0^{\pi^+ R_P[\overline{K}^0\pi^-] }(m_{D^0}^2) \right ].
\eqa

 As in the case of the $K^0_S \ [\pi^+\pi^-]_S$ channel, one aggregates the four CF and DCS amplitudes given in Eqs.~(\ref{A2Prho}), (\ref{A2PDCS}), (\ref{annrho}) and  (\ref{csannrho})  
to obtain the complete $K^0_S \ [\pi^+\pi^-]_P$ amplitude
\bqa
\label{S414}
\mathcal{M}_{4} &\equiv& \mathcal{T}_4+\mathcal{A}_4= \ \left (1 + \frac{\Lambda_2}{\Lambda_1} \right ) \ (T_4 + A_{4}) \nonumber \\
 &=& \frac{G_F}{2} \  a_2  \ (\Lambda_1 + \Lambda_2) \  \frac{1}{f_{{\rho}}} \  (s_- - s_+) \  F_1^{\pi^+\pi^-}(s_0) \nonumber \\
&\times & \left [f_{K^0} \ A_0^{D^0R_P[{\pi^+\pi^-}]}(m_{K^0}^2) \  + f_{D^0} \ 
A_0^{\overline{K}^0 R_P[{\pi^+\pi^-}]} (m_{D^0}^2)\right]. 
\eqa

\noi The combination
\be f_{K^0} \ A_0^{D^0R_P[\pi^+\pi^-]}(m_{K^0}^2) \  + f_{D^0} \ A_0^{\overline{K}^0 R_P[\pi^+\pi^-]} (m_{D^0}^2).
\ee
will be treated as a single real parameter (see Sec.~\ref{Brfrac}).

The $K^0_S[\pi^+\pi^-]_\omega$ 
amplitude results from Eqs.~(\ref{aomega}), (\ref{dcsomega}), (\ref{annomega}) and (\ref{csannomega})

\bqa
\label{S515}
\mathcal{M}_{5} &\equiv&  \mathcal{T}_5+\mathcal{A}_5 = \ \left (1 + \frac{\Lambda_2}{\Lambda_1} \right ) \ (T_5 + A_{5}) \nonumber \\
&=&  \frac{G_F}{2} \ (\Lambda_1 + \Lambda_2)   \  \frac{a_2 }{\sqrt{2}}  \ m_{\omega} \  (s_- - s_+) \nonumber \\ 
&\times& \left [ f_{K^0} \ A_0^{D^0 \omega}(m_{K^0}^2) \ - \ f_{D^0} \ A_0^{\overline{K}^0  [\pi^+\pi^-]_{\omega}}(m_{D^0}^2) \right ]
 \frac{g_{\omega\pi\pi}}{m_{\omega}^2 - s_0 - i \ m_{\omega} \ \Gamma_{\omega}
}.
\eqa

 The $[K^0_S\pi^-]_D \pi^+$ amplitude, which arises from Eqs.~(\ref{TCF}), (\ref{ANNCF}), (\ref{aK2}), (\ref{aK2cs}) and (\ref{annK2}), reads
\bqa 
\mathcal{M}_{6}   &\equiv& \mathcal{T}_6+\mathcal{A}_6 = T_6 + A_6   \nonumber \\
&=&  \frac{G_F}{2} \ \Lambda_1\  g_{K_2^{*-}  K^0_S\pi^-} \  B_{K_2^*} (s_+,s_-)\nonumber \\
& \times&
 \left [-  a_1 \ f_{\pi} \ F^{D^0R_D[\overline{K}^0\pi^-]}(s_-,m_{\pi}^2) 
+ \ a_2 \  f_{D^0}  \ F^{R_D[\overline{K}^0\pi^-] \pi^+  }(s_-,m_{D^0}^2)\right ]
\eqa 
where 
\be \label{BWK2}
B_{K_2^*} (s_+,s_-)= \frac{D(\bf{p_{1}},\bf{p_{+}})}
{m_{K_2^{*}}^2 - s_- - i \ m_{K_2^{*}} \ 
\Gamma_{K_2^{*}}
}.
\ee
Using 
\be \label{D0RD}
F^{D^0 R_D[\overline{K}^0\pi^-]}(s_-,m_{\pi}^2) = D_1 + E_1 (m_{D^0}^2 -s_-)\ee
and 
\be \label{PIPRD}
F^{ R_D[\overline{K}^0\pi^-] \pi^+ }(s_-,m_{D^0}^2) = d_1 + e_1 \ (m_{\pi}^2 - s_-),\ee
where $D_1$ and $E_1$ are real coefficients, related to the form factors in Eq.~(\ref{FD0K2}) by 
$$D_1=  k^{D^0K_2^{*-}}(m_{\pi}^2) +b^{D^0 K_2^{*-}}_-(m_{\pi}^2) \ m_{\pi}^2 
\hspace{1cm}{\rm and}\hspace{1cm} E_1= b^{D^0K_2^{*-}}_+(m_{\pi}^2) $$
while $d_1$ and $e_1$, related to the form factors in Eq. (\ref{TRFF}) by 
$$d_1=  k^{K_2^{*-}\pi^+}(m_{D^0}^2) +b^{K_2^{*-} \pi^+}_-(m_{D^0}^2) \ m_{D^0}^2 \hspace{1cm}{\rm and}\hspace{1cm}e_1= b^{K_2^{*-} \pi^+}_+(m_{D^0}^2) $$
are complex. 
\noi One  finally obtains
\be
\mathcal{M}_{6} =  \frac{G_F}{2} \ \Lambda_1 \ g_{K_2^{*-}  K^0_S\pi^-} \ 
\left (q_{6}\ m_{K_2^{*}}  +\frac{s_{6}}{m_{K_2^{*}}} s_-\right ) \ B_{K_2^*} (s_+,s_-)
\ee
with
\bqa 
q_{6} \ m_{K_2^{*}}   &=& - \ a_1 \ f_{\pi} \  (D_1 + E_1 \ m_{D^0}^2 )  + a_2 \  f_{D^0}  \ (d_1 + e_1 \ m_{\pi}^2), \\
\frac{s_{6}}{m_{K_2^{*}}} &=&  a_1 \ f_{\pi} \ E_1 -  a_2 \  f_{D^0}  \ e_1.
\eqa
The unknown complex parameters $q_6$ and $ s_6$ will be fitted.

For the $K^0_S [\pi^+\pi^-]_D$ amplitude dominated by the $f_2$ meson, we have, from Eqs.~(\ref{TCF}), (\ref{TDCS}) to (\ref{ANNDCS}), (\ref{af2}), (\ref{Af2cs}), (\ref{annf2}) and (\ref{csann2D}),
\bqa
\mathcal{M}_{7}  &\equiv& \mathcal{T}_7+\mathcal{A}_7 = \ \left (1 + \frac{\Lambda_2}{\Lambda_1} \right ) \ (T_7 + A_7)   \nonumber \\
&=& \frac{G_F}{2\ \sqrt{2}} \ a_2 \ (\Lambda_1 + \Lambda_2)\ g_{f_2  \pi^+\pi^-} \nonumber \\
&& \times 
\left [ -f_{K^0} F^{D^0 R_D[\pi^+\pi^-]}(s_0,m_{K^0}^2) + f_{D^0} \ F^{\overline{K}^0 R_D[\pi^+\pi^-]}(s_0,m_{D^0}^2) \right ] \  B_{f_2}(s_+,s_0) 
\eqa
with
\be \label{BWf2}
B_{f_2}(s_+,s_0) = \frac{D(\bf{p_{2}},\bf{p_{0}})}
{m_{f_2}^2 - s_0 - i \ m_{f_2} \
 \Gamma_{f_2}
(s_0)}. 
\ee
It is reexpressed as
\be
\mathcal{M}_{7} =  \frac{G_F}{2\ \sqrt{2}} \ (\Lambda_1+ \Lambda_2)  \  g_{f_2  \pi^+\pi^-}  \
\left (q_{7}\ m_{f2}  +\frac{s_{7}}{m_{f2}} \ s_0 \right ) \ B_{f_2}(s_+,s_0) ,
\ee
with
\bqa
q_{7}\ m_{f2}  &=& a_2 \ \big [- f_{K^0} \ (D_2 + E_2 \ m_{D^0}^2)  +   f_{D^0}  \ (d_2 + e_2 \ m_{K^0}^2 ) \big ]\\
\frac{s_{7}}{m_{f2}} &=&  a_2 \ \left (f_{K^0} \ E_2 -    f_{D^0}  \ e_2\right ).
\eqa
As for the $[K^0_S\pi^-]_D \ \pi^+$ amplitude, the coefficients $D_2$, $E_2$ are real but related to the form factors in Eq.~(\ref{FD0f2}) by
$$D_2= k^{D^0f_2}(m_{K^0}^2) +b_-^{D^0f_2}(m_{K^0}^2) \ m_{K^0}^2.
\hspace{1cm}{\rm and}\hspace{1cm} E_2= b_+^{D^0f_2}(m_{K^0}^2), $$
while $d_2$ and $e_2$,
arising from  the form factors of Eq.~(\ref{TRFF}), are complex. As $q_6$ and $s_6$,  $q_7$ and $s_7$ are
 unknown parameters that 
will be fitted.

The DCS $[K^0_S \pi^+]_S\pi^-$ amplitude results from Eqs.~(\ref{TDCS}), (\ref{ANNDCS}), (\ref{A1SDCS}) and (\ref{csann1S}) and reads 
\bqa
\label{S818}
\mathcal{M}_{8}  &\equiv& \mathcal{T}_8+\mathcal{A}_8 = z_8 \ (T_8 + A_8)\nonumber \\
& =&  \frac{G_F}{2} \ \Lambda_2 \ z_8 \left [a_1  \ (m_{D^0}^2-m_{\pi}^2) \ \frac{m_{K^0}^2 - m_{\pi}^2}{s_+} \ F_0^{D^0 {\pi^-}}(s_+) \right. \nonumber\\
& &-\left. \ a_2  \ \chi_1 \ f_{D^0} \ (m_{\pi}^2 - s_+)  \ F_0^{\pi^- R_S[K^0\pi^+]}(m_{D^0}^2) \right ] \ 
 F_0^{{K^0}\pi^+}(s_+)
\eqa

and the  DCS $[K^0_S \pi^+]_P\pi^-$ amplitude results from Eqs.~(\ref{TDCS}), (\ref{ANNDCS}), (\ref{A1PDCS}) and (\ref{csann1P})
\bqa
\label{S919}
\mathcal{M}_{9}  &\equiv& \mathcal{T}_9+\mathcal{A}_9 = z_9 \ (T_9 + A_9)\nonumber \\
&=& - \frac{G_F}{2} \ \Lambda_2 \ z_9\ \left [a_1  \  F_1^{D^0 \pi^-}(s_+) + a_2 \
 \frac{f_{D^0}}{f_{K^{*+}}} \ A_0^{R_P[{K^0}\pi^+] \pi^- }(m_{D^0}^2) \right ]  \nonumber \\
& & \times  \left [ s_0 - s_- + \frac{(m_{D^0}^2- m_{\pi}^2)(m_{K^0}^2- m_{\pi}^2)}{s_+} \right ] 
\ F_1^{K^0\pi^+}(s_+).
\eqa
\noi The unknown multiplicative complex constants $z_8$ and $z_9$, appearing in Eqs.~(\ref{S818})
 and (\ref{S919}), are introduced to allow some charge independence violation in the $[K\pi]_S\pi$ 
and $[K\pi]_P\pi$ amplitudes, as can be seen comparing, on the one hand, amplitudes $\mathcal{M}_{1} $ in Eq.~(\ref{S111}) and $\mathcal{M}_{8} $ in Eq.~(\ref{S818}) and, on the other hand, amplitudes  $\mathcal{M}_{3} $ in Eq.~(\ref{S313}) and $\mathcal{M}_{9}$ in Eq.~(\ref{S919}).
 They will be fitted. In the calculations that follow, we assume that  the $[K\pi]_{S,P}$ form 
factors fulfill the relation 
\be \label{F01Kpi}
F_{(0,1)}^{K^0\pi^+}(s) \equiv F_{(0,1)}^{\overline{K}^0\pi^-}(s).\ee

Finally, from Eq.~(\ref{csann1D}), the DCS annihilation  $[K \pi]_D\pi$ amplitude is 
$\mathcal{M}_{10}  \equiv A_{10}$. 
In analogy with the amplitudes $\mathcal{M}_{6}$ and  $\mathcal{M}_{7}$, we introduce the 
parametrization
\be
\label{q10s10} 
a_2 \ f_{D^0}  \ F^{R_D[K^0\pi^+] {\pi^-}}(s_+,m_{D^0}^2) = q_{10} \ m_{K_2^*} +
 \frac{s_{10}}{m_{K_2^{*}}} \ s_+,
\ee
where the unknown coefficients $q_{10}$ and $s_{10}$, related to the transition form factors in 
Eq.~(\ref{TRFF}), are free complex parameters that will be fitted. 
 We calculate practically 
\be
\mathcal{M}_{10} =  \frac{G_F}{2} \ \Lambda_2 \ \left ( q_{10} \ m_{K_2^{*}} + 
\frac{s_{10}}{m_{K_2^{*}}} \ s_+ \right ) \ \frac{g_{K_2^{*+}  K^0\pi^+} \ 
D({\bf p_{3}},{\bf p_{-}})}
 {m_{K_2^{*}}^2 - s_+ - i \ m_{K_2^{*}} \ 
\Gamma_{K_2^{*}}
}.
\ee

To summarize this subsection, the recombined amplitudes used in our calculations are given in Table~\ref{combin} (a similar table can be established for the conjugate $\overline  D^0$ decays).

\begin{table*}[h]
\caption{Summary of the Cabibbo favored, CF, and doubly Cabibbo suppressed, DCS, amplitudes associated 
to the different quasi two-body channels. For each channel, the dominant resonances are listed in 
column 3 and the total amplitudes, $\mathcal{M}_i, \ i=1,10$, are the sum of the CF and DCS 
amplitudes. The tree and annihilation amplitudes are denoted $T_i$ and $A_i$, respectively.}
\label{combin}
\begin{center}
\begin{tabular}{cllcc}
\hline
\hline
Amplitude& Quasi two-body & Dominant & CF & DCS  \\
 &   channel      &resonances  &         amplitudes       &        amplitudes         \\
\hline
$\mathcal{M}_1$ & $[K^0_S \,\pi^-]_S \,\pi^+$ & $K^*_0(800)^-$, $K^*_0(1430)^-$ & $T_1+A_1$ & ------ \\
$\mathcal{M}_2$  & $K^0_S \,[\pi^+ \pi^-]_S$ & $f_0(500)$, $f_0(980)$, $f_0(1400)$ & $T_2+A_2$ & $\frac{\Lambda_2}{\Lambda_1}\ (T_2+A_2)$ \\
$\mathcal{M}_3$  & $[K^0_S \,\pi^-]_P \,\pi^+$ & $K^*(892)^-$ & $T_3+A_3$ & ------ \\
$\mathcal{M}_4$  & $K^0_S \,[\pi^+ \pi^-]_P$ & $\rho(770)$ & $T_4+A_4$ & $\frac{\Lambda_2}{\Lambda_1}\ (T_4+A_4)$ \\
$\mathcal{M}_5$ & $K^0_S \,[\pi^+\pi^-]_{\omega}$ & $\omega(782)$ & $T_5+A_5$ & $\frac{\Lambda_2}{\Lambda_1}\ (T_5+A_5)$ \\
$\mathcal{M}_6$  & $[K^0_S \,\pi^-]_D \,\pi^+$ & $K^*_2(1430)^-$ & $T_6+A_6$ & ------ \\
$\mathcal{M}_7$  & $K^0_S \,[\pi^+ \pi^-]_D$ & $f_2(1270)$ & $T_7+A_7$ & $\frac{\Lambda_2}{\Lambda_1}\ (T_7+A_7)$ \\
$\mathcal{M}_8$  & $[K^0_S \,\pi^+]_S \,\pi^-$ & $K^*_0(800)^+$, $K^*_0(1430)^+$ & ------ & $z_8\ (T_8+A_8)$ \\
$\mathcal{M}_9$  & $[K^0_S \,\pi^+]_P \,\pi^-$ & $K^*(892)^+$ & ------ & $z_9\ (T_9+A_9)$ \\
$\mathcal{M}_{10}$  & $[K^0_S \,\pi^+]_D \,\pi^-$ & $K^*_2(1430)^+$ & ------ & $A_{10}$ \\
\hline
\hline
\end{tabular}
\end{center}
\end{table*}

\subsection{On branching fractions} {\label{Brfrac}}

The differential branching fraction or the Dalitz plot density distribution is defined as
\be
\label{d2Br}
\frac{d^2Br}{ds_-ds_+}=\frac{|{\cal M}|^2}{32(2\pi)^3m_{D^0}^3 \Gamma_{D^0}},
\ee
where $\Gamma_{D^0}$ is the $D^0$ width. The total branching fraction for the $D^0$ decay into 
$K^0_S \pi^+ \pi^-$ is obtained by integration of the differential branching fraction over the Dalitz diagram
surface. One can also obtain one dimensional densities by integration over one variable $s$, for example
the $s_-$ distribution reads
\be
\label{dBr-}
\frac{dBr}{ds_-}=\int_{(m_{\pi}+m_{K^0})^2}^{(m_{D^0}-m_{\pi})^2}\frac{d^2Br}{ds_-ds_+}ds_+.
\ee
We infer from Eq.~(\ref{d2Br}) that it is not possible to calculate all the phases of the amplitudes 
${\cal M}_i$ by knowing the differential branching fraction distribution only. Out of the 10 phases, one phase
cannot be determined. Let us call this particular phase $\phi_4$ and define the
modified partial amplitudes $\tilde{{\cal M}_i}$ as follows
\be
\label{Mtilda}
\tilde{{\cal M}_i}=e^{-i\phi_4} {\cal M}_i.
\ee
The phase $\phi_4$ is taken equal to the phase of the constant coefficient of the amplitude ${\cal M}_4$
defined in Eq.~({\ref{S414}}). By making this choice we proceed in the same way as in the isobar model analyses of 
Refs. [1,2,10].
Our basic amplitudes, which will be determined from the fit to the Dalitz plot density distributions,
are the $\tilde{{\cal M}_i}$ and $\mathcal{T}_i$ amplitudes.

The branching fraction distributions corresponding to the amplitudes ${\cal M}_i$ are defined as
\be
\label{Bri}
\frac{d^2Br_i}{ds_-ds_+}=\frac{|{\cal M}_i|^2}{32(2\pi)^3m_{D^0}^3 \Gamma_{D^0}}.
\ee
If one replaces ${\cal M}_i$ by $\tilde{{\cal M}_i}$ then the above branching fractions remain unchanged.
 It is instructive to define separately the branching fractions
corresponding to different tree and annihilation components $i$ of the decay amplitudes
\be  \label{BrTi}
\frac{d^2Br^{tree}_i}{ds_-ds_+}=\frac{|\mathcal{T}_i|^2}{32(2\pi)^3m_{D^0}^3 \Gamma_{D^0}}
\ee
and
\be  \label{BrAi}
\frac{d^2Br^{ann}_i}{ds_-ds_+}=\frac{|\mathcal{A}_i|^2}{32(2\pi)^3m_{D^0}^3 \Gamma_{D^0}}
= \frac{|\ e^{i\phi_4} \tilde{{\cal M}_i}-\mathcal{T}_i|^2}{32(2\pi)^3m_{D^0}^3 \Gamma_{D^0}}.
\ee
since from Eqs.~(\ref{T+A}), (\ref{Mtilda}) one has
\be \label{relA}
\mathcal{A}_i=e^{i\phi_4} \tilde{{\cal M}_i}-\mathcal{T}_i.
\ee 

While the branching fractions $d^2Br_i/{ds_-ds_+}$ and the tree branching fractions 
$d^2Br^{tree}_i/{ds_-ds_+}$ can be directly calculated
from the fitted amplitudes, the annihilation branching fractions $d^2Br^{ann}_i/{ds_-ds_+}$
cannot be evaluated since the phase $\phi_4$ is in general unknown.
 From Eq.~(\ref{relA}) we can, however, 
obtain the following inequality
\be
\label{ineqA}
|\tilde{{\cal M}_i}|^2+|\mathcal{T}_i|^2-2|\tilde{{\cal M}_i}|\ |\mathcal{T}_i| \leq |\mathcal{A}_i|^2 \leq 
|\tilde{{\cal M}_i}|^2+|\mathcal{T}_i|^2+2|\tilde{{\cal M}_i}| \ |\mathcal{T}_i|
\ee
from which the lower and upper limits of the annihilation branching fractions can be calculated.
For example, the lower limits of the integrated annihilation branching fractions are given by
\be
\label{low}
Br^{ann,~low}_i=Br_i +Br^{tree}_i-2\int \int ds_-ds_+|\tilde{{\cal M}_i}||T_i|,
\ee
where the double integration is performed over the Dalitz plot surface.

We introduce also the modified annihilation ($W$-exchange) amplitudes $A_i'$
\be
\label{Aprim}
\tilde{{\cal M}_i}= T_i+A_i'.
\ee 
As follows from Eqs.~(\ref{T+A}), (\ref{Mtilda}) these amplitudes are related to the tree and annihilation
amplitudes 
\be
\label{AprimA}
A_i'=T_i (e^{-i\phi_4} -1) + e^{-i\phi_4} A_i.
\ee
The formulae for the modified amplitudes $A_i'$ can be rewritten in the same way as the
corresponding formulae for the annihilation amplitudes if we introduce new coefficients replacing
the former form factors calculated at the momentum transfer squared $m_{D^0}^2$. Thus, for example,
the new coefficient $\tilde{A}_0^{\overline{K}^0 R_P[\pi^+\pi^-]}$ for the $A_4'$ amplitude is given by the 
formula
\be
\label{A0tilda}
\tilde{A}_0^{\overline{K}^0 [\pi^+\pi^-]_P}= (e^{-i\phi_4} -1)\frac{f_{K^0}}{f_{D^0}}
  A_0^{D^0 R_P[\pi^+\pi^-]}(m_{K^0}^2)+e^{-i\phi_4}A_0^{\overline{K}^0 R_P[\pi^+\pi^-]}(m_{D^0}^2).
\ee
Similar relations are valid for the new complex coefficients $\tilde{F}_0^{[\overline{K}^0 \pi^-]_S\pi^+}$,
$\tilde{F}_0^{\overline{K}^0 [\pi^+\pi^-]_S}$,  $\tilde{A}_0^{[\overline{K}^0 \pi^-]_P\pi^+}$ and 
$\tilde{A}_0^{\bar K^0 \omega}$, related to the amplitudes $A_1'$, $A_2'$, $A_3'$, and $A_5'$,
respectively. By definition, the $\tilde{A}_0^{\overline{K}^0 [\pi^+\pi^-]_P}$ coefficient is real.
All the six new coefficients, defined above, will be extracted by fitting the Dalitz density distributions.

Due to our poor knowledge of the form factor combinations, defined in Eqs.~(\ref{FD0K2}) and~(\ref{FD0f2})
  for the $D$ waves, we are unable to calculate separately the tree contributions ${\cal T}_6$ and ${\cal T}_7$.
Therefore in the following considerations leading to the possibly best determination of the lower limit
of the annihilation branching fraction we have to omit temporarily from the total sum the contributions
${\cal M}_6$ and ${\cal M}_7$. 

Denoting by ${\cal T}''$, ${\cal A}''$ and $\tilde{{\cal M}''}$ the sums of the tree, annihilation and 
modified partial amplitudes
\be
\label{sums}
{\cal T}''=\sum_{i\neq6,7}{\cal T}_i,~~~~{\cal A}''=\sum_{i\neq6,7}{\cal A}_i,
~~~~\tilde{\cal M}''=\sum_{i\neq6,7}{\cal M}_i 
\ee
and using Eq.~(\ref{relA}) we obtain
\be
\label{Acal}
{\cal A}''=e^{i\phi_4} \tilde{\cal M}''-{\cal T}''.
\ee
Then similar inequalities to those of Eq.~(\ref{ineqA}) are satisfied
\be
\label{ineq2}
|\tilde{\cal M}''|^2+|{\cal T}''|^2-2|\tilde{\cal M}''||{\cal T}''| \leq |{\cal A}''|^2 \leq 
|\tilde{\cal M}''|^2+|{\cal T}''|^2+2|\tilde{\cal M}''||{\cal T}''|,
\ee
from which we get the lower and upper limits of the total annihilation branching fractions
\be
\label{lowl}
Br''_{ann,~low}=Br'' +Br''_{tree}-2\int \int ds_-ds_+|\tilde{\cal M}''||{\cal T}''|
\ee
and
\be
\label{upl}
Br''_{ann,~up}=Br'' +Br''_{tree}+2\int \int ds_-ds_+|\tilde{\cal M}''||{\cal T}''|.
\ee
Here $Br''$ is the total branching fraction for the decay process considered by us with exclusion of 
the amplitudes ${\cal T}_6$ and ${\cal T}_7$
\be
\label{brbis}
Br''=\int \int ds_-ds_+|\tilde{\cal M}''|^2
\ee
and $Br''_{tree}$ is defined as
\be
\label{brtree}
Br''_{tree}=\int \int ds_-ds_+|{\cal T}''|^2.
\ee

\section{\bf {\large Input data and useful formulae}}\label{input}

The calculation of the full amplitude derived in the preceding section requires the input of many physical ingredients in addition to a number of parameters which will be considered as free. 

The Fermi coupling constant $G_F$  is taken to be equal to 1.16637$ \cdot 10^{-5}$ GeV$^{-2}$ \cite{PDG2012}. The values of the CKM coupling matrix elements of Eq.~(\ref{lambda}) are, 
to order $\lambda^4$, where $\lambda = 0.2253$ is the sine of the Cabibbo angle 
\cite{PDG2012}
$\Lambda_1 \approx  1 - \lambda^2$ 
 and
 $\Lambda_2\approx-\lambda^2.$
In the literature one can find  many different values for the effective coefficients $a_i$, $i=1,2$.
Reference~\cite{El-Bennich_PRD79} uses the leading order,
$a_1=1.1463$, $a_2=-0.2349$ while Ref.~\cite{BoitoPRD79_034020} approximates these by $a_1=1.15$, $a_2=-0.25$.
The phenomenological values $a_1=1.2\pm 0.1$, $a_2=-0.5\pm 0.1$ have been introduced in
 Ref.~\cite{BoitoPRD80_054007}. Reference~\cite{ChengPRD81_074021}, invoking a large $N_C$ approach,
 quotes the following  $a_1 \simeq C_1(\overline  m_c)=1.274$ and 
 $a_2 \simeq C_2(\overline  m_c)=-0.529$ with $\overline  m_c(m_c)=1.25 $~GeV, values
 extracted from Tables VI and VII of Ref.~\cite{Buchalla1996}.
In Refs.~\cite{Fu-Sheng_Yu_PRD84_074019},~\cite{ChengPRD81_074021} and \cite{ChengPRD81_074031}, 
the parameters $a_1$ and $a_2$ have been fitted to data for different kinds of two-body $D$-decays.
 Moreover, in Ref.~\cite{Fu-Sheng_Yu_PRD84_074019} two additional phenomenological coefficients $a_A$
 and $a_E$  have been included to account for the $W$-annihilation and $W$-exchange contributions.
Let us note that in the factorization approach the coefficient $a_E$ is equal to $a_2$ as follows from 
the derivation of our annihilation amplitudes in Sec. \ref{amplitudes}.  

All the annihilation amplitudes, proportional to $a_2$, can acquire strong phases related to the final state
interactions described by the relevant form factors fixed at the momentum transfer squared $m_{D^0}^2$
[see Eqs.~(\ref{ann1S}), (\ref{ann2S}), (\ref{ann1P})-(\ref{annomega}), (\ref{csann1S}),
 (\ref{csann1P}), (\ref{csann1D})].  Thus the $a_2$ phase cannot result from a fit to data. 
Furthermore, only the products of  $a_2$ with the above mentioned form factors can be well determined
 from the fit. Therefore in the present work we will 
adopt the real values 
\be \label{a_12}
a_1 = 1.1\hspace{2cm}{\rm and}\hspace{2cm} a_2 = -0.5.
\ee 

The amplitudes incorporate the $\pi$, $K^0$, $\rho$ and  $D^0$ mesons  decay constants as well as their masses and, when appropriate, their widths. They are respectively, following mainly
 Ref.~\cite{PDG2012} except when otherwise stated,
\bqa
f_{\pi}&=& 0.13041 \ {\rm GeV} \hspace{1cm}{\rm and}\hspace{1cm} m_{\pi} = 0.13957 \ {\rm GeV}, \label{mfpi}\\
f_{K^-}&=& 0.1561 \ {\rm GeV} \hspace{1cm}{\rm and}\hspace{1cm} m_{K^0} = 0.497614 \ {\rm GeV},\\
f_{\rho}&=&  0.209 \ {\rm GeV}\\
&& \hspace{4.6cm} m_{\omega}= 0.78265 \ {\rm GeV} \ {\rm and} \ \Gamma
_{\omega} = 0.00849 \ {\rm GeV},\\
f_{D^0} &= & 0.2067 \ {\rm GeV}\hspace{2.6cm}m_{D^0} = 1.86486  \ {\rm GeV}\ {\rm and}  \ 
 \Gamma_{D^0} = 1.605 \cdot 10^{-12} \ {\rm GeV}.
\eqa
The $\rho$ decay constant is extracted from Ref.~\cite{Beneke2003}. The $D^0$ decay constant is 
assimilated to the $D^+$  one, given in Ref.~\cite{PDG2012}. The mass and width of the $K^*(892)^{\mp}$
 are  considered as free parameters. 
Its decay constant, $f_{K^{*-}}= f_{K^{*+}}=0.2143 $ GeV, is taken from Ref.~\cite{Bppk}.

In addition, the mass and total width of the $f_2$ and $K_2^{*}$ mesons  read~\cite{PDG2012},
\bqa
 m_{f_2} &=& 1.2751\ {\rm GeV} \ {\rm and} \ \Gamma_{f_2} = 0.1851 \ {\rm GeV}, \label{Gamf2} \\
m_{K_2^{*}} &=& 1.4256\ {\rm GeV} \ {\rm and} \ \Gamma_{K_2^{*}} = 0.0985 \ {\rm GeV}, \label{GamK2}
\eqa
respectively.

We use   $ F_0^{D^0 R_S[\overline{K}^0\pi^-]}(m_\pi^2)= 0.48 $ following Ref.~\cite{ChengPRD81_074031}  and   $F_0^{D^0 R_S[\pi^+\pi^-]} (m_{K^0}^2)= 0.18$ according to Ref.~\cite{El-Bennich_PRD79}.
  We extract  $A_0^{D^0 R_P[\overline{K}^0\pi^-]}(m_\pi^2 )~=~0.76$ from Table 9 of 
Ref.~\cite{Melikhov}. Although the values given in Table 14 of Ref.~\cite{Bauer1987} are at zero momentum transfer,
we assume here that  $A_0^{D^0 R_P[\pi^+\pi^-]}(m_{K^0}^2)~=~0.7$ and 
$A_0^{D^0\omega}(m_{K^0}^2) = 0.669$.

\noi Finally, from Eq.~(4.12) and Table 12 of Ref.~\cite{Melikhov}, we have : 
\be \label{F0D0pi}
F_0^{D^0 {\pi^-}}(s_+) = \frac{F_0}{ 1 - \sigma_1 \ \frac{s_+}{M^2} + \sigma_2 \ \frac{s_+^2}{M^4}}\ee 
with $M = 2.01  \ {\rm GeV},\ \sigma_1 = 0.54,\  \sigma_2 = 0.32 \ {\rm and}  \ F_0 = 0.69$, 
and, from Eq.~(4.10) and Table 12 of the same reference, 
\be \label{F1D0pi}
F_1^{D^0 {\pi^-}}(s_+) = \frac{F_0}{ \left(1-\frac{s_+}{M^2}\right )\ \left (1 - \sigma_1 \ \frac{s_+}{M^2}\right)} \ee
with $M = 2.01 \  {\rm GeV},\ \sigma_1 = 0.30\   {\rm and}  \ F_0 = 0.69$.\\

The coupling constant $g_{\omega\pi\pi}$ is given by 
\be \label{gompipi}g_{\omega\pi\pi} = m_{\omega} \sqrt{\frac{ 24 \ \pi \ \Gamma_{\omega\pi\pi}}{p^3}}\hspace{1cm}
{\rm with}\hspace{1cm}
p= \frac{1}{2} \ \sqrt{m_{\omega}^2 - 4 \ m_{\pi}^2}\ee
and, using  $\Gamma_{\omega\pi\pi}=0.0153\ \Gamma_{\omega}
 = 1.299 \cdot 10^{-4}$ GeV, we have 
$g_{\omega\pi\pi} = 0.3504$. 

 The coupling constant $g_{f_2\pi^+\pi^-}$ in Eqs.~(\ref{af2}) and (\ref{annf2})  is defined as
\be \label{gf2pipi}
g_{f_2\pi^+\pi^-}= m_{f_2}\ \sqrt{ \frac{60 \ \pi \  \Gamma_{f_2\pi^+\pi^-}}{q_{f_2}^5}}.
\ee
The partial width $\Gamma_{f_2\pi^+\pi^-}$ is given by 
\be \label{gamaf2pipi}
 \Gamma_{f_2\pi^+\pi^-} = \frac{2}{3} \  \   0.848\ \Gamma_{f_2}
\ee
with $\Gamma_{f_2}$ from Eq.~(\ref{Gamf2}), so that 
 $\Gamma_{f_2\pi^+\pi^-} = 0.1046 \  {\rm GeV}$
 and 
$g_{f_2\pi^+\pi^-} = 18.55 \ {\rm GeV}^{-1}$.

The total width $\Gamma_{f_2}
(s_0)$ reads (see, \textit{e.g.} Eqs.~(A.29) and (A.30) of Ref.~\cite{DedonderPol})
\be \label{gamaf2s0}
\Gamma_{f_2}
 (s_0)=  \left (\frac{q}{q_{f_2}} \ \right )^5
     \ \frac{m_{f_2}}{\sqrt{s_0}} \ \frac{(q_{f_2} r)^4 + 3 \ (q_{f_2} r)^2 + 9}
     {(q r)^4 + 3 \ (q r)^2 +9} \ \Gamma_{f_2},
\ee
 with   $r=4.0$ GeV$^{-1}$.
 
\noi The centre of mass pion momenta that enter those expressions are respectively 
\be \label{qf2q}
  q_{f_2} = \frac{1}{2} \sqrt{m_{f_2}^2 - 4 m_{\pi}^2} 
     \hspace{1cm}{\rm and} \hspace{1cm} 
     q = \frac{1}{2} \sqrt{s_0- 4 \ m_{\pi}^2}.
\ee   
 The coupling constant $g_{K_2^{*-}K^0_S\pi^-}$ appearing in Eqs.~(\ref{aK2}) and (\ref{annK2})
     is fixed at 
\be \label{gk2*}
g_{K_2^{*-} K^0_S\pi^-}
= m_{K_2^{*-}}\ \sqrt{ \frac{60 \ \pi \  \Gamma_{K_2^{*-} K_S^0\pi^-}}{q_{K_2^{*}}^5}}= 11.72\ 
{\rm GeV}^{-1}
\ee
with
\be \label{qk2*}
q_{K_2^{*}} = \frac{1}{2 m_{K_2^{*}}} \ \sqrt{\big [m_{K_2^{*}}^2 - (m_{\pi} + m_{K^0})^2 \big ] 
 \big [m_{K_2^{*}}^2 - (m_{\pi} - m_{K^0} )^2\big ]}
\ee
and 
\be \label{gamak2kpi}
\Gamma_{K_2^{*-} K^0_S\pi^-} = \frac{2}{3} \ 0.489 \ \Gamma_{K_2^{*}}
 = 0.0321\ {\rm GeV}.
\ee
We take $g_{K_2^{*+} K^0_S\pi^+}=g_{K_2^{*-} K^0_S\pi^-}$.

To  summarize this section, we have 33 free parameters: 14 complex parameters, namely,  $\chi_1$, $\chi_2$, 
$F_0^{R_S[\overline{K}^0\pi^-]\pi^+}(m^2_{D^0})$, $F_0^{\overline{K}^0 R_S[\pi^+\pi^-]}(m^2_{D^0})$,
 $A_0^{R_P[\overline{K}^0\pi^-] \pi^+}(m^2_{D^0})$, $A_0^{\overline{K}^0 \omega}(m^2_{D^0})$, $q_6$, $s_6$, $q_7$, $s_7$,
 $q_{10}$, $s_{10}$, $z_8$, $z_9$  and 5 real parameters, 
$A_0^{\overline{K}^0 R_P[\pi^+\pi^-]}(m^2_{D^0})$, $\kappa$,
 $c$, $m_{K^{*\mp}}$, $\Gamma_{K^{*}}$. The parameters $\kappa$ and $c$ enter the pion scalar form factor
 (see Eqs.~(28) and (39) in Ref.~\cite{DedonderPol}). The dominating $P$- and $S$-wave amplitudes 
require 9 and  12 parameters,  respectively, while the $D$-amplitudes, whose magnitudes are much smaller,  depend on 12 parameters.

In addition to $a_1$ and $a_2$ fixed at the values given in Eq.~(\ref{a_12}), and to the masses,
 widths and decay constants listed in Eqs.~(\ref{mfpi}-\ref{GamK2}), Table~\ref{fixed} sums up the values 
of the fixed form factors and of the coupling constants needed in the calculations that follow. 

\begin{table*}[h]
\caption{Values  of the fixed form factors and  coupling constants.}
\label{fixed}
\begin{center}
\begin{tabular}{lc}
\hline
\hline
\hspace{0.0cm}parameter & \multicolumn{1}{c}{value} \\ 
\hline
 $F_0^{D^0 R_S[\overline{K}^0\pi^-]}(m_\pi^2)$  & $0.48$     \\
$F_0^{D^0 R_S[\pi^+\pi^-]} (m_{K^0}^2)$  &  $0.18$   \\
 $A_0^{D^0 R_P[\overline{K}^0\pi^-]}(m_\pi^2 )$  &   $0.76$  \\
 $A_0^{D^0 R_P[\pi^+\pi^-]}(m_{K^0}^2)$  &   $0.7$   \\
 $A_0^{D^0\omega}(m_{K^0}^2)$ & $0.669$\\
 $g_{\omega\pi\pi}$ & $0.3504$ \\
 $g_{f_2\pi^+\pi^-}$ &  $18.55 \ {\rm GeV}^{-1}$ \\
 $ g_{K_2^{*-} K^0\pi^-}$ & $11.72\ {\rm GeV}^{-1}$\\
\hline
\hline
\end{tabular}
\end{center}
\end{table*}

\section{Results and discussion} \label{results}

The free parameters of the $D^0 \to K^0_S \pi^+ \pi^-$ decay amplitudes
described in the preceding section
are fitted to the 2010 Belle Collaboration
 data~\cite{A.Poluektov_PRD81_112002_Belle,A.Poluektov_private2013}. 
We have calculated the two-dimensional effective mass distribution
corrected for background and efficiency variation as a function of Dalitz plot position.
 A grid of $125\times125$ squared cells covering the Dalitz plot
in $s_-$ and $s_+$ variables is constructed. 
For each cell a corresponding number
of events is evaluated. The width of each cell is chosen to be $\Delta s=$ 0.02055 GeV$^2$. 
If the number of events in a given cell is smaller than 5 then the
adjacent cells with the same $s_-$ value are combined. If necessary, in the vicinity of the Dalitz plot 
edge, cells corresponding to $s_-$ and
 $s_-+\Delta s$ values are grouped in order to accumulate more than 5 events. This allows a better 
application of mathematical methods to estimate the statistical errors $\Delta N^{exp}$ of the experimental
event numbers $N^{exp}$. 
The total number of effective cells with $N^{exp}$ greater than 5 is 6321. 
The total number of signal events in these cells is equal to 453876.
The corresponding theoretical number of events $N^{th}_j$ is calculated using the model density 
distribution integrated over the surface of a given cell $j$. 
The experimental finite effective mass resolution is taken into account by calculating
the convolution of the theoretical distribution with the Gaussian function using its resolution 
parameter equal to 0.0055 GeV$^{2}$~\cite{A.Poluektov_private2013}.
The total number of events in the theoretical distribution is normalized to the experimental one.
The parameter fitting procedure is based on the following definition of the
$\chi^2_D$ function:
\begin{equation}
\label{chiD}
 \chi^2_D= \sum_{j} \left [\frac{N^{th}_j-N^{exp}_j}{\Delta N^{exp}_j}\right ]^2.
\end{equation}
The statistical errors have been calculated as $\Delta N^{exp}_j = \sqrt{N^{exp}_j}$.

In the fitting procedure, as indicated in Sec.~\ref{input}, the mass and width of the $K^*(892)$ meson 
are free parameters.
These parameters enter also in the $K\pi$ vector form factor 
taken from the Belle Collaboration
fit to the $\tau^- \to K^0_S \pi^-\nu_\tau$ decays~\cite{EpifanovPLB654}.
The contributions of $K^*(892)$ and $K^*(1410)$ resonances are taken into account but without that of the
$K ^*(1680)$ resonance. Including
that resonance cannot improve the quality of the fit because its large mass is close to the upper limit 
of the $K\pi$ effective mass in the $D^0~\to~K^0_S \pi^+ \pi^-$ decay.  The parameters of the $K^*(1410)$
resonance are fixed to the values given in the middle column of Table 3 in Ref.~\cite{EpifanovPLB654}.

\begin{figure}[h] 
\begin{center}
\includegraphics[scale = 0.4]{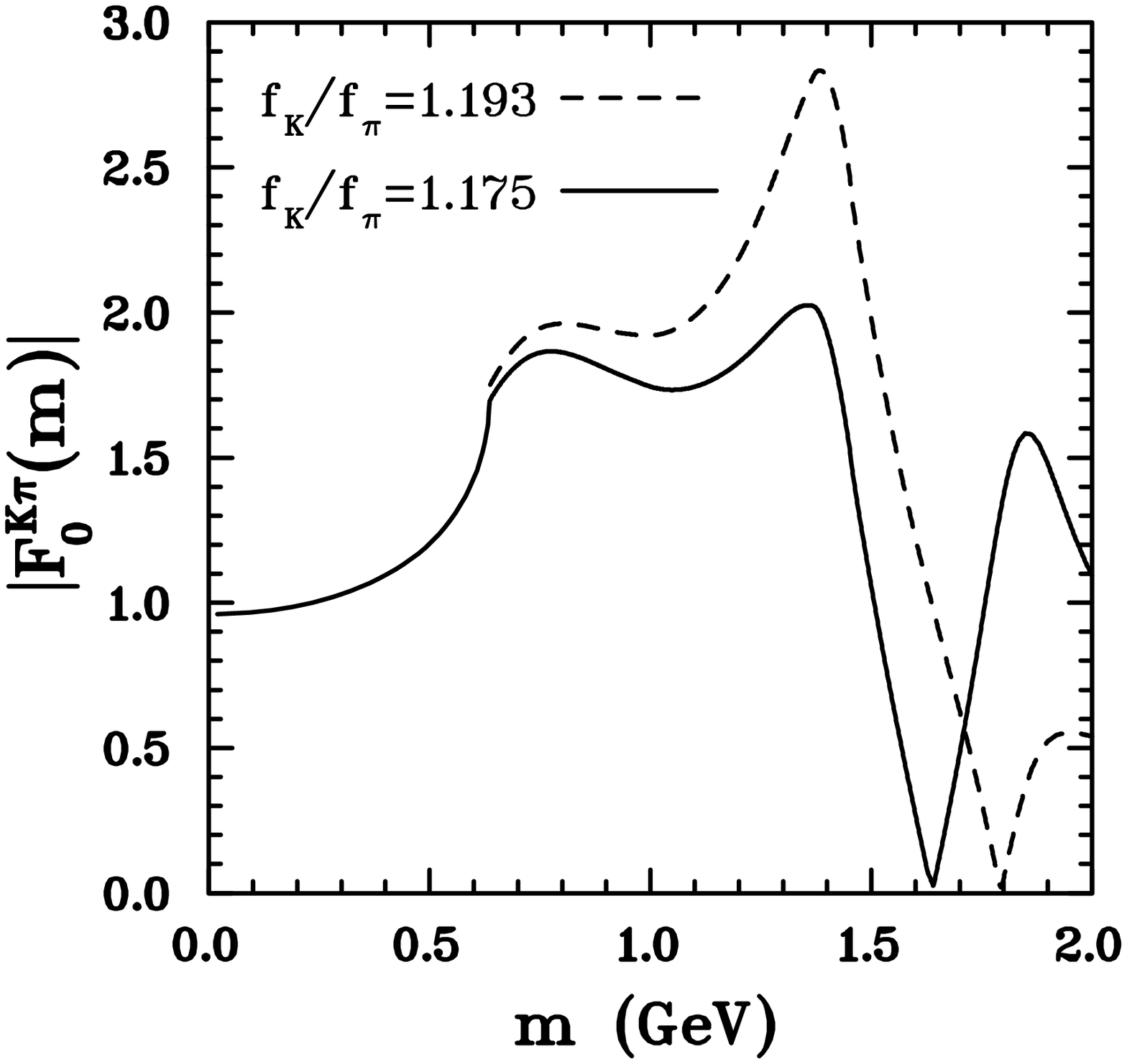}~~~~
\includegraphics[scale = 0.4]{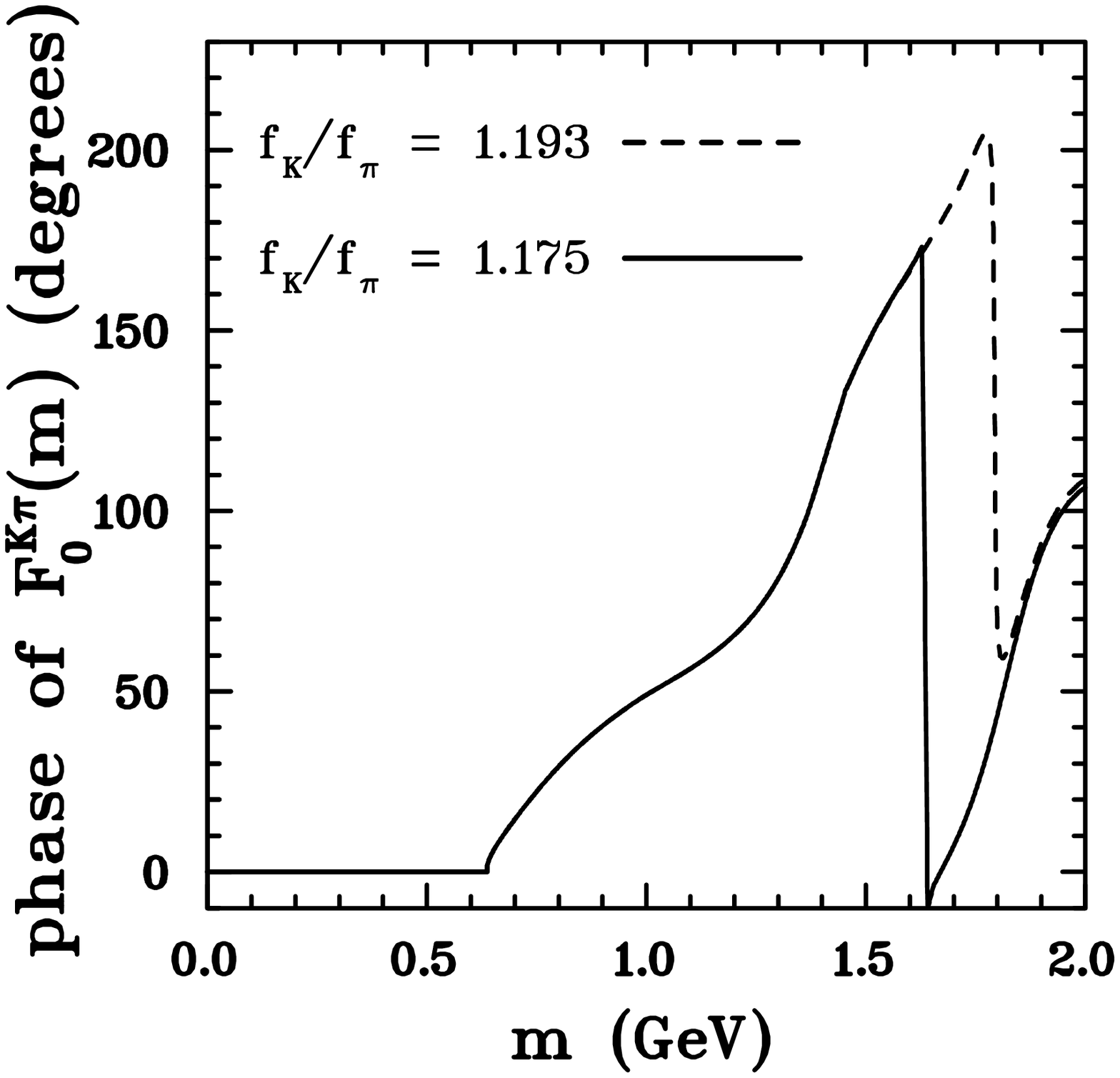}~~~~
\caption{The modulus (left panel) and the phase (right panel) of the 
 $K \pi$ scalar form factor  $F^{K\pi}_0$ as function of the
$K\pi$ effective mass for two values of the $f_K/f_{\pi}$ ratio.
}
\label{F0Kpiff}
\end{center}
\end{figure}

 In order to have consistent $K^*(892)$ parameters
we perform a simultaneous fit of the $D^0$ and $\tau$ decay data.  
The $\chi^2_{\tau}$ function is defined similarly to the $\chi^2_D$ function of Eq.~(\ref{chiD}). 
We use the first 89 experimental points up to the $K\pi$ effective mass equal to 1.65 GeV covering 
a range where the statistical errors are not too large~\cite{EpifanovPLB654}.
The $K\pi$ mass distribution is calculated with Eq.~(2) of this reference.
Alternatively to the experimental parameterization of Ref.~\cite{EpifanovPLB654} we use the model of 
the
$K\pi$ vector form factor of Boito \textit{et al.}~\cite{BoitoEPJC} in which some constraints from analyticity and 
elastic unitarity are incorporated. 
We also found that the unitary $K\pi$ vector form factor derived and used in 
Ref.~\cite{Bppk} to fit the $B \to K \pi^+ \pi^- $ decay data gives $K^*(892)$ parameters in 
disagreement with those required here to fit well the present high statistics  
$D^0 \to K^0_S \pi^+ \pi^-$ data.
As mentioned in Section IIA the scalar $K\pi$ form factor is calculated as in Ref.~\cite{Bppk}. 
Its functional form in the $K\pi$ effective mass range close to the position of the $K^*_0(1430)$ 
resonance
depends sensitively on the $f_K/f_{\pi}$ ratio of the kaon to pion coupling constants
~\cite{Moussallam_private2013}. It is illustrated in Fig.~\ref{F0Kpiff}. 
We find that the best fit is obtained with the $K\pi$ scalar form factor calculated with 
a $f_K/f_{\pi}$ value of 1.175.

As pointed out below Eq.~(\ref{A2Prho}) two types of the pion vector form factor have been tested, namely the experimental parameterization used by
 the Belle Collaboration in the data analysis of the $\tau^- \to \pi^-\pi^0 \nu_\tau$ 
decays~\cite{Belletau2008}
and the Hanhart model presented in Ref.~\cite{Hanhart}.

\begin{figure}[h] 
\begin{center}
\includegraphics[scale = 0.4]{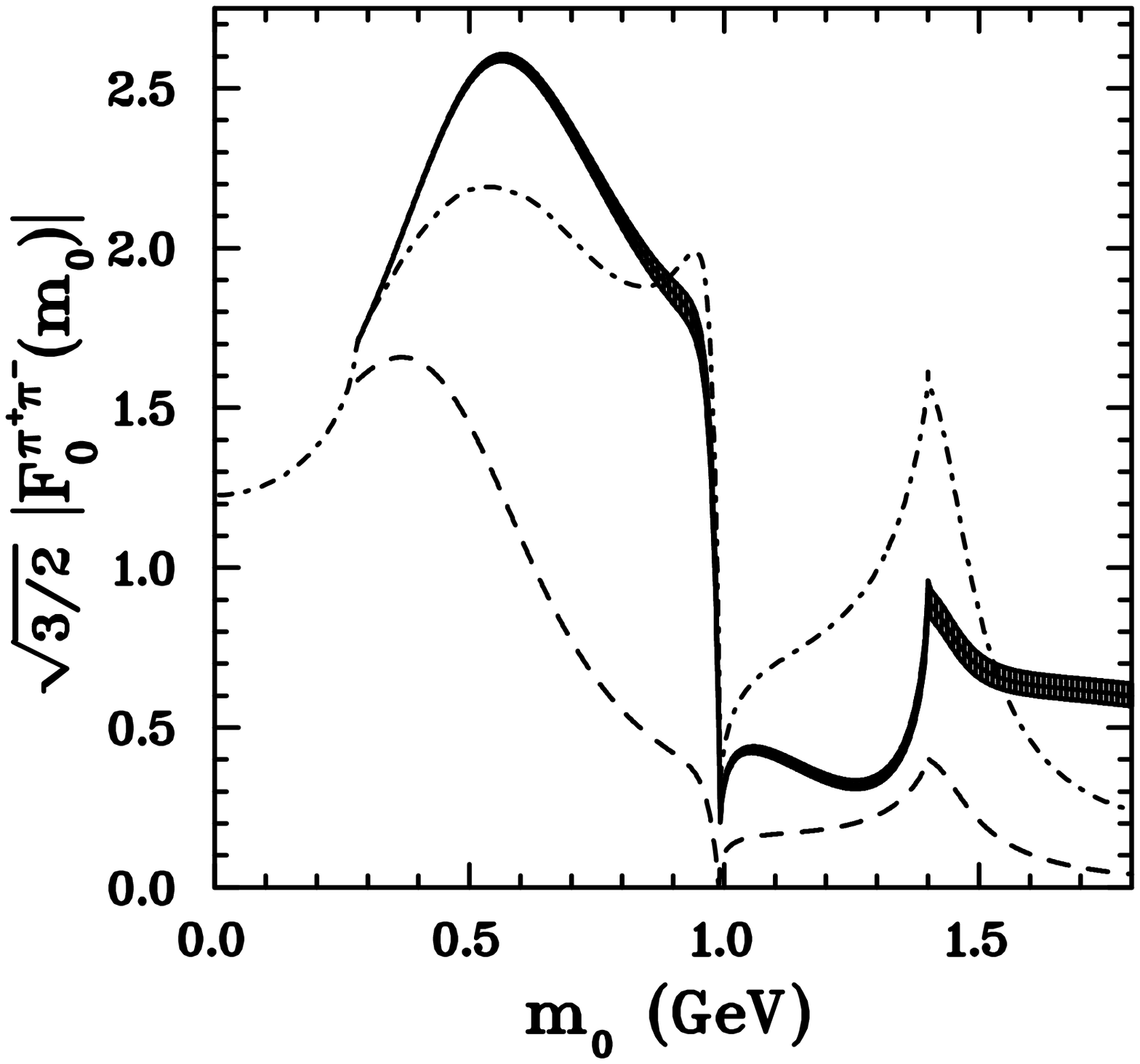}~~~~
\includegraphics[scale = 0.4]{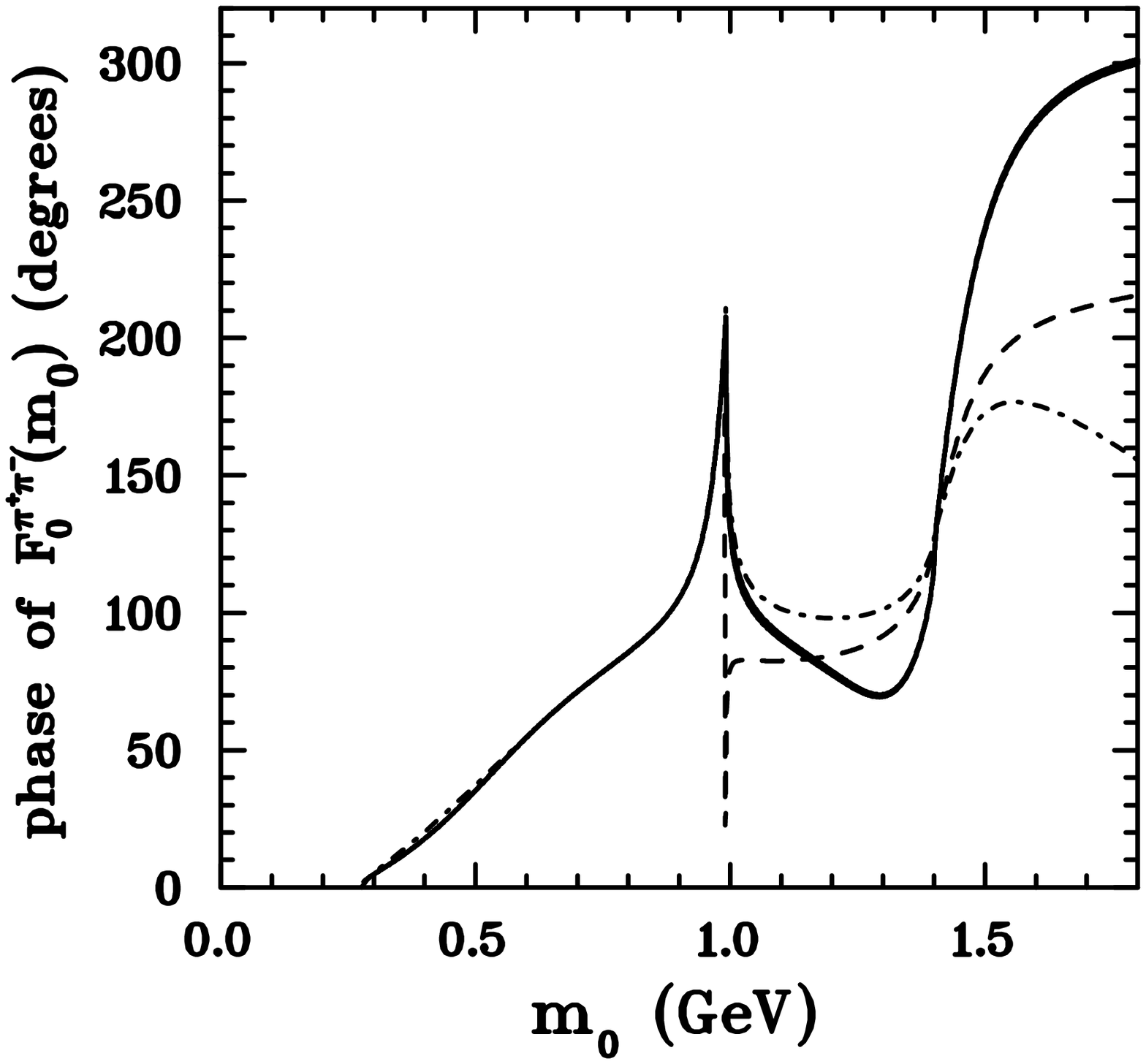}

\caption{The modulus (left panel) and the phase (right panel) of the pion scalar form factor 
$F^{\pi^+\pi^-}_0(m_0)$, obtained in the 
fit to the Belle data, is plotted as the dark band which represents its variation when the parameters 
$\kappa$ and $c$ vary within their errors given in Table~\ref{C}. It is compared with 
the same form factor introduced in Ref.~\cite{DedonderPol} with the parameters $\kappa=2$ GeV and 
$c= 19.5$ GeV$^{-4}$ (dashed line) and with that calculated using the Muskhelishvili-Omn\`es 
equations~\cite{Moussallam_2000} (dotted-dashed line).}
\label{fig7}
\end{center}
\end{figure} 

We fit also the total experimental branching fraction  of the 
$D^0~\to~K^0_S \pi^+ \pi^-$ decay, $Br^{tot}_{exp}=(2.82 \pm 0.19)~ \%$~\cite{PDG2012}. Denoting its contribution to the
 total $\chi^2$ function as $\chi^2_{Br}$ we define:
\begin{equation}
\label{chi}
 \chi^2= \chi^2_D + \chi^2_{\tau} + w\ \chi^2_{Br},
\end{equation}
where the weight $w$, in principle equals to 1, will be set so as to obtain reasonable value of the 
total branching fraction (see below).
The total number of free parameters in our model being equal to 33, the number of degrees of freedom, 
$ndf$, in the fit is $ndf = 6321+89+1-33 = 6378$. 
The combined $D^0$ and $\tau$ decay data fit leads, with $w=1$, to $\chi^2 = 9451$ which gives 
$\chi^2/ndf = 1.48$. The values of $\chi^2_D$, $\chi^2_{\tau}$ and $\chi^2_{Br}$ are equal to 
9328, 123 and 0.04, respectively. The calculated total branching fraction is 
$Br^{tot}= 2.78~\%$.
This fit is obtained for the pion vector form factor calculated according to Hanhart's model with
the 2C fit parameters shown in Table 1 of Ref.~\cite{Hanhart}. 
For the $K\pi$ vector form factor we have used the Belle parameterization of 
Ref.~\cite{EpifanovPLB654}.
The results quoted above have been obtained for the value of $f_K/f_{\pi}=1.175$ which belongs to input 
parameters in the $K\pi$ scalar form factor as described in Ref.~\cite{Bppk}. 
In studies of the $B$ decays into $K\pi^+\pi^-$~\cite{Bppk} the value $f_K/f_{\pi}=1.193$ has been used 
although it has already been noticed that 
the lower value of this ratio, $1.183$, gave an improved $\chi^2$.
Here, for the $D^0 \to K^0_S \pi^+ \pi^-$ decays, we have checked that with $f_K/f_{\pi}=1.193$ one obtains 
a much worse fit with $\chi^2=10045$.
 However, if we lower the $f_K/f_{\pi}$ value down to 1.165 the $\chi^2$ rises again to 9979, being 
by 528 units higher than the minimum of $\chi^2=9451$ for $f_K/f_{\pi}=1.175$.
Thus the functional dependence of the scalar $K\pi$ form factor on the $K\pi$ 
effective mass plays a major role in finding the $\chi^2$ minimum. Taking the vector $K\pi$ form factor
of Boito {\it et al.}~\cite{BoitoEPJC} instead of that  from Belle parametrization~\cite{EpifanovPLB654} leads to sligthly higher
$\chi^2=9488$. The two sets of parameters obtained for $\chi^2=9451$ and for $\chi^2=9488$ will be 
discussed in more detail below. However, for the sake of completeness we quote the corresponding 
$\chi^2$ values when the Hanhart's pion vector form factor is replaced by the Belle form factor of 
Ref.~\cite{Belletau2008}. Then one gets still higher $\chi^2$ values equal to 9514 and 9522, respectively. 
    
The resulting values of parameters for the best fit are shown in Table~\ref{C}.
As in the experimental analyses we fix the phase of the term multiplying the pion vector form 
factor $F_1^{\pi^+\pi^-}(s_0)$ to be zero. Consequently the parameter 
${\tilde A}_0^{\overline{K}^0R_P[\pi^+ \pi^-]}(m_{D^0}^2)$ is real as explained in Sec.~\ref{Brfrac}.
This forces us to introduce a tilda on the other form factor parameters appearing in Table~\ref{C}
to differentiate them from the physical form factors.
The value of $\chi_1$ can be estimated from a Breit-Wigner amplitude representation for the strange scalar
 meson $K^*_0(1430)$ whose decay into $K\pi$ dominates the $K\pi$ $S$-wave.  Using a formula similar to 
Eq.~(18) of Ref.~\cite{fkll} with 
 $\left\vert F_0^{\overline{K}^0\pi^-}(m_{K^*_0(1430)}^2)\right\vert = 1.73$~\cite{Bppk} 
for $f_K/f_{\pi}=1.175$
 one obtains $\chi_1=5.6$ GeV$^{-1}$ which is close to the value (5.43$\pm0.22$) GeV$^{-1}$ given in 
Table~\ref{C}. 
It is also  comparable to the $\chi_S^{eff}=(4.9 \pm 0.4)$ GeV$^{-1}$ obtained in the Dalitz plot analysis of 
the $D^+ \to K^- \pi^+ \pi^+$ decay performed in Ref.~\cite{BoitoPRD80_054007}, as can be seen from their 
Eq.~(38). 
A similar estimation of $\chi_2$   
for the $[\pi^+ \pi^-]$~$S$-wave is
unfeasible since in that channel  one has three scalar resonances which cannot be properly approximated by 
Breit-Wigner functions so the $\chi_2$ value represents an effective coupling. However its value is 
compatible with the $\chi_{f_0}$ value of ($26\pm9$)~GeV$^{-1}$ obtained in Ref.~\cite{BoitoPRD79_034020}
for the  $D^+ \to \pi^+ \pi^- \pi^+$ decays, 
as seen from their Eq.~(46). 

The parameters $q_6, s_6, q_7, s_7, q_{10} ,s_{10}$ are related to the $D$-wave contributions. As noted
in Sec. III, the  
multiplicative complex parameters $z_8$ and $z_9$ entering the doubly Cabibbo suppressed 
$\mathcal{M}_8$ and $\mathcal{M}_9$ 
amplitudes can be interpreted in terms of some charge independence violation in the $[K\pi]_{S,P}\ \pi$ 
systems [see Eqs.~(\ref{S818}) and~(\ref{S919})].

 The parameters $c$ and $\kappa$ enter the calculation
 of the pion scalar form factor as described in chapter 3 of Ref.~\cite{DedonderPol}. 
Figure~\ref{fig7} displays this form factor, obtained in the present fit to the Belle data 
compared to that calculated in the fit to the $B \to \pi \pi \pi$ data with $\kappa=2$ GeV and $c= 19.5$ 
GeV $^{-4}$ in Ref.~\cite{DedonderPol}. In spite of the seemingly large differences observed, we have 
checked 
that with the form factor fitted here to achieve the lowest $\chi^2_D$ for the $D^0\to K^0_S\pi^+\pi^-$ 
decay, the main conclusions drawn in Ref.~\cite{DedonderPol} for the $B\to \pi\pi\pi$ were
not altered.
This is due
 to the interplay between $\kappa$ and  $c$  with the parameter 
$\chi_S$ in Ref.~\cite{DedonderPol} and to the fact that the $B\to \pi\pi\pi$ data 
(see Ref.~\cite{AubertPRD79}) are 
statistically
less restricting than the $D^0\to K^0_S\pi^+\pi^-$ data.
We also 
want to point out that the modulus of the pion scalar form factor is presently closer to that of the form 
factor calculated by Moussallam solving the Muskhelishvili-Omn\`es equations~\cite{Moussallam_2000},
notably below $1$ GeV.
Moussallam's form factor has been calculated for the meson-meson amplitudes taken from 
the
three-channel model of Ref.~\cite{KLL} under an additional assumption that the off-diagonal matrix
 elements
$T_{13}$ and $T_{23}$ are set equal to zero in the region below the third threshold ($m_0 < 1.4$ GeV).
Moreover the cut-off energy $E_0$ defined in ~\cite{Moussallam_2000} has been chosen equal to 2 GeV.
  
\begin{figure}
\centering
\includegraphics[width=7cm,angle=0]{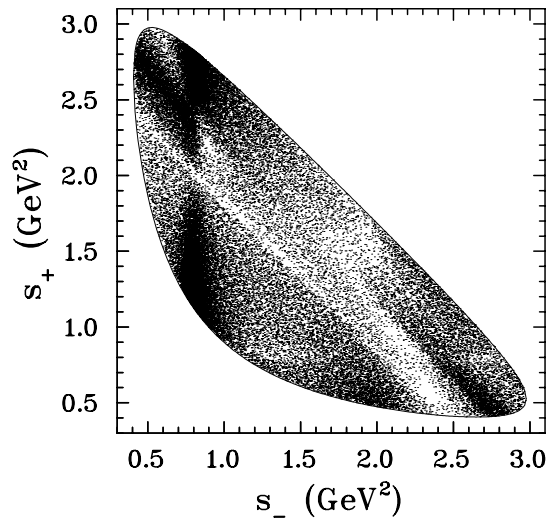}
\caption{ Dalitz plot distribution from the fit to the Belle 
data~\cite{A.Poluektov_PRD81_112002_Belle}.} \label{densfitPol}
\end{figure}

\begin{figure}[h]
\centering
\includegraphics[width=7cm,angle=0]{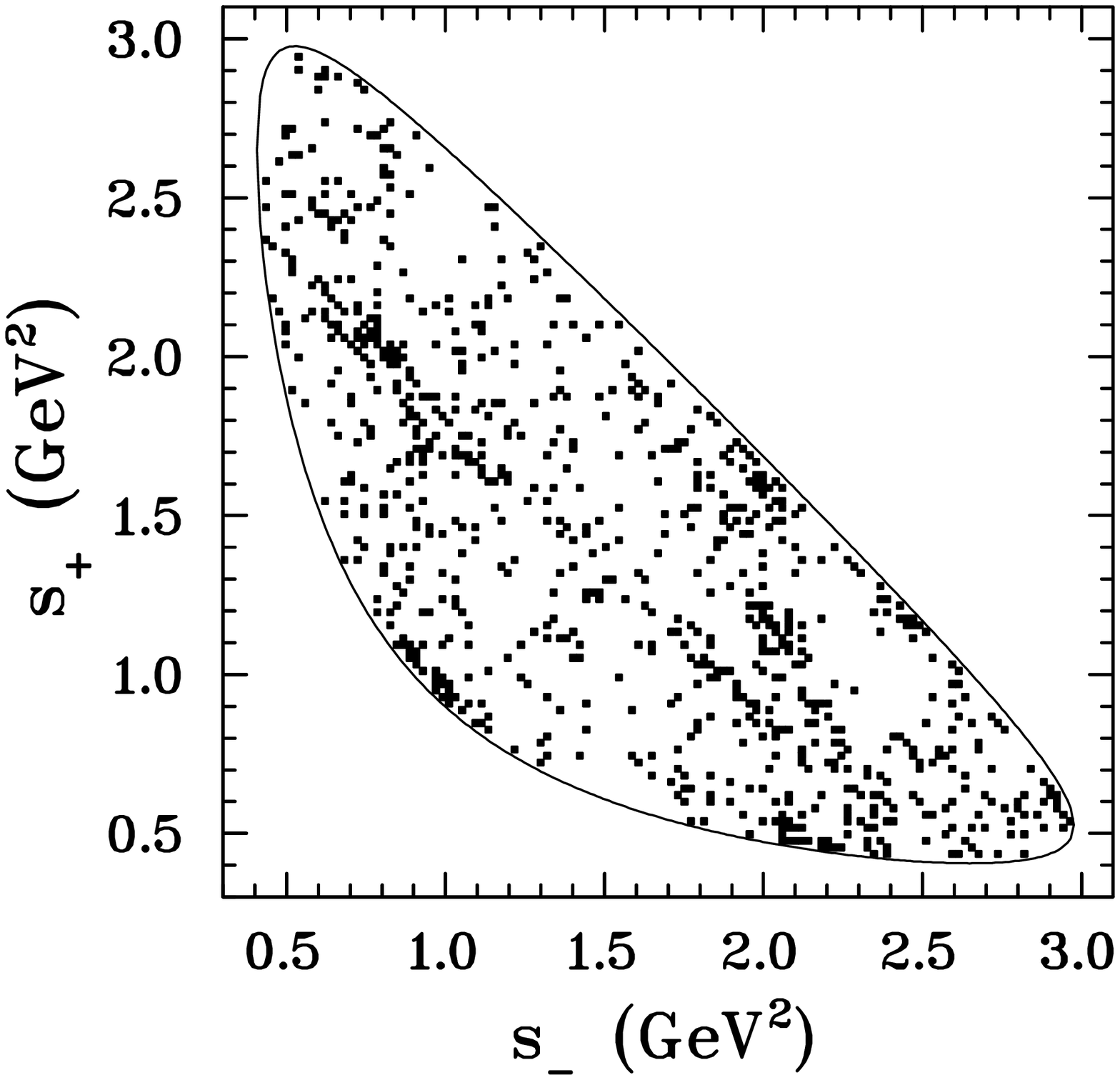}
\caption{ Distribution of the $\chi^2$ values inside the Dalitz plot contour
drawn as a solid line.
Black squares correspond to $\chi^2$ values larger than 4. }\label{denschi2}
\end{figure}

The Dalitz plot density distribution that emerges from the fit of our model to the Belle data is
plotted in Fig.~\ref{densfitPol}. It displays a very rich interference pattern dominated by the 
presence of the $K^*(892)$ resonance. Figure~\ref{denschi2} illustrates the distribution of $\chi^2$ in the Dalitz plot. It shows that there is only a limited number of regions where the $\chi^2$ exceeds 4 and, thus, that 
a good overall agreement of our model with the experimental density distribution of 
Ref.~\cite{A.Poluektov_PRD81_112002_Belle} is achieved.
The mass and width of the charged $K^*(892)$ that come out of the minimization process are in very good 
agreement with the determination of the Belle Collaboration for $\tau^- \to K^0_S \pi^-\nu_{\tau}$ 
decays~\cite{EpifanovPLB654}.

\begin{table*}[h]
\caption{Parameters obtained from the best fit to the Belle 
data~\cite{A.Poluektov_PRD81_112002_Belle} ($\chi^2 = 9451$). The first error is statistical and the second
one shows the modulus of the difference between the parameter value obtained in the fit 
using the $K\pi$ form factor of Boito \textit{et al.}~\cite{BoitoEPJC} ($\chi^2=9488$)
and that of the best fit performed with the Belle parametrization~\cite{EpifanovPLB654} for this form
factor.}
\label{C}
\begin{center}
\begin{tabular}{lrr}
\hline
\hline
\hspace{0.0cm}parameter & \multicolumn{1}{c}{modulus} & phase (deg) \\ 
\hline
$\chi_1$  &    5.43 $\pm$ 0.22 $\pm$ 0.00  &   248.1 $\pm$ 1.3 $\pm$  2.0   \\
$\chi_2$  & 32.50 $\pm$ 1.21 $\pm$ 0.09  &   221.9 $\pm$ 0.9 $\pm$  0.7  \\
$\tilde{F}_0^{ \pi^+ R_S[\overline{K}^0 \pi^-],}(m_{D^0}^2)$  &   1.94 $\pm$ 0.03 $\pm$ 0.00  &   245.6 $\pm$ 1.1 $\pm$  1.1 \\
$\tilde{F}_0^{ \overline{K}^0 R_S[\pi^- \pi^+]}(m_{D^0}^2)$  &   1.36 $\pm$ 0.02 $\pm$ 0.00  &    37.7 $\pm$ 0.4 $\pm$  0.2  \\
$\tilde{A}_0^{\pi^+ R_P [\overline{K}^0\pi^-]}(m_{D^0}^2)$  &   0.95 $\pm$ 0.05 $\pm$ 0.06  &   294.2 $\pm$ 2.2 $\pm$ 11.9  \\
$\tilde{A}_0^{ \overline{K}^0 R_P[\pi^- \pi^+]}(m_{D^0}^2)$  &   0.66 $\pm$ 0.04 $\pm$ 0.01  &     0.0 (fixed)\,\,\,\,\,\,\,\,\, \\
$\tilde{A}_0^{ \overline{K}^0 \omega}(m_{D^0}^2)$    &   1.23 $\pm$ 0.04 $\pm$ 0.03  &   319.1 $\pm$ 1.1 $\pm$  0.2  \\
$q_6$  &  1.44 $\pm$ 0.07 $\pm$ 0.15  &    26.2 $\pm$ 1.6 $\pm$  3.8   \\
$s_6$   &   1.84 $\pm$ 0.09 $\pm$ 0.16  &   199.2 $\pm$ 1.3 $\pm$  1.5  \\
$q_7$   &   0.68 $\pm$ 0.03 $\pm$ 0.02  &   245.9 $\pm$ 1.6 $\pm$  4.9  \\
$s_7$  &   1.01 $\pm$ 0.05 $\pm$ 0.03  &   102.3 $\pm$ 1.7 $\pm$  4.1  \\
$z_8$  &   2.09 $\pm$ 0.12 $\pm$ 0.04  &   206.1 $\pm$ 3.1 $\pm$  3.5  \\
$z_9$  &   1.64 $\pm$ 0.09 $\pm$ 0.31  &   135.3 $\pm$ 1.9 $\pm$  0.3  \\
$q_{10}$   &  23.19 $\pm$ 1.26 $\pm$ 3.10  &   220.8 $\pm$ 3.1 $\pm$ 15.6  \\
$s_{10}$   &  24.26 $\pm$ 1.33 $\pm$ 3.74  &    40.3 $\pm$ 3.0 $\pm$ 14.5  \\
$c$ (GeV$^{-4}$)  &   0.29 $\pm$ 0.02 $\pm$ 0.02  &                               \\
$\kappa$ (MeV)  & 305.61 $\pm$ 2.74 $\pm$ 1.33  &                               \\
$m_{K^{*\mp}}$   (MeV) & 894.74 $\pm$ 0.08             &                               \\
$\Gamma_{K^*}$   (MeV)  &  46.98 $\pm$ 0.18             &                               \\
\hline
\hline
\end{tabular}
\end{center}
\end{table*}

In Ref.~\cite{SanchezPRL105_081803} the BABAR Collaboration has reported results of their Dalitz plot 
analysis containing 540800 signal events for the $D^0 \to K^0_S \pi^+ \pi^-$ decays.
The Dalitz plot density distribution has been fitted using the isobar model with 43 free parameters.
In the present work the values of the density distribution are calculated starting from 
a $1000\times1000$ grid
tabulating the values of the BABAR model decay amplitude~\cite{F.Martinez_private2013}. 
Summing these values in adjacent cells one gets a set of pseudo-data on a
$125\times125$ grid with 7286 cells.
Then the 33 free parameters of our model are fitted to these data using the same method as 
described above 
for the Belle data.
The weight $w$ of $\chi^2_{Br}$ in Eq.~(\ref{chi}) is increased by a factor 10 since with $w=1$ one 
obtains a much
 too low value of $Br^{tot}$  in comparison with the experimental value.
Then, the total $\chi^2$ equals to 6687 for $ndf=7286+89+1-33=7343$ which gives $\chi^2/ndf=0.91$. 
The values of $\chi^2_D$, $\chi^2_{\tau}$ and $\chi^2_{Br}$ are 
6533, 151 and 0.3, respectively ($Br^{tot}= 2.71~\%$).
Taking as previously the alternative vector $K\pi$ form factor
from Ref.~\cite{BoitoEPJC} instead of that from Ref.~\cite{EpifanovPLB654} leads to a much higher
$\chi^2=6951$. 

\begin{figure}
\includegraphics[height=.35\textheight]{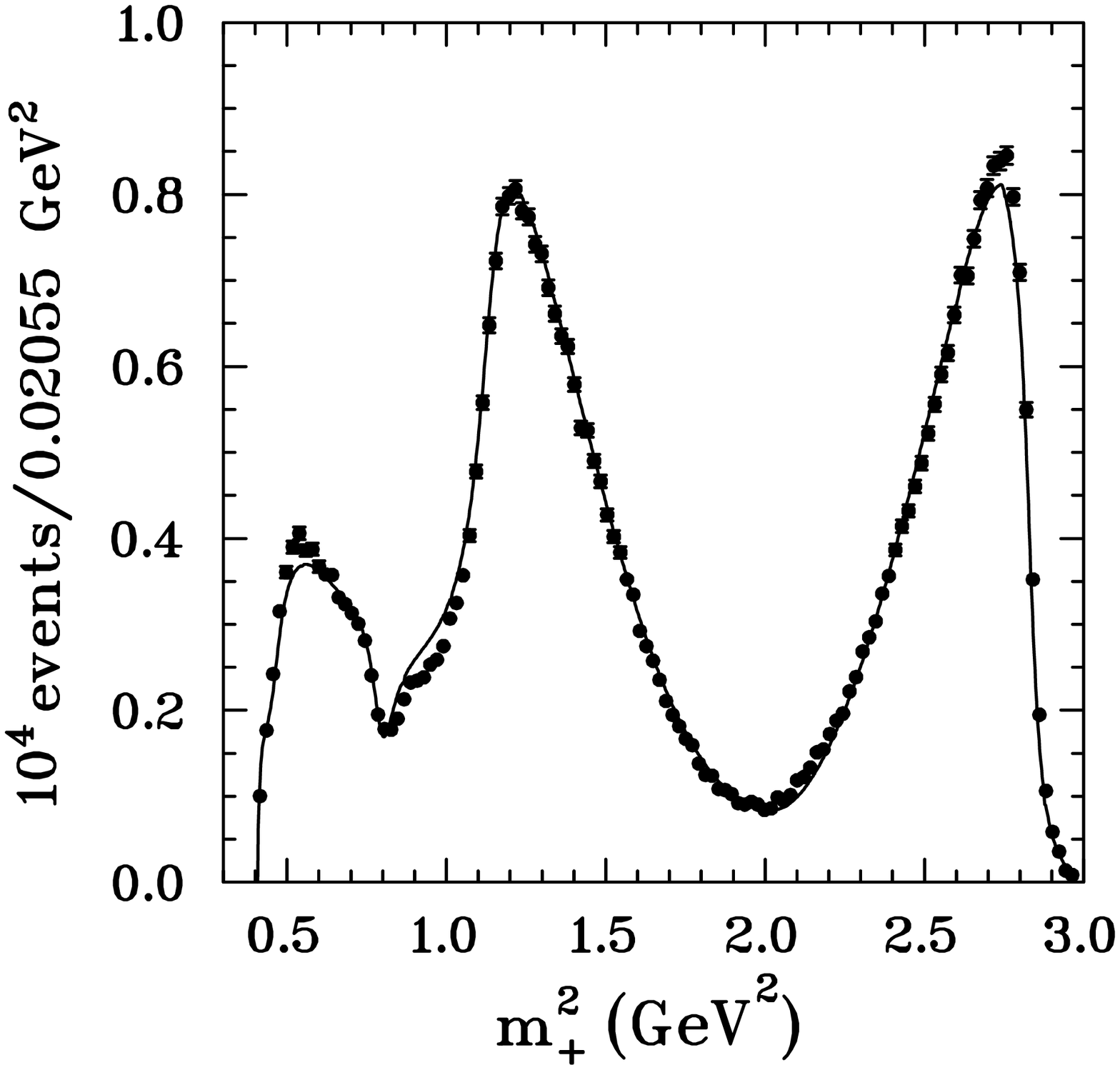}~~~~
\includegraphics[height=.35\textheight]{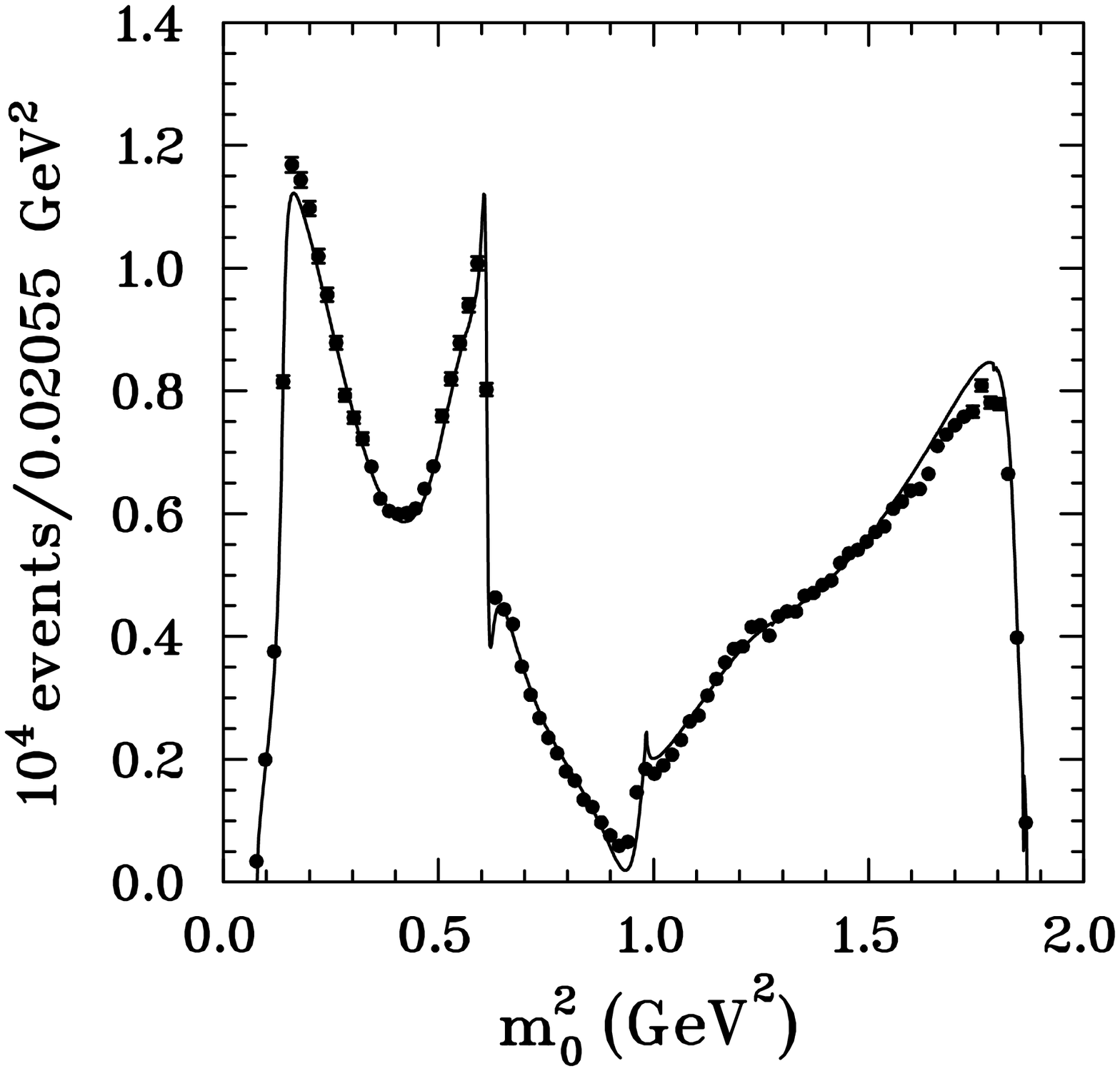}~~~~
\caption{Left panel: comparison of the $K^0_S\pi^+$ effective mass squared distributions
for the Belle data~\cite{A.Poluektov_PRD81_112002_Belle}  (black dots) with the BABAR 
model~\cite{F.Martinez_private2013} (solid curve), normalized to the number of events of the Belle 
experiment. Right panel: as in left panel but  for the $\pi^+\pi^-$ effective mass 
squared.}
\label{compPolmart}
\end{figure}

Compared to Table~\ref{C}, Table~\ref{parMartinez} reveals that the numerical values of the parameters fitted to the Belle data and to
the BABAR model are quite close. 
Somehow indirectly this means that the Dalitz density distributions measured by both collaborations are 
very similar.
  Some noticeable differences between parameters are seen, mostly for the amplitudes whose 
contributions are small.
In Fig.~\ref{compPolmart} two one-dimensional projections of the Dalitz density distributions are shown
as an illustration of an overall agreement of the Belle data and the BABAR model.

\begin{table*}[h]
\caption{Parameters obtained from the best fit to the BABAR model 
data~\cite{F.Martinez_private2013} ($\chi^2 = 6687$). The first error is statistical and the second
one shows the modulus of the difference between the parameter value obtained in the fit using the
$K\pi$ vector form factor of Boito \textit{et al.}~\cite{BoitoEPJC} ($\chi^2=6951$) and that of the 
best fit
performed with the Belle parametrization~\cite{EpifanovPLB654} for this form factor.}
\label{parMartinez}
\begin{center}
\begin{tabular}{lrrrr}
\hline
\hline
\hspace{0.0cm}parameter & \multicolumn{1}{c}{modulus} & phase (deg) \\ 
\hline
$\chi_1$  &     5.08 $\pm$ 0.10 $\pm$ 0.03  &   229.0 $\pm$ 1.1 $\pm$  2.0 \\
$\chi_2$  & 32.89 $\pm$ 0.46 $\pm$ 0.13  &   214.1 $\pm$ 0.6 $\pm$  0.1 \\
$\tilde{F}_0^{ \pi^+ R_S[\overline{K}^0 \pi^-],}(m_{D^0}^2)$  &   1.99 $\pm$ 0.03 $\pm$ 0.00  &   262.8 $\pm$ 1.0 $\pm$  1.2 \\
$\tilde{F}_0^{ \overline{K}^0 R_S[\pi^- \pi^+]}(m_{D^0}^2)$  &   1.41 $\pm$ 0.01 $\pm$ 0.00  &    41.0 $\pm$ 0.3 $\pm$  0.4 \\
$\tilde{A}_0^{\pi^+ R_P [\overline{K}^0\pi^-]}(m_{D^0}^2)$  &  0.96 $\pm$ 0.02 $\pm$ 0.05  &   287.5 $\pm$ 0.9 $\pm$ 10.8 \\
$\tilde{A}_0^{ \overline{K}^0 R_P[\pi^- \pi^+]}(m_{D^0}^2)$  &  0.61 $\pm$ 0.01 $\pm$ 0.00  &    0.0 (fixed)\,\,\,\,\,\,\,\,\, \\ 
$\tilde{A}_0^{ \overline{K}^0 \omega}(m_{D^0}^2)$    &  1.12 $\pm$ 0.02 $\pm$ 0.01  &   318.9 $\pm$ 0.6 $\pm$  0.1 \\
$q_6$  &   1.24 $\pm$ 0.03 $\pm$ 0.05  &    50.2 $\pm$ 1.7 $\pm$  6.3 \\
$s_6$   &    1.50 $\pm$ 0.04 $\pm$ 0.10  &   217.4 $\pm$ 1.3 $\pm$  3.8 \\
$q_7$   &   0.74 $\pm$ 0.02 $\pm$ 0.02  &   227.2 $\pm$ 1.0 $\pm$  4.4 \\
$s_7$  &   0.82 $\pm$ 0.03 $\pm$ 0.02  &    69.4 $\pm$ 1.5 $\pm$  5.3 \\
$z_8$  &   2.84 $\pm$ 0.08 $\pm$ 0.06  &   182.5 $\pm$ 1.9 $\pm$  3.8 \\
$z_9$  &   1.53 $\pm$ 0.04 $\pm$ 0.26  &   126.9 $\pm$ 1.0 $\pm$  0.3 \\
$q_{10}$   &   21.17 $\pm$ 0.69 $\pm$ 4.15  &   199.6 $\pm$ 2.2 $\pm$ 11.8 \\
$s_{10}$   &  22.36 $\pm$ 0.74 $\pm$ 4.81  &    17.9 $\pm$ 2.2 $\pm$  9.6 \\
$c$ (GeV$^{-4}$)  &    0.19 $\pm$ 0.01 $\pm$ 0.02  &                              \\
$\kappa$ (MeV)  &  306.09 $\pm$ 1.78 $\pm$ 0.72  &                              \\
$m_{K^{*\mp}}$   (MeV) &  894.31 $\pm$ 0.07             &                              \\
$\Gamma_{K^*}$   (MeV)  &  46.90 $\pm$ 0.15             &                              \\
\hline
\hline
\end{tabular}
\end{center}
\end{table*}

The total branching fractions for different quasi two-body channel amplitudes are given in 
tables~\ref{TabBr} and ~\ref{TabBrMartinez}. The contribution of the $[K^0_S \pi^-]_P \,\pi^+$ amplitude is clearly dominant
as was also found in the isobar model analysis for the $K^*(892)^-\pi^+$ of the 
Belle~\cite{ZhangPRL99_131803} and BABAR~\cite{SanchezPRL105_081803} Collaborations. 
The four amplitudes ${\cal M}_1$, ${\cal M}_2$, ${\cal M}_3$ and ${\cal M}_4$ 
give sizable contributions while the branching fractions of the remaining amplitudes are small. 
Our branching fraction for the ${\cal M}_3$ and ${\cal M}_4$ 
amplitudes compare well with the $K^*(892)\pi$ and $K^0_S\rho$ determinations of the experimental analyses 
\cite{ZhangPRL99_131803,SanchezPRL105_081803,A.Poluektov_PRD81_112002_Belle}.

The amplitudes ${\cal M}_1$ and ${\cal M}_2$, corresponding to the $S$-wave $K^0_S\ \pi$ and
 $\pi^+\pi^-$ subchannels, merge contributions from several resonances. Then, if one 
wishes, for example, to compare the branching fraction
 $(16.92\pm1.27)$ \% obtained for the amplitude ${\cal M}_2$ (see Table~\ref{TabBr}) with the 
results of the Belle Collaboration~\cite{A.Poluektov_PRD81_112002_Belle} one
has to combine in the latter case the branching fractions for the following intermediate states:
 $K^0_S\ \sigma_1$,
$K^0_S\ f_0(980)$, $K^0_S\ \sigma_2$ and $K^0_S\ f_0(1370)$. The sum of these four contributions,
 $18.16$~\% compares well with the above  value of our fit. 
Because of interferences between amplitudes the sum of the partial branching fractions differs
from $100$~\%. For example, for the fit to the Belle data it is equal to $132.8$~\%, so that  the total sum of 
the interference terms with respect to the total branching fraction amounts to $-32.8$~\%. 
The most important negative interference terms are equal to $-26.4$~\% for the amplitudes ${\cal M}_1$ and
 ${\cal M}_2$ and $-10.1$~\% for the amplitudes ${\cal M}_3$ and ${\cal M}_4$, respectively.
There is also a positive interference term of $10.5$ \% for the ${\cal M}_2$ and ${\cal M}_3$ amplitudes.
Other interference contributions are much smaller.

As a consequence of the arbitrary choice of the ${\cal M}_4$
amplitude phase, one can only calculate the lower or upper limits of the branching fractions of the annihilation amplitudes   (see derivation in Sec.~\ref{Brfrac}). 
Their lower limits are displayed in Tables~\ref{TabBr}
and ~\ref{TabBrMartinez}.
 These are sizable for the ${\cal M}_1$, ${\cal M}_2$, ${\cal M}_3$ and ${\cal M}_4$  
 cases. This points out to the 
 importance of the annihilation-diagram contributions. 
As can be seen from Eq.~(\ref{ineqA}) in Sec.~\ref{Brfrac}, the upper limits are larger than the sum of 
the branching fractions $Br_i$ and $Br_i^{tree}$. Therefore they are not shown in Table~\ref{TabBr}.

Lower limits, $Br^{''}_{ann.~low}$, of the summed annihilation amplitudes with the
exclusion of the small components ${\cal M}_6$ and ${\cal M}_7$ can be calculated using Eq.~(\ref{lowl}).
These divided by 
the fitted total branching fraction $Br^{tot}$ are $(20.0\pm 2.5)$~\% and 
$(20.5\pm2.1)$~\%, for the fits to the Belle data and to the BABAR model, respectively. 
The corresponding values of the tree branching fractions defined in Eq.~(\ref{brtree}) are $45.9$~\% and 
$46.7$~\% for the two cases considered 
here. Taking into account the above large values of the lower limits of the annihilation branching fractions,
 close to $20$~\%, one must conclude that the annihilation contributions are important when compared with the tree amplitude terms. 

The importance of the annihilation diagrams has also been pointed out in Refs.~\cite{ Fu-Sheng_Yu_PRD84_074019}, \cite{ChengPRD81_074021} and~\cite{ChengPRD81_074031}.
In Ref.~\cite{Fu-Sheng_Yu_PRD84_074019} a calculation of branching ratios for two-body hadronic decays 
of $D$ and $D_s$ mesons into pseudoscalar-pseudoscalar and pseudoscalar-vector mesons has been 
performed in a factorization approach for the \textquotedblleft emission"-type diagrams and in a 
pole-dominance model for the annihilation-type diagrams. Relative strong phases between the different
 diagrams were introduced to obtain a better reproduction of the experimental data.
As in  our model, the contribution of the annihilation diagrams were found to be relatively large.
An analysis of experimental data on branching fractions of charmed meson decays into 
pseudoscalar-pseudoscalar and pseudoscalar-vector mesons has been performed in 
Ref.~\cite{ChengPRD81_074021} using a quark-diagram approach.
It  
suggests that $W$-exchange topology 
must play an important role.
A comparison with the factorization procedure allowed to extract information on the effective Wilson coefficients and to discriminate between different solutions obtained in the diagrammatic scheme.
The flavor-diagram approach has also been used in Ref.~\cite{ChengPRD81_074031} to study $D$ and $D_s$ decays into a pseudoscalar meson and an even-parity scalar or axial vector or tensor meson.
It was found that the contribution of annihilation diagrams could be important. The factorization formalism has also been used as a complementary tool to calculate some decay rates and again the inclusion of weak annihilation processes was found to be necessary to account for the data. 

\begin{table}[h]
\caption{Branching fractions ($Br$) for different quasi two-body channels calculated for the best 
fit to the Belle data~\cite{A.Poluektov_PRD81_112002_Belle} ($\chi^2 = 9451$).
The sum of branching fractions is 132.81~\%.
The branching fractions for the tree amplitudes (tree), and the lower limits for the 
annihilation amplitudes (ann. low) are also given. The first error of $Br$ is statistical. The second error
 of $Br$ and the errors of the tree and annihilation parts show the 
difference between the branching fractions obtained for the fit with $\chi^2=9488$ and those for the best 
fit (see Table~\ref{C} caption). All numbers are in per cent.}
\label{TabBr}
\begin{center}
\begin{tabular}{clrrr}
\hline
\hline
Amplitude & channel & \multicolumn{1}{c}{Br} & \multicolumn{1}{c}{tree} & \multicolumn{1}{c}{ann. low} \\ 
\hline
${\cal{M}}_1$ & $[K^0_S \,\pi^-]_S \,\pi^+$ & 25.03 $\pm$  3.61 $\pm$  0.18 &  8.24 $\pm$  0.10 &  7.88 $\pm$  0.11 \\
${\cal{M}}_2$ &  $K^0_S[\pi^-\pi^+]_S$ & 16.92 $\pm$  1.27 $\pm$  0.02 & 14.70 $\pm$  0.17 &  2.92 $\pm$  0.09 \\
${\cal{M}}_3$ &   $[K^0_S \,\pi^-]_P \,\pi^+$ & 62.72 $\pm$  4.45 $\pm$  0.15 & 24.69 $\pm$  5.65 &  8.74 $\pm$  2.97 \\
${\cal{M}}_4$ &   $K^0_S[\pi^-\pi^+]_P$ & 21.96 $\pm$  1.55 $\pm$  0.06 &  4.36 $\pm$  0.06 &  6.74 $\pm$  0.04 \\
${\cal{M}}_5$ & $K^0_S \omega$         &  0.79 $\pm$  0.07 $\pm$  0.04 &  0.24 $\pm$  0.01 &  0.16 $\pm$  0.02 \\
${\cal{M}}_6$ &  $[K^0_S \,\pi^-]_D\, \pi^+$ &  1.41 $\pm$  0.11 $\pm$  0.04 &  &  \\
${\cal{M}}_7$ &  $K^0_S[\pi^-\pi^+]_D$ &  2.15 $\pm$  0.19 $\pm$  0.10 &  &  \\
${\cal{M}}_8$ &  $[K^0_S \,\pi^+]_S \,\pi^-$ &  0.56 $\pm$  0.07 $\pm$  0.03 &  0.07 $\pm$  0.00 &  0.29 $\pm$  0.02 \\
${\cal{M}}_9$ & $[K^0_S \,\pi^+]_P \,\pi^-$ &  0.64 $\pm$  0.06 $\pm$  0.02 &  0.77 $\pm$  0.15 &  0.01 $\pm$  0.01 \\
${\cal{M}}_{10}$ & $[K^0_S \,\pi^+]_D \,\pi^-$ &  0.63 $\pm$  0.07 $\pm$  0.11 & 0\,\,\,\,\,\,\,\,\,\,\,\,\,  & 0.63 $\pm$  0.11 \\
\hline
\hline
\end{tabular}
\end{center}
\end{table}

\begin{table}[h]
\caption{Branching fractions ($Br$) for different quasi two-body channels calculated for the best 
fit to the BABAR model data~\cite{SanchezPRL105_081803} ($\chi^2 = 6687$). 
The sum of branching fractions is 138.77~\%.
The branching fractions for the tree amplitudes (tree), and the lower limits for the 
annihilation amplitudes (ann. low) are also given. The first error of $Br$ is statistical. The second error
 of $Br$ and the errors of the tree and annihilation parts show the 
difference between the branching fractions obtained for the fit with $\chi^2=6951$ and those for the best 
fit (see Table~\ref{parMartinez} caption). All numbers are in per cent.}
\label{TabBrMartinez}
\begin{center}
\begin{tabular}{clrrr}
\hline
\hline
Amplitude & channel & \multicolumn{1}{c}{Br} & \multicolumn{1}{c}{tree} & \multicolumn{1}{c}{ann. low} \\ 
\hline
${\cal{M}}_1$ & $[K^0_S \,\pi^-]_S \,\pi^+$ & 30.11 $\pm$  1.25 $\pm$  0.03 &  7.40 $\pm$  0.13 & 10.64 $\pm$  0.04 \\
${\cal{M}}_2$ &  $K^0_S[\pi^-\pi^+]_S$ & 21.57 $\pm$  0.55 $\pm$  0.25 & 16.25 $\pm$  0.12 &  4.20 $\pm$  0.16 \\
${\cal{M}}_3$ &   $[K^0_S \,\pi^-]_P \,\pi^+$ & 60.36 $\pm$  1.39 $\pm$  0.28 & 25.33 $\pm$  5.60 &  7.53 $\pm$  2.77 \\
${\cal{M}}_4$ &   $K^0_S[\pi^-\pi^+]_P$ & 20.79 $\pm$  0.21 $\pm$  0.11 &  4.48 $\pm$  0.03 &  5.96 $\pm$  0.03 \\
${\cal{M}}_5$ & $K^0_S \omega$         &  0.64 $\pm$  0.02 $\pm$  0.01 &  0.25 $\pm$  0.00 &  0.09 $\pm$  0.00 \\
${\cal{M}}_6$ &  $[K^0_S \,\pi^-]_D\, \pi^+$ &  1.38 $\pm$  0.04 $\pm$  0.06 &  &  \\
${\cal{M}}_7$ &  $K^0_S[\pi^-\pi^+]_D$ &  1.75 $\pm$  0.07 $\pm$  0.12 &  &  \\
${\cal{M}}_8$ &  $[K^0_S \,\pi^+]_S \,\pi^-$ &  0.99 $\pm$  0.06 $\pm$  0.06 &  0.13 $\pm$  0.00 &  0.50 $\pm$  0.03 \\
${\cal{M}}_9$ & $[K^0_S \,\pi^+]_P \,\pi^-$ &  0.64 $\pm$  0.03 $\pm$  0.02 &  0.68 $\pm$  0.11 &  0.00 $\pm$  0.00 \\
${\cal{M}}_{10}$ & $[K^0_S \,\pi^+]_D \,\pi^-$ &  0.54 $\pm$  0.03 $\pm$  0.15 & 0\,\,\,\,\,\,\,\,\,\,\,\,\,  & 0.54 $\pm$  0.15  \\
\hline
\hline
\end{tabular}
\end{center}
\end{table}

\vspace{5pt}

\begin{figure}
  \includegraphics[height=.35\textheight]{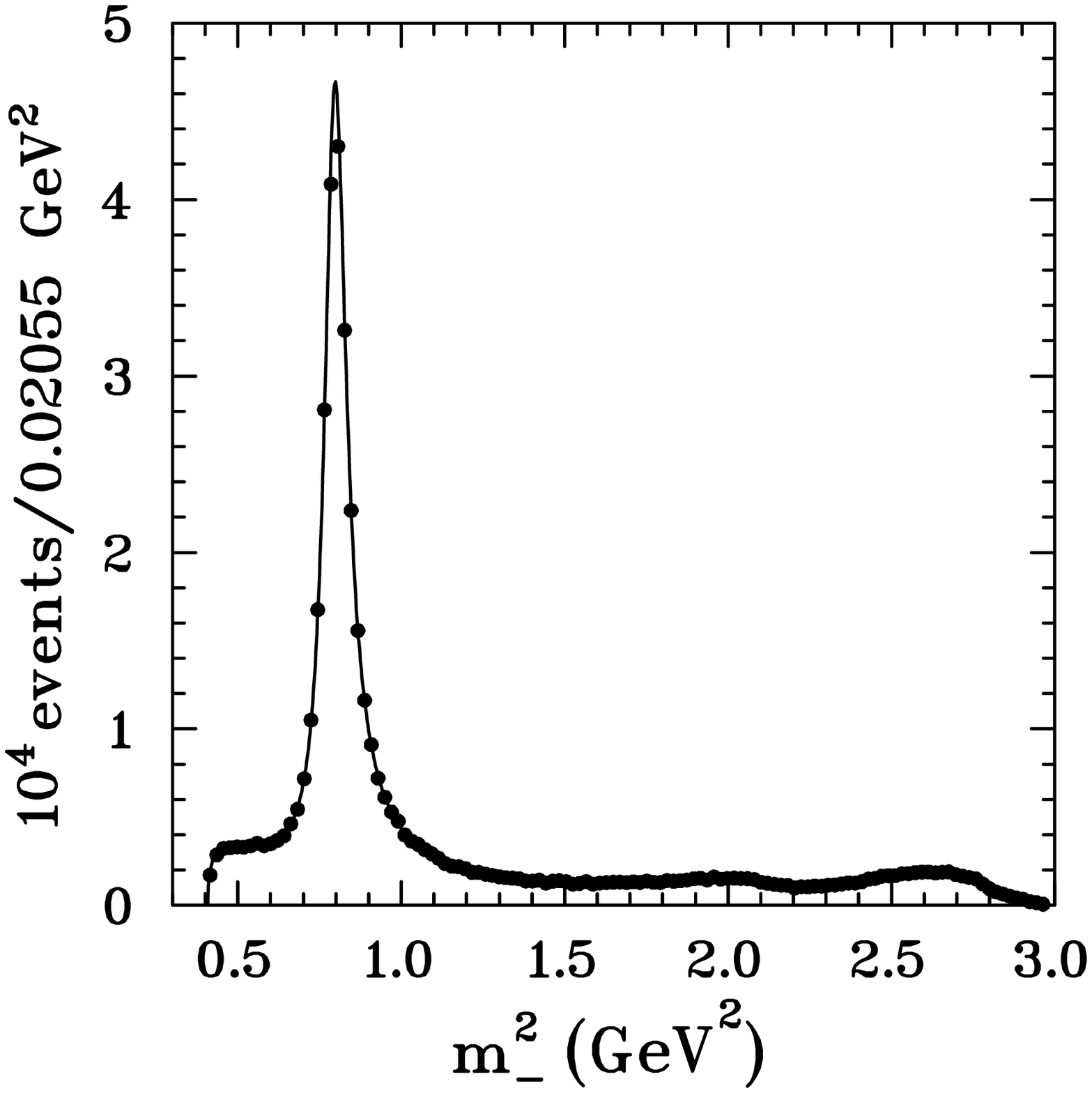}~~~~
   \includegraphics[height=.35\textheight]{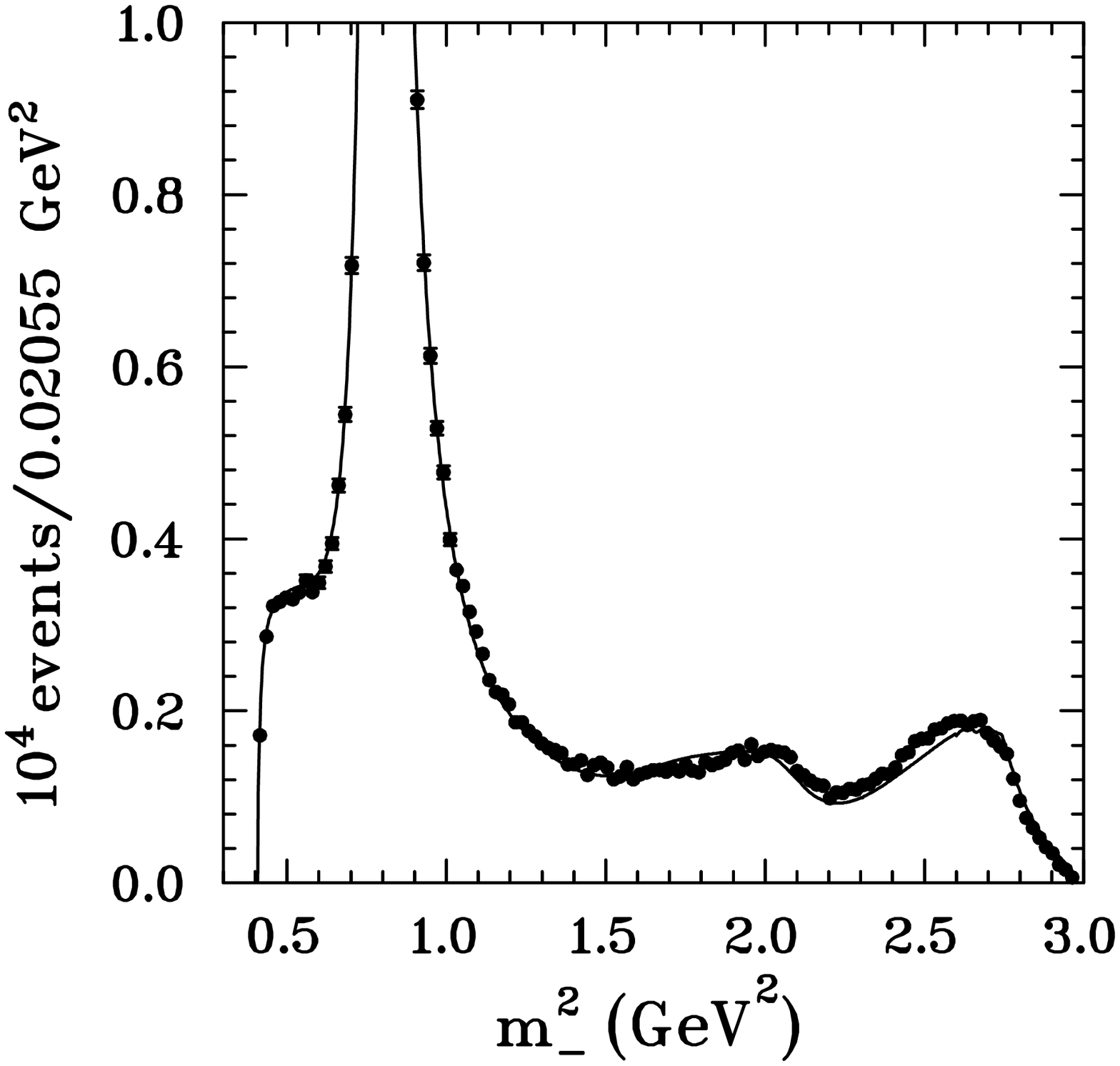}~~
\caption{Comparison of the $K^0_S\pi^-$ effective mass squared distributions
for our model (solid curve) with the Belle data~\cite{A.Poluektov_PRD81_112002_Belle} (points with error 
bars). In the right panel the vertical scale is enlarged by a factor of 5 in order to enforce
the differences at higher $K^0_S\pi^-$ masses. 
}\label{Polmmin}
\end{figure}

\begin{figure}
  \includegraphics[height=.35\textheight]{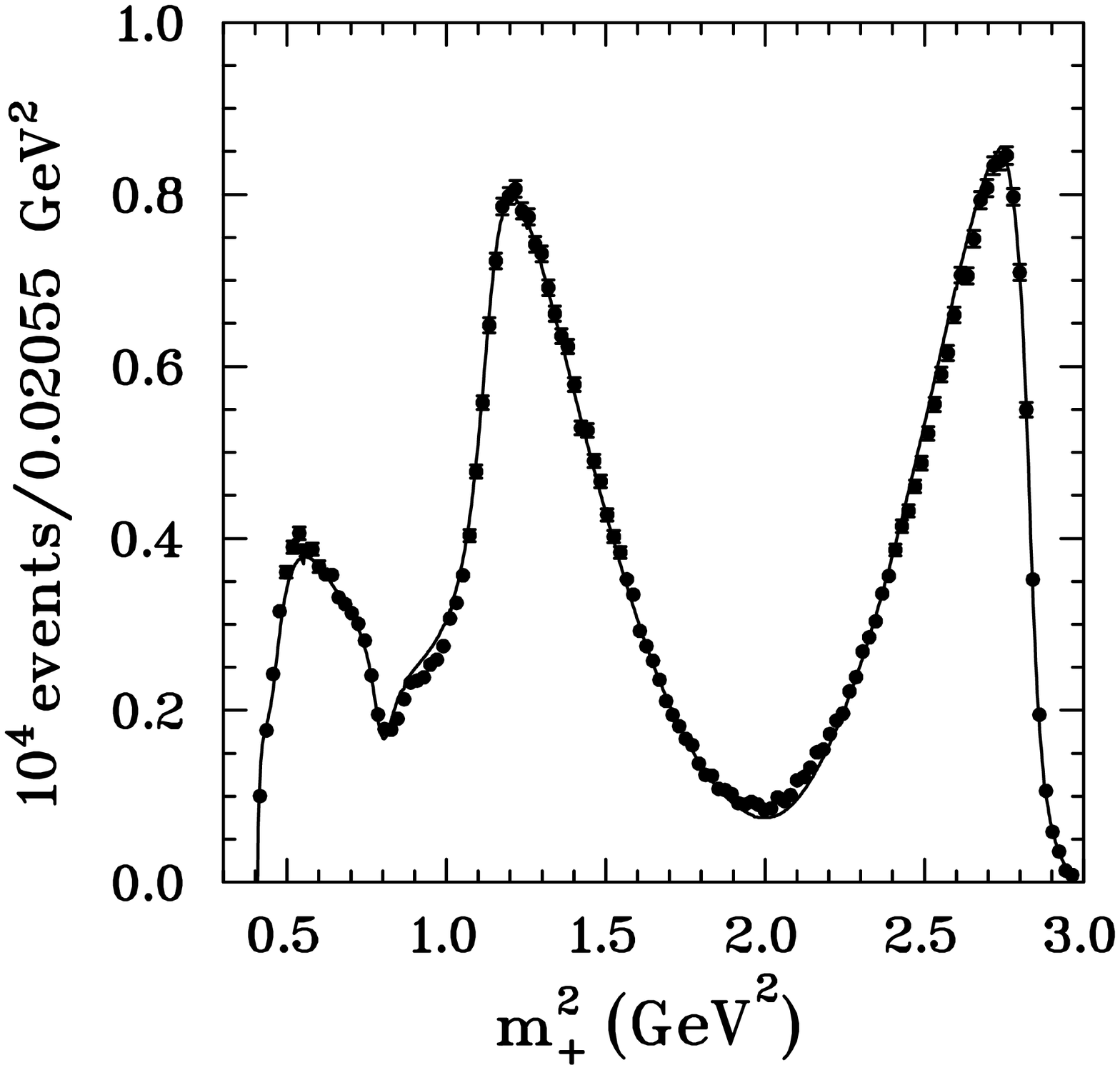}~~~~
   \includegraphics[height=.35\textheight]{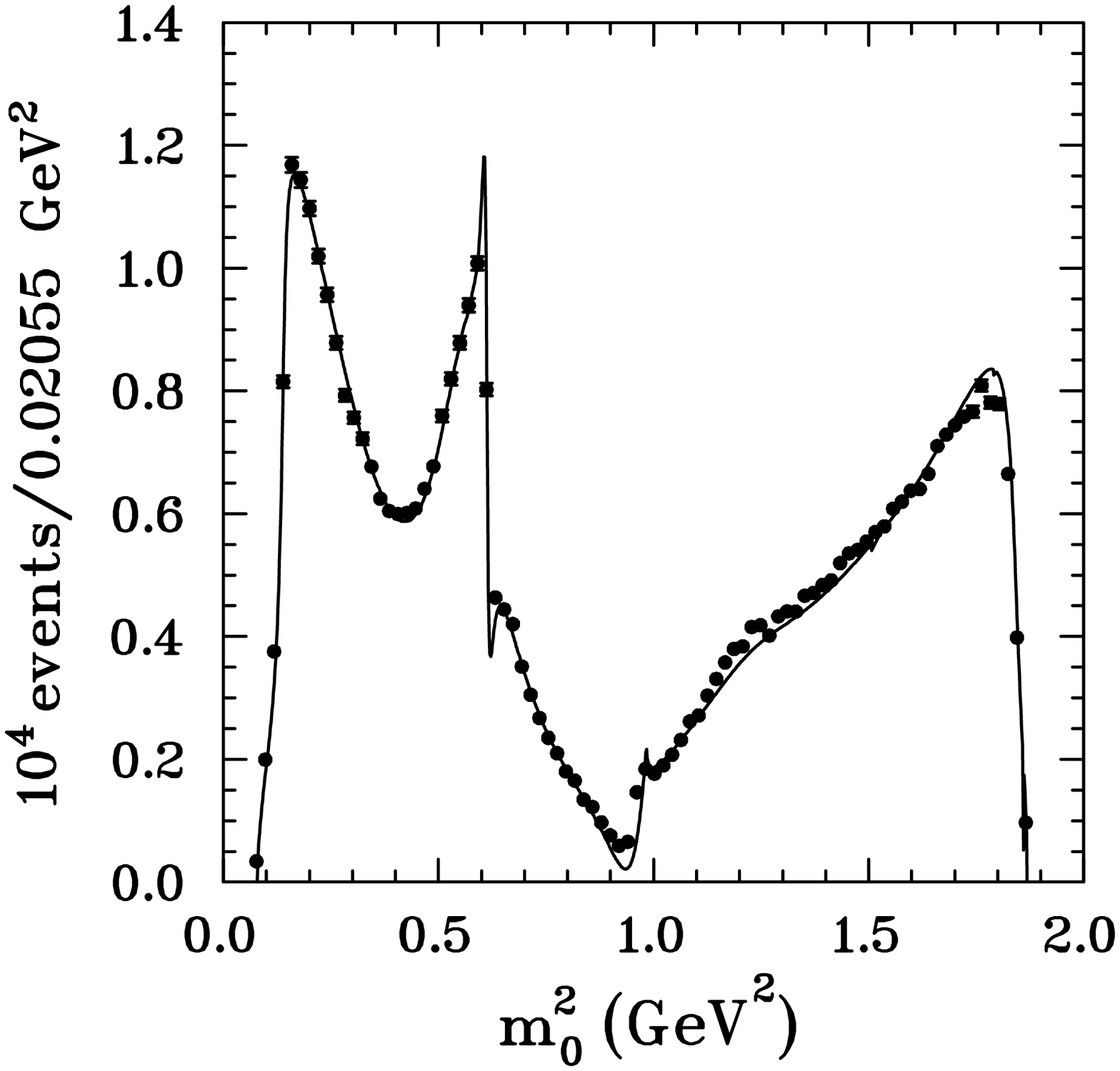}~~
\caption{Left panel: comparison of the $K^0_S\pi^+$ effective mass squared distributions
for the best fit (solid curve) with the Belle data~\cite{A.Poluektov_PRD81_112002_Belle} (points with error bars). 
Right panel: as in left panel but for the $\pi^+\pi^-$ effective mass squared.} \label{Polmplus}
\end{figure}

\begin{figure}
  \includegraphics[height=.35\textheight]{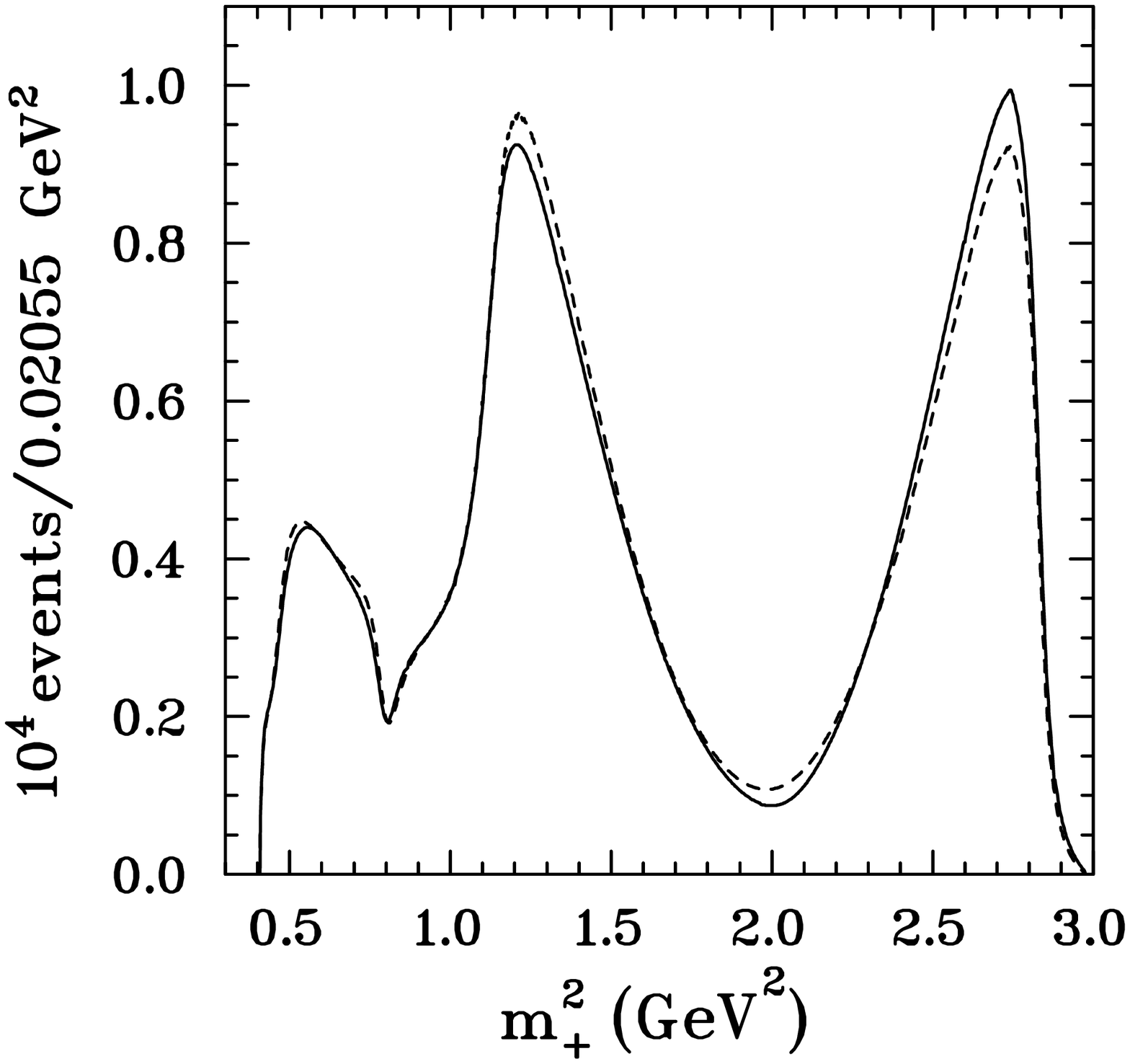}~~~~
   \includegraphics[height=.35\textheight]{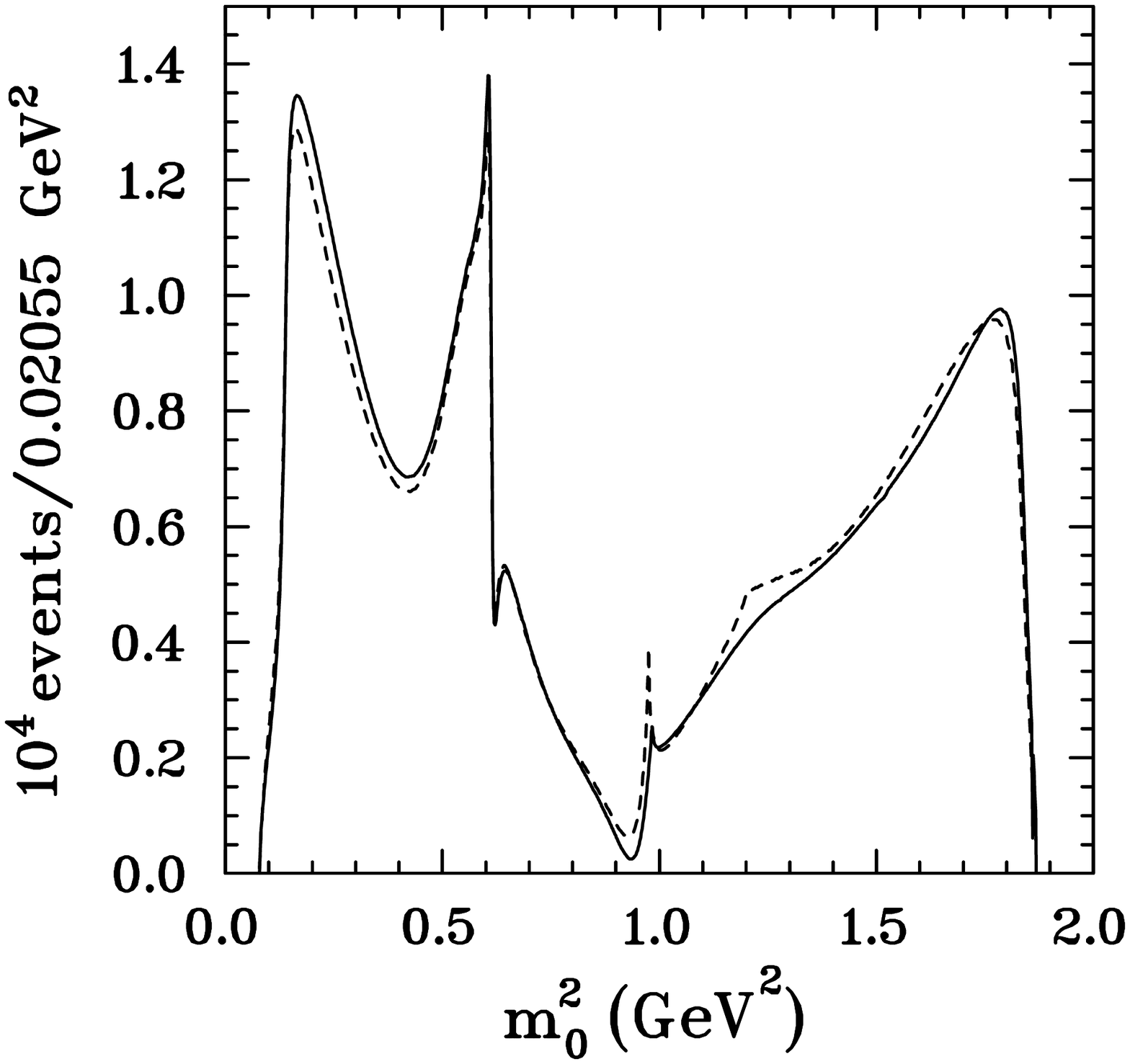}~~~~
\caption{Left panel: comparison of the $K^0_S\pi^+$ effective mass squared distributions
for the best fit (solid curve) with the BABAR model~\cite{F.Martinez_private2013} (dashed curve).
 Right panel: as in left panel but for the $\pi^+\pi^-$ effective mass squared.}\label{Polp0BABAR}
\end{figure}

\vspace{5pt}
Dalitz plot projections or one dimensional effective mass distributions are obtained
by proper integration of the Dalitz plot density distributions. 
They are shown in Figs.~\ref{Polmmin} to~\ref{Polp0BABAR}. 
The experimental $K^0_S\pi^-$ mass distribution in Fig.~\ref{Polmmin}, dominated by the $K^*(892)$
resonance, is well reproduced by our model. 
In the right panel of this figure, where the vertical scale is expanded, some discrepancies above 
2~GeV$^2$ are apparent.
A good agreement between the model and data is seen in the left panel Fig.~\ref{Polmplus}
 showing the $K^0_S\pi^-$ distributions.
The two prominent peaks, together with the minimum separating them, arise from the $K^*(892)^-$ resonance
contribution. The left maximum is mainly associated with the $\rho(770)^0$ while the minimum, in the vicinity of 0.8~GeV$^2$, comes from interferences with the $K^*(892)^+$ resonance. 
The maxima at 1.2 GeV$^2$ and at 2.75 GeV$^2$, and the deep minimum at about 2 GeV$^2$ are due to a typical $P$-wave dependence of the ${\cal M}_3$ amplitude dominated by the $K^*(892)^-$ resonance.

The right panel of Fig.~\ref{Polmplus} shows the very rich structure of the Belle data which is well 
reproduced by our model.  It exhibits clearly the $\pi^+\pi^-$ $S$-, $P$- and $D$-wave resonance effects.
The first peak comes mainly
from the $K^*(892)^+$ and $f_0(500)$, the second one from the $\rho(770)^0$, the strong decrease on its 
right being due
to its interference with the narrow $\omega(782)$, the $f_0(980)$ being responsible for the deep minimum
near~1~GeV$^2$, the $f_2(1270)$ contributes to the rise around~1.5~GeV$^2$,
the right-hand side bump being dominated once more by the $K^*(892)^+$.
 
In Fig.~\ref{Polp0BABAR} our $m_+^2$ and $m_0^2$ distributions are compared with the distributions 
calculated for the BABAR model.
A noticable deviation is seen for values of
 $m_0^2$ around 1.2 GeV$^2$ where the BABAR model shows a shoulder. The corresponding shoulder  
is also observed in the right panel of 
Fig.~\ref{Polmplus} for the Belle data.
To account for the presence of such a structure near~1.2~GeV$^2$, a scalar resonance term called $\sigma_1$, with 
a mass of $(1033\pm7)$ MeV and a width of $(88\pm7)$ MeV, has been introduced in Ref.~\cite{A.Poluektov_PRD81_112002_Belle}.
In Ref.~\cite{SanchezPRL105_081803} 
the K-matrix parametrization of the $\pi\pi$ S-wave state with a coupling to the $\eta\eta$
channel is introduced. The threshold mass squared corresponding to opening of the $\eta\eta$
channel is indeed equal to 1.201 GeV$^2$ and coincides with localization of the structure seen in 
Fig.~\ref{Polp0BABAR}  (dashed line). However, as seen in Fig. 3 of Ref.~\cite{SanchezPRL105_081803} 
this structure is rather wide.
So, on the basis of experimental data for the $m_0^2$ distributions it is difficult to identify clearly the
origin of this rather wide structure seen by both collaborations at 1.2 GeV$^2$. In our pion scalar form 
factor shown in Fig.~\ref{fig7} one does not observe a sharp structure near 1.1 GeV. 
Further studies of different coupled channel production processes are needed to resolve this structure question. 

\vspace{5pt}

\vspace{5pt}

\section{Summary, conclusions and perspectives} \label{concl}

We have used  
the quasi two-body factorization 
to analyze 
the high-statistics data of the $D^0_S \to K^0_S \pi^+\pi^-$ decay process measured by the Belle ~\cite{ZhangPRL99_131803}  and BABAR~\cite{SanchezPRL105_081803} Collaborations. 
The three-meson final states are assumed to be the combinations of a meson pair in $S$-, $P$- and 
$D$-waves and an isolated meson, leading to the quasi two-body channels,
 $[K^0_S \pi^+]_{S, P, D}\ \pi^-$,  $[K^0_S \pi^-]_{S, P, D}\ \pi^+$ and
$K^0_S \ [\pi^+\pi^-]_{S, P, D}$. 
The decay amplitudes, built from the weak effective Hamiltonian, consist of Cabibbo favored
(proportional to $V_{cs}^* V_{ud}$)
 and doubly Cabibbo suppressed (proportional to $V_{cd}^* V_{us}$)
 tree 
and $W$-exchange parts.
All amplitudes are given in terms of superpositions of the effective Wilson coefficients and of product of two
 transition matrix elements.
The CF tree amplitudes are proportional to the product of the pion or kaon decay constant by the transition
 matrix element between the $D^0$ and $[K\pi]_{S,P, D}$ or $[\pi^+ \pi^-]_{S,P, D}$  states, respectively.
One DCS tree amplitude is proportional to the scalar or vector $K\pi$ form factor multiplied by the $D^0$ 
transition to the pion.
The other DCS tree amplitude is proportional to the kaon decay constant times the $D^0$ transition to the
 $[\pi \pi]_{S,P, D}$ states.
The 
W-exchange (or annihilation) amplitudes are proportional to the product of the $D^0$ decay constant by
the form factor of the meson pair transition to a pion or a kaon.

We calculate the different transition matrix elements assuming that the meson pair  involved goes first
 through the dominant intermediate resonance of this pair.
The  $K^*_0(1430)$, $K^*(892)$ and $K^*_2(1430)$ are the dominant resonances for the $S$-, $P$- and 
$D$-waves 
of the $[K\pi]_{S,P, D}$ states, respectively and the $f_0(980)$, $\rho(770)^0$ and $f_2(1270)$ for those 
of the $[\pi \pi]_{S,P, D}$ states.
We then introduce the relevant vertex function to describe the decays of the resonant meson-pair state into the final meson pair.
We further express this vertex function as being proportional to the kaon-pion or pion-pion scalar, vector
 or tensor form factors.
We use the unitary $K \pi$ and $\pi \pi$ scalar form factors calculated with analyticity and chiral symmetry constraints in Ref.~\cite{Bppk} and~\cite{DedonderPol}, respectively.
These functions describe the $K_0^*(800)$, $K^*_0(1430)$ and the $f_0(500)$, $f_0(980)$ and $f_0(1400)$ 
scalar resonances contributions to the $K\pi$ and $\pi \pi$ final state interactions.
The Belle analysis of the $\tau^- \to K^0_S \pi^- \nu_\tau$~\cite{EpifanovPLB654} 
and Hanhart's model~\cite{Hanhart} of the  
$\tau^- \to  \pi^- \pi^0 \nu_\tau$~\cite{Belletau2008} decays yield the vector form factors.
The $D^0 \to \omega(782) [\to \pi^+\pi^-] K^0_S$  decay amplitude is also added. 
The tensor 
vertex functions are parametrized by relativistic Breit-Wigner formulae.

Our 27 non-zero amplitudes are then combined into 10 effective independent amplitudes.
The reduction in the number of effective amplitudes, as compared to the isobar analyses, results from the 
factorization hypothesis. This leads to parametrization in terms
 of transition matrix elements which can be form factors or chosen to be proportional to form factors in 
which resonances are grouped together.

A  $\chi^2$ fit to a Dalitz plot data sample of the Belle Collaboration 
analysis~\cite{A.Poluektov_private2013}
 is performed to determine the 33 free parameters of our $D^0 \to K^0_S \pi^+ \pi^-$ decay amplitude. 
Our  parameters are mainly related to the strength of the $[{K} \pi]_{S}$ and $[\pi \pi]_{S}$ scalar form 
factors and to the unknown meson to meson transition form factors at a large momentum transfer squared
equal to $m_{D^0}^2$.

The fit to the data 
is very sensitive to the values of the mass and width of the $K^*(892)$ resonance.
We include them in the fit, performing a combined analysis of the $D^0 \to K^0_S \pi^+ \pi^-$ and 
$\tau^- \to K^0_S \pi^- \nu_\tau$ decay data.
The total experimental branching fraction is also fitted.
An overall good fit, with a  $\chi^2/ndf=1.48$ for a number of degree of freedom, $ndf=6378$, is carried 
out.
Another set of amplitudes fits the  BABAR Collaboration Dalitz plot model of 
Ref.~\cite{F.Martinez_private2013} with a  $\chi^2/ndf=0.91$ for $ndf=7343$.
The parameters of both fits are close, which indicates similar Dalitz density distribution measurements 
for both collaborations.

The Dalitz plot distribution of our fit to the Belle data~\cite{A.Poluektov_PRD81_112002_Belle} 
exhibits a very rich interference pattern governed by the $K^*(892)^-$ resonance.
A good overall agreement  with the experimental density distribution of 
Ref.~\cite{A.Poluektov_PRD81_112002_Belle} has been achieved.
The corresponding  one dimensional effective mass distributions compare well those of 
Belle~\cite{A.Poluektov_PRD81_112002_Belle} or BABAR~\cite{F.Martinez_private2013} and show the 
contributions of the different $K\pi$ [$K^*_0(800)$, $K^*(892), K_0^*(1430)$] and $\pi \pi$ [$f_0(500), f_0(980),  \rho(770)^0, \omega(782), f_2(1270)$] resonances and of their interferences.
The small bulge in the slope of the $\pi^+ \pi^-$ effective mass squared distribution seen in the Belle
 and BABAR 
data at 1.2~GeV$^2$ might be associated with the coupling of the $\pi \pi$ channel to the $\eta \eta$ one.
Our model, which does not include this coupling, does not exhibit such a behavior. 
Investigations on this matter would be worthwhile.

The branching fraction calculations show the dominance of the quasi two-body channel
$[K^0_S \,\pi^-]_P \,\pi^+$ with a branching fraction Br = ($62.7\pm4.5$)~\% 
close to the values found in the isobar Belle~\cite{ZhangPRL99_131803} or BABAR~\cite{SanchezPRL105_081803}
 models for the $K^*(892)^- \pi^+$ amplitude.
The next important contributions come from the  $[K^0_S \,\pi^-]_S \,\pi^+$  amplitude with a Br of 
(25.60$\pm$3.6)~\%, from the  $K^0_S \ [\pi^-\pi^+]_P$ one, with a Br of (22.0$\pm$1.6)~\% and from the 
$K^0_S \ [\pi^-\pi^+]_S$ one with a Br of (16.9$\pm$1.3)~\%.
Branching fractions for the other amplitudes, $K^0_S \ [\pi^-\pi^+]_\omega$, 
$[K^0_S \,\pi^-]_D\, \ \pi^+$,  $K^0_S \ [\pi^-\pi^+]_D$, $[K^0_S \,\pi^+]_S \,\pi^-$, 
$[K^0_S \,\pi^+]_P \,\pi^-$ and $[K^0_S \,\pi^+]_D \,\pi^-$ are small.
The importance  of the interference contributions \mbox{(-32.8~\%)} is seen in the fact that the total sum of all the branching fractions is larger than $100$~\%.

The branching fractions corresponding to the quasi two-body channel tree amplitudes  give sizable contributions.
The knowledge of the branching fractions does not allow to calculate all phases of our amplitudes, as it is the modulus square of the amplitudes which appears in the branching fraction formula.
One of the phases of our 10 amplitudes cannot be determined.
We proceed as in the isobar model analysis in requiring the phase  of the term multiplying  the pion vector
 form factor  in the $K^0_S \ [\pi^-\pi^+]_P$ amplitude to be zero. 
Consequently we can predict only lower or upper limits of the branching fraction contributions of the annihilation amplitudes.
We find that these lower limits can be sizable for the important quasi two-body channels, 
$[K^0_S \,\pi^-]_P \,\pi^+$,   $[K^0_S \,\pi^-]_S \,\pi^+$~, $K^0_S \ [\pi^-\pi^+]_P$  and 
$K^0_S \ [\pi^-\pi^+]_S$  and we can say that, compared to the tree amplitudes, the annihilation ones 
have a significant contribution.
The analyses of the two-body hadronic decays of $D$ and $D_s$ mesons in 
Refs.~\cite{ Fu-Sheng_Yu_PRD84_074019}, \cite{ChengPRD81_074021} and~\cite{ChengPRD81_074031} have also
 pointed out the importance of the annihilation diagrams.

As we do not know the $\overline K_0$ to $\rho(770)^0$ transition form factor value at the $D^0$ mass 
squared, our fit cannot be used to estimate the physical unknown $\pi$ or $K$ meson to $K \pi$ or $\pi \pi$ meson pair transition form factors entering the annihilation amplitudes.
The full knowledge of the strong  interaction meson-meson form factors can be obtained only if the strong meson-meson interaction is known at all
 energies~\cite{Barton65}.
Consequently  some information on the $\overline K_0\  \rho(770)^0$ strong interaction would be required to estimate  the $\overline K_0$ to $\rho(770)^0$ transition form factor.
It would be of interest if the unknown form factors entering the present model could be evaluated.

\vspace{0.5cm}
\textit{Concluding remarks and perspectives}
\vspace{0.5cm}

In our quasi two-body factorization approach the $CP$ asymmetry, proportional to the very small imaginary part
 of $V_{cd}^* V_{us}$, is found to be of the order of $10^{-4}$.
This is in agreement with present observations~\cite{D.M.AsnerPRD70_0911018CLEO, T.Aaltonen_CDF2012} 
and 
values predicted by the standard model in the charm sector.
Our $D^0 \to K^0_S \pi^+ \pi^-$ decay amplitudes could be useful input for calculations of
 $D^0$-$\overline{D}^0$ mixing~\cite{ZhangPRL99_131803,SanchezPRL105_081803} and 
determination~\cite{J.Libby_PRD82_CLEO} -~\cite{ A.Poluektov_PRD81_112002_Belle}
 of the CKM angle $\gamma$ (or $\phi_3$). Upon request we can provide numerical values of our 
amplitudes.
The kaon-pion and pion-pion scalar form factors, entering our quasi two-body factorization decay 
amplitude
     and built using other experimental data, are constrained by the present Dalitz plot analysis of 
the the weak process $D^0 \to K^0_S \pi^+ \pi^-$.
In principle our analysis could  also give constraints on $\pi K$ and $\pi \pi$ tensor resonances.
There have been recent observations (see \textit{e.g.} Refs.~\cite{D*PRD82, LHCbDK})
of $D$ and $D_s$ excited states which can be formed due to
 the $\pi D$ and $K D$ strong interactions, respectively.
Their properties could be used to constrain theoretical $\pi D$ and $K D$ scattering models and
possibly also $\pi D$ and $K D$ transition form factors.

Taking advantage of the coupling between the $\pi \pi$ and  the $K K$ channels and extending the derivation
 of the unitary pion form factor~\cite{DedonderPol} to that of the kaon, two of the present authors, LL and
 RK, together with two collaborators, have recently studied, in the quasi two-body QCD factorization 
approach, the $B^\pm \to K^+K^- K^\pm$ decays~\cite{PLB699_102}. 
We could also extend our present work to study,  in the quasi two-body factorization framework, 
the $D^0 \to K_S^0 K^+K^-$ data analysed by the BABAR~\cite{SanchezPRL105_081803}, 
CLEO~\cite{J.Libby_PRD82_CLEO},  and, more recently, by the LHCb~\cite{R.Aaij_LHCb2012} Collaborations.
A good knowledge of the  $D^0 \to K_S^0 K^+K^-$ decay amplitudes will also help in the determinations of  
the $D^0$-$\overline{D}^0$ mixing~\cite{SanchezPRL105_081803} and of the the CKM angle
 $\gamma$~\cite{J.Libby_PRD82_CLEO, R.Aaij_LHCb2012}.
 
 \vspace{3cm}
Acknowledgments
\\
 
 We are deeply indebted to Anton Poluektov from the Belle Collaboration and Fernando 
Martinez-Vidal from the BABAR Collaboration who provided vital information for this study. We thank 
them for many fruitful exchanges. Anze Zupanc must be thanked for useful exchanges about the Belle
 data. We appreciate the help of Bachir Moussallam who supplied various numerical tables for 
the $K \pi$ scalar form factors used in this work. 
We would like to thank Christoph Hanhart for sending us tables of the pion vector
 form factor. The authors are obliged to Diogo Boito for useful correspondence and
sending numerical values of his $K\pi$ vector form factor. 
We also thank Agnieszka Furman for her contribution in an early stage of this work. 
Fruitful discussions with Pascal David are 
gratefully recognized. This work has been partially supported by a grant from the French-Polish 
exchange program COPIN/CNRS-IN2P3, collaboration 08-127.\\

\newpage

\appendix

\section{On kinematics}
\label{kine}

In this Appendix, we recall some kinematic formulae useful for the calculation of our amplitudes.
These kinematic expressions can also be found in the Appendix {\bf A} of Ref.~\cite{Bonvicini}. 
For the $[K^0_S\pi^-]_L\  \pi^+$ amplitudes, in the $[K^0_S \pi^-]$ center of mass system 
defined by ${\bf p}_0 + {\bf p}_- = 0$, using Eqs.~(\ref{3sa}) and (\ref{pD0}), one finds,
 
\be \label{cmkpimp0}
 \bf{p_{1}} =\bf{p_{0}} =- \bf{p_{-}}~~~~ \rm{and} ~~~~ \vert {\bf p}_1\vert 
 = \frac{\sqrt{[s_- - (m_{K^0} + m_{\pi})^2]\  
[s_- - (m_{K^0} - m_{\pi})^2]}}{2\ m_-}
\ee
and

\be  \label{cmkpimpp}
 \vert {\bf p}_+\vert = \frac{ \sqrt{[m_{D^0}^2 - (m_- + m_{\pi})^2]\  [m_{D^0}^2 - 
(m_- - m_{\pi})^2]}} {2\ m_-}.
\ee
From Eqs.~(\ref{3sa}) one obtains, 

\be  \label{cmkpimpopp}
4 \  {\bf p}_1 \cdot {\bf p}_+=  s_0 - s_+ + \frac{(m_{D^0}^2 - m_{\pi}^2) \ 
(m_{K^0}^2 - m_{\pi}^2)}{s_-},
\ee
factor which enters the $[\overline{K}^0 \pi^-]_P \ \pi^+$ amplitude, Eq.~(\ref{A1P}).

In the $[\pi^+\pi^-]$ center of mass system, defined by ${\bf p}_+ + {\bf p}_- = 0$, one has

\be \label{cmpipipp}
{\bf p_{2}} =\bf{p_{+}} = - {\bf p_{-}}~~~~~ \rm{and}~~~~~~ 
\vert {\bf p_{2}}\vert =\frac{1}{2} \  \sqrt{s_0 - 4 \ m_{\pi}^2}\ee
and
\be \label{cmpipipo} 
\vert {\bf p}_0 \vert = \frac{\sqrt{[m_{D^0}^2 - (m_0 + m_{K^0})^2] \ [m_{D^0}^2 - 
(m_0 - m_{K^0})^2]}}{2 \ m_0}.
\ee
The scalar product  $ {\bf p}_2 \cdot {\bf p}_0 $,  given by 
\be \label{cmpipipopp}
4\ {\bf p}_2 \cdot {\bf p}_0   =  s_- - s_+,
\ee
enters the $\overline{K}^0 [\pi^+ \pi^-]_P$ amplitude, Eq.~(\ref{A2Prho}).

The analogous formulae for the $[K^0_S\pi^+]_L \ \pi^-$ amplitudes, in the $[K^0_S\pi^+]$ center of mass
 system, are obtained by exchanging subscripts $-$ and $+$ in Eqs.~(\ref{cmkpimp0}), (\ref{cmkpimpp})
 and~(\ref{cmkpimpopp}). 
Then $\bf{p_{1}}$ becomes $\bf{p_{3}}$ and $\bf{p_{+}}$ is changed into $\bf{p_{-}}$
[see \textit{e.g.} the corresponding $[{K}^0\pi^+]_P \ \pi^-$ amplitude, Eq.~(\ref{A1PDCS})].


\begin{thebibliography}{99}
\bibitem{Moussallam_private2013} 
   Bachir Moussallam has provided us with numerical values of the kaon-pion scalar form factor for 
various values of the ratio $f_K/f_{\pi}$.
\end{thebibliography}
\end{document}